\documentclass[twocolumn]{aastex63}

\submitjournal{ApJ}

\shorttitle{KN Energy injection}
\shortauthors{Fraija et al.}
\usepackage{epstopdf}
\usepackage{ulem}
\usepackage{amssymb}
\usepackage{times}
\usepackage{graphicx}
\usepackage[usenames,dvipsnames]{pstricks}
\usepackage{psfrag}
\usepackage{lipsum}
\usepackage{natbib}
\usepackage{textcase}

\def\apjl{ApJ}
\def\apjs{ApJS}
\def\ao{Appl.~Opt.}
\def\aap{A\&A}

\newcommand{\be}{\begin{equation}}
\newcommand{\ee}{\end{equation}}
\newcommand{\bary}{\begin{eqnarray}}
\newcommand{\eary}{\end{eqnarray}}

\DeclareUnicodeCharacter{2212}{-}

\begin{document}


\title{GRB Afterglow of the Sub-relativistic Materials with Energy Injection}

\correspondingauthor{Fraija, N.}
\email{nifraija@astro.unam.mx}

\author[0000-0002-0173-6453]{N. Fraija}
\affiliation{Instituto de Astronom\' ia, Universidad Nacional Aut\'onoma de M\'exico,\\ Circuito Exterior, C.U., A. Postal 70-264, 04510 M\'exico City, M\'exico}

\author{B. Betancourt Kamenetskaia}
\affiliation{Instituto de Astronom\' ia, Universidad Nacional Aut\'onoma de M\'exico,\\ Circuito Exterior, C.U., A. Postal 70-264, 04510 M\'exico City, M\'exico}
\affiliation{TUM Physics Department, Technical University of Munich, James-Franck-Str, 85748 Garching, Germany}

\author{A. Galvan-Gamez}
\affiliation{Instituto de Astronom\' ia, Universidad Nacional Aut\'onoma de M\'exico,\\ Circuito Exterior, C.U., A. Postal 70-264, 04510 M\'exico City, M\'exico}

\author{M. G. Dainotti}
\affiliation{National Astronomical Observatory of Japan, Division of Science, Mitaka, 2-chome}
\affiliation{Space Science Institute, Boulder, Colorado}

\author{R.~L.~Becerra}
\affiliation{Instituto de Ciencias Nucleares, Universidad Nacional Aut\'onoma de M\'exico, Apartado Postal 70-264, 04510 M\'exico, CDMX, Mexico}

\author{S. Dichiara}
\affiliation{Department of Astronomy, University of Maryland, College Park, MD 20742-4111, USA}
\affiliation{Astrophysics Science Division, NASA Goddard Space Flight Center, 8800 Greenbelt Rd, Greenbelt, MD 20771, USA}

\author{P. Veres}
\affiliation{Center for Space Plasma and Aeronomic Research (CSPAR), University of Alabama in Huntsville, Huntsville, AL 35899, USA}

\author{A. C. Caligula do E. S. Pedreira}
\affiliation{Instituto de Astronom\' ia, Universidad Nacional Aut\'onoma de M\'exico,\\ Circuito Exterior, C.U., A. Postal 70-264, 04510 M\'exico City, M\'exico}

\begin{abstract}
Sub-relativistic materials launched during the merger of binary compact objects and the core-collapse of massive stars acquire velocity structures when expanding in a stratified environment.  The remnant (either a spinning magnetized neutron star (NS) or a central black hole) from the compact-object or core-collapse could additionally inject energy into the afterglow via spin-down luminosity or/and by accreting fall-back material, producing a refreshed shock, modifying the dynamics, and leading to rich radiation signatures at distinct timescales and energy bands with contrasting intensities.  We derive the synchrotron light curves evolving in a stratified environment when a power-law velocity distribution parametrizes the energy of the shock, and the remnant continuously injects energy into the blastwave.  As the most relevant case, we describe the latest multi-wavelength afterglow observations ($\gtrsim 900$ days) of the GW170817/GRB 170817A event via a synchrotron afterglow model with energy injection of a sub-relativistic material.  The features of the remnant and the synchrotron emission of the sub-relativistic material are consistent with a spinning magnetized NS and  the faster ``blue" kilonova afterglow, respectively. Using the multi-band observations of some short-bursts with evidence of kilonova, we provide constraints on the expected afterglow emission.

\end{abstract}

\keywords{Gravitational wave Astronomy --- Compact binary stars  --- Non-thermal radiation sources --- Gamma-rays bursts}

\section{Introduction} \label{sec:intro}

Gamma-ray bursts (GRBs) are among the most powerful gamma-ray sources in the Universe. They could be generated from the merger of binary compact objects \citep[BCOs;][]{1992ApJ...392L...9D, 1992Natur.357..472U, 1994MNRAS.270..480T, 2011MNRAS.413.2031M} or the death of massive stars \citep{1993ApJ...405..273W,1998ApJ...494L..45P, Woosley2006ARA&A}.   The merger of BCOs; a black hole (BH) - a neutron star (NS) or NS-NS, leading to kilonovae (KNe), is correlated with short-duration gamma-ray bursts \citep[sGRBs; $T_{90}\footnote{$T_{90}$ is defined as the time during which the cumulative number of collected counts above background rises from 5\% to 95\%.}\lesssim 2\,{\rm s}$;][]{1998ApJ...507L..59L, 2005ApJ...634.1202R, 2010MNRAS.406.2650M, 2013ApJ...774...25K, 2017LRR....20....3M}.  On the other hand, long-duration gamma-ray bursts \citep[lGRBs; $T_{90}\gtrsim 2\,{\rm s}$;][]{1993ApJ...413L.101K} are associated with the core-collapse (CC) of dying massive stars \citep{1993ApJ...405..273W, 1998Natur.395..670G} leading to supernovae \citep[SNe;][]{1999Natur.401..453B, 2006ARA&A..44..507W}.  It is believed that in both scenarios large quantities of materials with a wide range of velocities are ejected. In the framework of CC-SNe (depending on the type of SN association), several materials ejected with sub-relativistic velocities less than $\beta\lesssim0.4$\footnote{Hereafter, we adopt natural units $c=\hbar=1$.} have been reported \citep[e.g., see][]{2020ApJ...892..153M, 2020arXiv200405941I, 2020NatAs.tmp...78N, 2019Natur.565..324I, 2017hsn..book..195G,2008MNRAS.383.1485V, 1998Natur.395..663K,1999Natur.401..453B, 2006ARA&A..44..507W}. Regarding the merger of two NSs, sub-relativistic materials such as the cocoon, the shock breakout, the dynamical and the wind ejecta are launched with velocities in the range $0.03\lesssim \beta\lesssim 0.8$\footnote{Some authors have considered the shock breakout material in the sub-, trans- and ultra-relativitic regimes \citep[e.g., see][]{2014MNRAS.437L...6K, 2015MNRAS.446.1115M, 2019ApJ...871..200F}.} \citep[e.g., see][]{2009ApJ...690.1681D, 2014MNRAS.441.3444M, 2015MNRAS.446..750F,2014MNRAS.437L...6K, 2015MNRAS.446.1115M, 2014ApJ...784L..28N, 2014ApJ...788L...8M,2017ApJ...848L...6L, 2018PhRvL.120x1103L, 2011ApJ...738L..32G, 2013ApJ...778L..16H,2013ApJ...773...78B, 2014ApJ...789L..39W}.  While the mass and velocity inferred for the first GRB/KN association\footnote{GRB 170817A/AT2017gfo} was $M_{\rm ej}\approx(10^{-4} - 10^{-2})\,M_{\odot}$ and $\beta\approx (0.1- 0.3)$, respectively \citep{2017Sci...358.1556C, 2017Natur.551...64A, 2017ApJ...848L..17C, 2017ApJ...848L..18N, 2019LRR....23....1M}, the mass and velocity inferred for the first GRB/SN association\footnote{GRB 980425/SN1998bw} was $M_{\rm ej}\approx 10^{-5}\,M_{\rm \odot}$ and $\beta\approx (0.2-0.3)$, respectively \citep{1998Natur.395..663K}.\\

In the sub-relativistic regime, the interaction of the decelerated material with the surrounding circumburst medium has been explored to interpret the multi-wavelength  observations in timescales from days to several years as synchrotron afterglow models \citep[e.g., see][]{1997MNRAS.288L..51W, 1999ApJ...519L.155D, 1999MNRAS.309..513H, 2000ApJ...538..187L, 2003MNRAS.341..263H, 2013ApJ...778..107S, 2015MNRAS.454.1711B}. In most of the cases, the isotropic-equivalent kinetic energy of the materials launched during the coalescence of BCOs and the CC-SNe has been described by a power-law (PL) velocity distribution \citep[e.g., see][and the references therein]{2001ApJ...551..946T}.

The canonical X-ray light curve exhibits a typical shape that consists of four distinct PL segments $\propto t^{-\alpha}$ with a great flare \citep[e.g., see][]{2006ApJ...642..354Z, 2006ApJ...642..389N}.   The initial steep decay with a temporal decay index $3\lesssim\alpha\lesssim\,$5, the normal decay phase with $0.9\lesssim\alpha\lesssim$1.5 and the late abrupt decay with $1.9\lesssim\alpha\lesssim\,$2.4 have been explained in terms of the end of main episode, the standard synchrotron forward-shock model \citep{1998ApJ...497L..17S} and the post-jet-break decay phase \citep{2006ApJ...638..920V}, respectively. There is, however, one segment that occurs between the end of the prompt phase and the normal decay, a shallower than usual decay with  $-0.1\lesssim\alpha\lesssim$0.7.  This so-called ``plateau" phase has been explained in several scenarios such as continuous energy injection from the central engine (either a spinning magnetized NS or a central BH) into the blastwave \citep{2005ApJ...635L.133B,  2005ApJ...630L.113K, 2006Sci...311.1127D, 2006ApJ...636L..29P, 2006MNRAS.370L..61P, 2005Sci...309.1833B, 2007ApJ...671.1903C, 2017MNRAS.464.4399D, 2019ApJ...872..118B, 2019ApJ...887..254B}, stratified ejecta \citep{2006ApJ...640L.139T, 2007ApJ...656L..57J, 2017MNRAS.472L..94H}, ejecta with a wide range of Lorentz factors \citep{1998ApJ...496L...1R,2000ApJ...532..286K, 2000ApJ...535L..33S, 2002ApJ...566..712Z, 2019ApJ...871..200F} and  variation on microphysical parameters \citep{2006MNRAS.369..197F, 2006A&A...458....7I, 2020ApJ...905..112F}.  Several modeling efforts of multi-wavelength afterglows evoking energy injection by central engine have been widely explored \citep[e.g., see][]{2015ApJ...814....1L, 2021ApJ...906...60Z, 2019ApJ...872..118B, 2022MNRAS.511.6205P, 2021ApJ...918...12F}. For instance, \cite{2015ApJ...814....1L} described GRB 100418A, GRB 100901A, GRB 120326A and GRB 120404A and found that 
the majority of the kinetic energy of the relativistic jet in each burst was carried by slow-moving ejecta, thus indicating a correlation between the injection rates and the Lorentz factor distribution.\\

On August 17, 2017, a gravitational wave (GW) event \citep[GW170817;][]{PhysRevLett.119.161101,2041-8205-848-2-L12} was linked with a faint gamma-ray prompt emission of GRB 170817A \citep{2017ApJ...848L..14G, 2017ApJ...848L..15S}. Immediately,  GRB 170817A was followed by an extensive observational campaign covering radio, optical, and X-ray bands \citep[e.g., see][and references therein]{troja2017a, 2041-8205-848-2-L12, 2017arXiv171100243K, 2018arXiv180106164D}. The observations of the non-thermal spectrum of GRB 170817A gathered during the first $\approx 900$ days after the initial merger were analyzed by several authors. It was shown that they were consistent with synchrotron forward-shock emission generated by the deceleration of an off-axis structured jet with an opening angle $\theta_{j}\approx5^{\circ}$ that was observed from a viewing angle in the range of $15^{\circ}\leq \theta_{\mathrm{obs}}\leq 25^{\circ}$  \citep{troja2017a, 2017Sci...358.1559K, 2017MNRAS.472.4953L, 2018MNRAS.478..733L, 2018ApJ...867...57R, 2017ApJ...848L..20M, 2017ApJ...848L...6L, 2018MNRAS.479..588G, 2019ApJ...884...71F, 2018MNRAS.479..588G, 2018ApJ...867...95H, 2019ApJ...871..200F}. In some proposed models, the off-axis structured jet is formed with an off-axis jet with a cocoon \citep{2017ApJ...848L...6L, 2018MNRAS.479..588G, 2019ApJ...884...71F} and a shock breakout \citep{2018MNRAS.479..588G, 2018ApJ...867...95H, 2019ApJ...871..200F,2021MNRAS.503.4363U}. Recently, \cite{2021arXiv210402070H} analyzed the latest X-ray and radio observations of GRB 170817A collected with the Chandra X-ray Observatory, the Very Large Array (VLA), and the MeerKAT radio interferometer about 3.3 years after the initial merger. These new observations did not agree with the best-fit synchrotron curves from the off-axis jet model, thus reporting evidence of a new X-ray emission component. Given these contrasting observations, the authors offered the solution to explain this phenomenon in the context of either radiation from accretion processes on the compact-object remnant or a KN afterglow.

The study of properties of KNe has a great impact for the present-day field of study, especially given the link between GWs, short-duration gamma-rays, and KN emission. The merger of two NS associated with GW170817,  GRB 170817A, and AT 2017gfo have provided the needed tools to predict the KN emission and its characteristics. While the prompt episode and the early afterglow are produced in internal and external shocks by an ultra-relativistic and extremely collimated jet, the KN transient is associated with a quasi-isotropic emission easier to detect at angles far away from those emitted from a collimated jet \citep{2012ApJ...746...48M}. Despite the advantageous prospects for detection, only four transient events with different brightness to AT 2017gf have been classified as KNe. They are associated to  GRB 050709 \citep{2016NatCo...712898J}, GRB 060614 \citep{2015NatCo...6.7323Y}, GRB 130603B \citep{2013Natur.500..547T, 2013ApJ...774L..23B} and GRB 160821B \citep{2017ApJ...843L..34K, 2019MNRAS.489.2104T}.


Recently, \cite{2021ApJ...907...78F} presented the afterglow light curves generated by the deceleration of sub-relativistic masses ejected from the merger of BCOs and the death of massive stars. The authors assumed that a PL velocity distribution describes the isotropic-equivalent kinetic energy of these masses and that the sub-relativistic ejected masses were decelerated, in turn, by a stratified-density environment.  As a particular case, to explain the multi-wavelength observations of the gravitational event GW170817/GRB 170817A at $\sim 900\,{\rm days}$, they constrained the parameter space of the synchrotron light curves of a sub-relativistic mass ejected during the merger of two NSs and decelerated in a constant-density environment.  The synchrotron radiation of the sub-relativistic material was consistent with the faster ``blue" KN afterglow.    Inspired by the new observations of this GW event at 3.3 years after the initial merger \citep{2021arXiv210402070H}, in this paper, we extend the synchrotron model presented in \cite{2021ApJ...907...78F} including the continuous energy injection from the central engine (either a spinning magnetized NS or BH remnant) into the blastwave through a numerical approach and analytic arguments. In addition, we apply the current model to potential candidates of sGRB events  with evidence of a KN.   The paper is organized as follows: Section 2 presents the dynamical evolution of the afterglow when the central engine continuously injects energy into the blastwave. We show an analytical solution and numerical approach. In Section 3, we show a synchrotron model with energy injection from a spinning magnetized NS and BH remnants.  Section 4 shows the analysis of the multi-wavelength light curves using typical values of the GRB afterglow.   In Section 5, we apply our model to several potential candidates including GW170817/GRB 170817A, and finally, in Section 6, we summarize.  We consider the convention $Q_{\rm x}=\frac{Q}{10^{\rm x}}$ in c.g.s. units and assume  for the cosmological constants a spatially flat universe $\Lambda$CDM model with  $H_0=69.6\,{\rm km\,s^{-1}\,Mpc^{-1}}$, $\Omega_{\rm M}=0.286$ and  $\Omega_\Lambda=0.714$ \citep{2016A&A...594A..13P}.\\

\section{Afterglow light curves with energy injection} \label{sec:LC}

\subsection{Synchrotron radiation}
We consider electrons accelerated in the forward shock which evolves in a stratified external environment with a density profile described by $\rho(r) =A_{\rm k}\, r^{\rm -k}= \frac{\dot{M}_{\rm W}}{4\pi v_{\rm W}}\, r^{\rm -k}$, where $\dot{M}_{\rm W}$ is the mass-loss rate and $v_{\rm W}$ is the wind velocity. There are two common choices of stratification, the value of ${\rm k=0}$ corresponds to the constant-density medium ($A_{\rm 0}=n$), and ${\rm k = 2}$ is associated to the density of the stellar wind ejected by its progenitor ($A_{\rm 2}\simeq A_{\rm W}\,3\times 10^{35}{\rm cm}^{-1}$) with $A_{\rm W}$ the density parameter. We assume that the shocked-accelerated electrons in the forward shocks can be described by a single PL energy distribution $\frac{dn}{d\gamma_e}\propto \gamma_e^{-p}$ for $\gamma_{\rm e}\geq \gamma_{\rm m}$ with $p$ the spectral index and $\gamma_{\rm m}$ the Lorentz factor of the lowest-energy electrons.\\

The post-shock magnetic field evolves as {\small $B'\propto A^\frac12_{\rm k}$ $\beta^{\frac{2-k}{2}}\,t^{-\frac{k}{2}}$}.\footnote{Hereafter, we use prime and unprimed quantities for the comoving and observer frames, respectively.}   The Lorentz factors of the lowest-energy electrons and of the higher energy electrons, which are efficiently cooled by synchrotron emission evolve as {\small $\gamma_{\rm m}\propto  \,\beta^2$} and {\small $\gamma_{\rm c}\propto\,A^{-1}_{\rm k} \beta^{k-2}\,t^{k-1}$}, respectively.  It is worth noting that due to the synchrotron process, the effect that $\gamma_{\rm c}$ has on the electron energy distribution is to introduce a break.   Given the evolution of the synchrotron frequency and the electron Lorentz factors, the corresponding spectral breaks can be written as   {\small $\nu_{\rm m}\propto A^\frac{1}{2}_{\rm k} \beta^\frac{10-k}{2}\,t^{-\frac{k}{2}}$} and {\small $\nu_{\rm c}\propto A^{-\frac32}_{\rm k} \beta^{\frac{3k-6}{2}}\,t^{\frac{3k-4}{2}}$}.  The terms $\nu_{\rm m}$ and $\nu_{\rm c}$ correspond to the characteristic and cooling spectral breaks, respectively.  For $\nu_{\rm c} < \nu_{\rm m}$ the synchrotron spectrum lies in the fast-cooling regime and for $\nu_{\rm m} < \nu_{\rm c}$ this spectrum lies in the slow-cooling regime. In the self-absorption regime, the synchrotron spectral breaks evolve as  {\small $\nu_{\rm a,1}\propto A^{\frac45}_{\rm k} \beta^{-\frac{4k+5}{5}}\,t^{\frac{3-4k}{5}}$} for {\small $\nu_{\rm a,1}\leq \nu_{\rm m} \leq \nu_{\rm c}$},  {\small $\nu_{\rm a,2}\propto  A^{\frac{p+6}{2(p+4)}}_{\rm k}\beta^{\frac{10p-kp-6k}{2(p+4)} }\,t^{\frac{4-kp-6k}{2(p+4)}}$} for 
{\small $\nu_{\rm m} \leq\nu_{\rm a,2}\leq \nu_{\rm c}$, and {\small $\nu_{\rm a,3}\propto  A^{\frac95}_{\rm k} \beta^{\frac{15-9k}{5}}\,t^{\frac{8-9k}{5}}$} for {\small $\nu_{\rm a,3}\leq \nu_{\rm c} \leq  \nu_{\rm m}$}.  Taking into account that the peak spectral power evolves as {\small $P_{\rm \nu, max} \propto \,A^\frac12_{\rm k} \beta^{\frac{2-k}{2}}\,t^{-\frac{k}{2}}$}  and that the number of swept-up electrons in the post-shock develop as {\small $N_{\rm e}\propto A_{\rm k}\beta^{3-k}\, t^{3-k}$},  the spectral peak flux density varies as {\small $F_{\rm \nu,max}\propto  A^{\frac32}_{\rm k}\, \beta^{\frac{8-3k}{2}}\,t^{\frac{3(2-k)}{2}}$} \citep[details of the derivation are explicitly written in][]{2021ApJ...907...78F}.\\

Given the synchrotron spectral breaks and the maximum flux density,  the synchrotron light curves in each cooling condition evolve as
{\small
\begin{eqnarray}
\label{fc_coast}
F_{\rm \nu}\propto \cases{ 
A_{k}^{-1}\beta^{k}t^{1+k}\, \nu^{2},\hspace{2.9cm} \nu<\nu_{\rm a,3}, \cr
A_{k}^2\beta^{5-2k}t^{\frac{11-6k}{3}}\, \nu^{\frac13},\hspace{2.2cm} \nu_{\rm a,3}<\nu<\nu_{\rm c}, \cr
A_{k}^{\frac34}\beta^{\frac{10-3k}{4}}t^{\frac{8-3k} {4}}\, \nu^{-\frac{1}{2}},\hspace{1.9cm} \nu_{\rm c}<\nu<\nu_{\rm m},\,\,\,\,\, \cr
A_{k}^{\frac{p+2}{4}}\beta^{\frac{10p-k(p+2)}{4}}t^{\frac{8-k(p+2)} {4}}\,\nu^{-\frac{p}{2}},\,\,\,\,\hspace{0.2cm}   \nu_{\rm m}<\nu\,, \cr
}
\end{eqnarray}
}
{\small
\begin{eqnarray}
\label{sc_coast1}
F_{\rm \nu}\propto \cases{ 
A_{k}^{0}\beta^{4}t^{2}\, \nu^{2},\hspace{4.2cm} \nu<\nu_{\rm a,1}, \cr
A_{k}^{\frac43}\beta^{\frac{7-4k}{3}}t^{\frac{9-4k}{3}}\, \nu^{\frac13},\hspace{2.9cm} \nu_{\rm a,1} <\nu<\nu_{\rm m}, \cr
A_{k}^{\frac{p+5}{4}}\beta^{\frac{2(3+5p)-k(p+5)}{4}}t^{\frac{12-k(p+5)} {4}}\, \nu^{-\frac{p-1}{2}},\hspace{0.3cm} \nu_{\rm m}<\nu<\nu_{\rm c},\,\,\,\,\, \cr
A_{k}^{\frac{p+2}{4}}\beta^{\frac{10p-k(p+2)}{4}}t^{\frac{8-k(p+2)} {4}}\,\nu^{-\frac{p}{2}},\,\,\,\,\hspace{1.0cm}   \nu_{\rm c}<\nu\,, \cr
}
\end{eqnarray}
}
and 
{\small
\begin{eqnarray}
\label{sc_coast2}
F_{\rm \nu}\propto \cases{ 
A_{k}^{0}\beta^{4}t^{2}\, \nu^{2},\hspace{4.2cm} \nu<\nu_{\rm m}, \cr
A_{k}^{-\frac14}\beta^{\frac{6+k}{4}}t^{\frac{8+k}{4}}\, \nu^{\frac52},\hspace{2.9cm} \nu_{\rm m} < \nu<\nu_{\rm a,2}, \cr
A_{k}^{\frac{p+5}{4}}\beta^{\frac{2(3+5p)-k(p+5)}{4}}t^{\frac{12-k(p+5)} {4}}\, \nu^{-\frac{p-1}{2}},\hspace{0.2cm} \nu_{\rm a,2}<\nu<\nu_{\rm c},\,\,\,\,\, \cr
A_{k}^{\frac{p+2}{4}}\beta^{\frac{10p-k(p+2)}{4}}t^{\frac{8-k(p+2)} {4}}\,\nu^{-\frac{p}{2}},\,\,\,\,\hspace{1.cm}   \nu_{\rm c}<\nu\,. \cr
}
\end{eqnarray}
}

We want to emphasize that the synchrotron spectrum is always in the slow-cooling regime ($\nu_{\rm m} < \nu_{\rm c} $) and the spectrum  in the fast-cooling regime ($\nu_{\rm c} < \nu_{\rm m} $) is derived for completeness since it is not relevant for the timescales investigated here.  It is worth noting that the peak flux density is always at the peak of the spectrum.

The velocity $\beta$ in the case of the coasting and deceleration phase without energy injection was derived in \cite{2021ApJ...907...78F}. In the following, we derive the evolution of the velocity when the central engine continuously injects energy into the blastwave.

\subsection{Dynamical evolution}
Energy injection by the central engine on the GRB afterglow can produce refreshed shocks.  The luminosity injected from the central engine into the blastwave  can be described by \citep[e.g.,][]{2006ApJ...642..354Z}
\be\label{Lin}
\dot{E}_{\rm inj}(t)= L_{\rm inj}\,\left(\frac{t}{t_{\rm c}}  \right)^{-q}\,,
\ee
where $q$ is the energy injection index, $L_{\rm inj}$ is the initial luminosity and $t_{\rm c}$ is the characteristic timescale.    The isotropic-equivalent kinetic energy can be estimated as 

\be \label{E_inj}
E_{\rm t}=\int \dot{E}_{\rm inj}\, dt\propto L_{\rm inj}\, t^{-q+1}\,.
\ee

Given $q=1$, the energy does not evolve with time and the standard synchrotron light curves are recovered \citep{2013ApJ...778..107S, 2015MNRAS.454.1711B, 2021ApJ...907...78F}, and for $q>1$, the decreasing value of  isotropic-equivalent kinetic energy is not considered.  The energy injection could be due to magnetic spin-down from a spinning magnetized NS \citep[$q=0$;][]{1997A&A...319..122R, 1998A&A...333L..87D, 2000ApJ...537..803D, 2001ApJ...552L..35Z} and a fall-back material onto a central BH ($q\leq 1$) \citep[][]{2006MNRAS.370L..61P,2005ApJ...635L.133B,  2005ApJ...630L.113K, 2006Sci...311.1127D, 2006ApJ...636L..29P, 2006MNRAS.370L..61P, 2005Sci...309.1833B, 2007ApJ...671.1903C, 2013ApJ...765..125L, 2013ApJ...767L..36W, 2017MNRAS.464.4399D}. It is relevant to mention that the spinning magnetized NS could also accrete material \citep[e.g., see][]{2018ApJ...857...95M}.   The luminosity due to magnetic spin-down ($L_{\rm sd}$) or accreting BH ($L_{\rm BH}$) scenario  can be converted  into  flux through the efficiency in converting its spin-down/accreting energy to radiation ($\eta$) and the beaming factor of the wind ($f_b=1-\cos\theta_j$) with $\theta_j$ the half-opening angle. In both scenarios, the initial luminosity can be written as $L_{\rm inj}=\frac{\eta}{f_b} L_{\rm j}$, and for typical and similar values of $\eta$ and $f_b$, $L_{\rm inj}\approx L_{\rm j}$ with ${\rm j=sd\,{\rm or}\, BH}$.\\

During the deceleration phase, the ejected mass acquires a velocity structure,  the velocity of matter in the front of the ejected mass is faster than the one that moves in the back \citep{2000ApJ...535L..33S}.  \cite{2001ApJ...551..946T} studied  the acceleration of the ejected mass with relativistic and sub-relativistic velocities. They found that the isotropic-equivalent kinetic energy in the sub- and ultra-relativistic limit  can be expressed as a PL velocity distribution given by {\small $E_{\rm k} (\geq \beta) \propto \beta^{-5.2}$ for   $\beta\ll 1$ and  $E_{\rm k} (\geq \beta\Gamma) \propto \left( \beta\Gamma \right)^{-1.1}$} for $\beta\Gamma\gg 1$ (with $\Gamma=\sqrt{1/1-\beta^2}$), respectively.\footnote{The polytropic index $n_p=3$ is used.}    Here,  we consider the sub-relativistic regime, so the isotropic-equivalent kinetic energy distribution is given by 

\be\label{E_beta}
E_{\rm \beta} (\geq \beta)= \tilde{E}\,\beta^{-\alpha}\,,
\ee

where  $\tilde{E}$ is the fiducial energy.  We consider the power-law velocity distribution in the sub-relativistic regime with the values of $\alpha$ in the range $3 \leq \alpha \leq 5.2$ presented in \cite{2001ApJ...551..946T}. It is worth noting that these values of $\alpha$ in the sub-relativistic regime have been widely used \cite[e.g., see][]{2015MNRAS.450.1430H, 2017LRR....20....3M, 2019ApJ...886L..17H}.\\

The total isotropic-equivalent kinetic energy is given by the superposition of the energy injection (Eq. \ref{E_inj}) and the energy distribution (Eq. \ref{E_beta}).   In the sub-relativistic regime, the ejected material is described by the Sedov–Taylor solution as
\small{
\begin{equation}\label{sedov}
    E_\beta + E_t=  \left(\frac{5-{\rm k}}{2}\right)^{3-{\rm k}}\left(\frac{2 \pi m_p}{3-{\rm k}}\right)^{5-{\rm k}}\, (1+z)^{{\rm k}-3}\,A_{\rm k}\,\beta^{5-k}\,t^{3-k}\,,
\end{equation}
}

where $E_\beta=\tilde{E}\beta^{-\alpha}$ and $E_t=\hat{E}\left(\frac{t}{t_{\rm c}}\right)^{1-q}$ with $\hat{E}=\frac{1}{1-q}\,t_{\rm c}\,L_{\rm inj}$ for $q<1$ and $m_p$ the proton mass.

\subsubsection{Analytical solution}

The Sedov-Taylor solution can be solved analytically in the asymptotic cases; {\small $ E_t \ll E_\beta$} and {\small $E_\beta \ll E_t$}.   Each limiting case leads to a different velocity; they are given by

{\small
\begin{eqnarray}
\beta \propto \cases{
(1+z)^{-\frac{k-3}{\alpha+5-k}}\,A_{\rm k}^{-\frac{1}{\alpha+5-k}}\,\tilde{E}^{\frac{1}{\alpha+5-k}}\,t^{\frac{k-3}{\alpha+5-k}}\,{\rm for}\,\, {\rm E_t\ll E_\beta}\,\,\,\,\, \cr
(1+z)^{\frac{k-3}{k-5}}\,A_{\rm k}^{\frac{1}{k-5}}\,\hat{E}^{-\frac{1}{k-5}}\,t^{\frac{q+2-k}{k-5}}\,\hspace{1.2cm} \,{\rm for}\,\, {\rm E_\beta \ll E_t}.
}
\end{eqnarray}
}

Both cases may be written with just one expression:
{\small
\be\label{beta_dec}
\beta\propto \,(1+z)^{-\frac{k-3}{\alpha+5-k}}\,A^{-\frac{1}{\alpha+5-k}}_{\rm k}\,E^{\frac{1}{\alpha+5-k}}\, t^{\frac{k-(q+2)}{\alpha+5-k}}\,,
\ee
}
where the case  {\small $ E_t \ll E_\beta$} is obtained by setting $q=1$ with $E=\tilde{E}$, while the case \small $E_\beta \ll E_t$}  is obtained by setting $\alpha=0$ with $E=\hat{E}$. It is worth mentioning that the deceleration time can be obtained from Eq. \ref{beta_dec}, and is presented in Section \ref{DecelerationTimescales}.  The blastwave radius ($r=\beta t/(1+z)$) can be written as

{\small
\be\label{R_dec}
r\propto (1+z)^{-\frac{\alpha+2}{\alpha+5-k}}\,A^{-\frac{1}{\alpha+5-k}}_{\rm k}\, E^{\frac{1}{\alpha+5-k}}\,t^{\frac{\alpha+3-q}{\alpha+5-k}}\,.
\ee
}

The standard equations in constant-density medium are recovered when $q=1$ and $\alpha=0$ (i.e., $\beta\propto t^{-\frac35}$ and $r\propto t^{\frac25}$; \cite{2013ApJ...778..107S}). The dynamics, spectral breaks and synchrotron light curves derived in \cite{2021ApJ...907...78F} are recovered for {\small $ E_t \ll E_\beta$}. Using Eqs. (\ref{beta_dec}) and (\ref{R_dec}), we report in the Appendix the equations of the dynamics, the synchrotron spectral breaks,  the flux density and the light curves in the fast- and slow-cooling regime. It is worth noting that the synchrotron spectrum can lie in the slow- or fast-cooling regime, depending on the parameter values. 

\subsubsection{Numerical approach: Comparison with analytic solution}

\begin{figure}
{\centering
\resizebox*{\textwidth}{0.45\textheight}
{\includegraphics{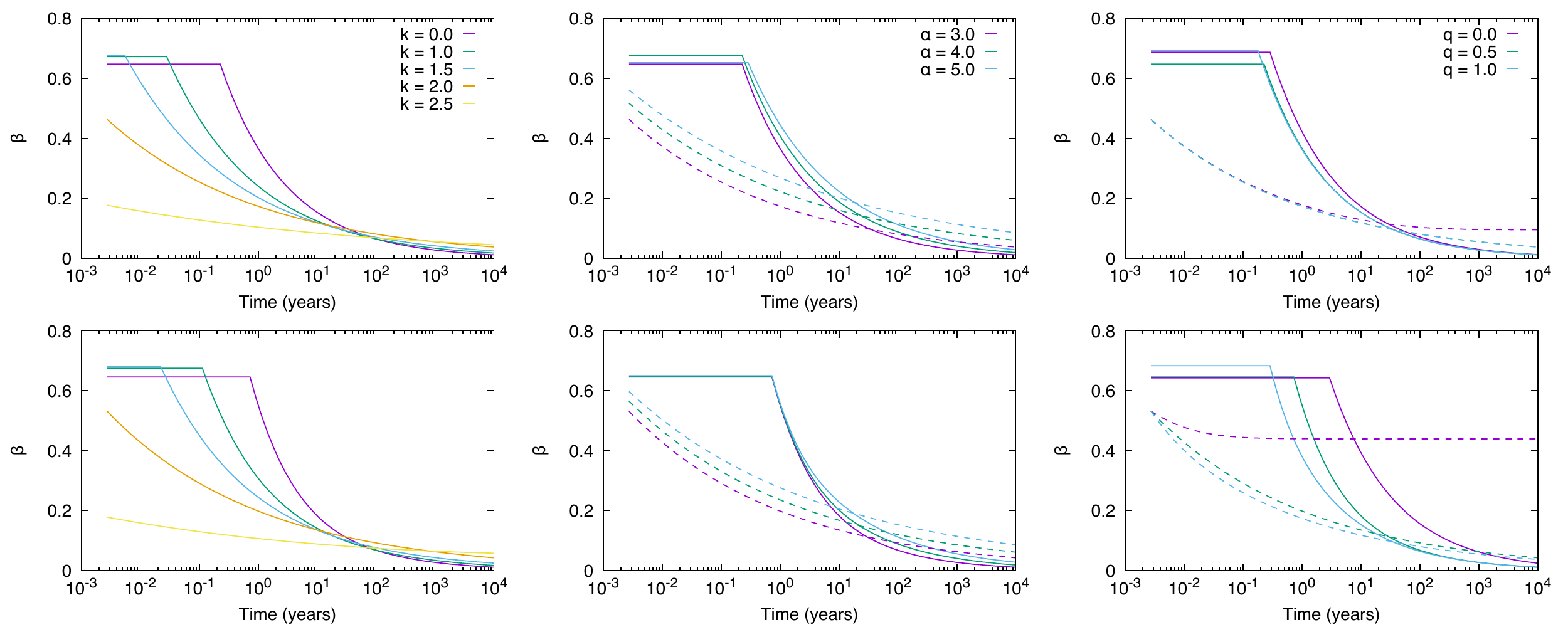}}
}   
\caption{Evolution of the shock's velocity in time for different choices of parameters. The top row is equivalent to an energy injection luminosity of $L_{\mathrm{inj}}=10^{43}\ \mathrm{erg}/\mathrm{s}$ and the lower one to $L_{\mathrm{inj}}=10^{45}\ \mathrm{erg}/\mathrm{s}$. Panels from left to right correspond to variation of the circumburst density distribution, the shock's velocity distribution and the energy injection parameter, respectively. The solid lines in the middle and rightmost columns represent the case for a constant-density medium, while the dashed lines correspond to a wind-like one. The following parameters $\tilde{E}= 10^{49}\,{\rm erg}$ and $z=0.1$ are used.}
\label{beta}
\end{figure}

We solve Eq. (\ref{sedov}) numerically using the bisection method\footnote{https://docs.scipy.org/doc/scipy/reference/generated/scipy.optimize.bisect.html} and plot  the evolution of the shock's velocity for different parameter values, as shown in Figure \ref{beta}. The rows represent different choices of luminosity, namely the top one corresponds to a value of $L_{\mathrm{inj}}=10^{43}\ \mathrm{erg}\,\mathrm{s}^{-1}$ and the lower one to $L_{\mathrm{inj}}=10^{45}\ \mathrm{erg\,s^{-1}}$. The column on the left presents the velocity's development for different choices of the circumburst density profile $\propto r^{-\rm k}$ with ${\rm k}=0$, $1$, $1.5$, $2$ and $2.5$. The center column displays its transformation according to different values of the isotropic-equivalent kinetic energy of the outermost matter's PL distribution index with $\alpha=3.0$, $4.0$ and $5.0$. On the rightmost column, we show the evolution of the velocity for varying possibilities of the energy injection index, namely $q=0$, $0.5$ and $1$. For the middle and right-hand panels, the solid lines represent a choice of constant-density medium (${\rm k=0}$), while the dashed lines correspond to stellar wind (${\rm k=2}$).\\

All panels show two distinct behaviours. At early times (before approximately $10$ years), the forward shock expands into the circumburst medium uninhibited with a constant velocity $\beta_0$, the so-called ``coasting phase'' which is represented in the panels from Figure \ref{beta}, as horizontal lines. Once the shock has interacted with enough material, the coasting phase comes to an end and the deceleration phase commences. It is governed by the solution of equation (\ref{sedov}) and can be seen in Figure \ref{beta} as the decrease in the velocity after its constant segment.\\

Upon comparison of the middle panels between both rows, it can be seen that when the dominant contribution to the energy of the shock is the fiducial energy $\tilde{E}$ (represented by the top row) the behaviour of the shock's velocity depends more strongly on the velocity distribution's PL index $\alpha$. This is made apparent from the separation of the curves for different values of this parameter in the deceleration phase, which is more pronounced in the top row for both constant-density and wind-like environment.\\

On the other hand, the opposite behaviour is noticed when the same comparison is performed between the rightmost panels. In this case, the energy injection parameter $q$'s variation is more easily observed when $\hat{E}$ is the dominant component, as is made clear by the lower row in both types of medium considered. It is also apparent that when the energy that is injected into the shock is substantial, it may reach a quasi-constant value, where the energy injected equals the energy lost by interaction with the surrounding medium. This behaviour is best exemplified by the dashed $q=0$ curve in the lower panel, where it can be seen that the shock's velocity becomes quasi-constant even at late times.\\

\section{Energy injection from a spinning magnetized NS and an accreting BH}\label{sec:2}

A spinning magnetized NS and BH remnants are created from the merger of a BCO system \citep{2006Sci...312..719P, 2013PhRvD..87l1302S, 2014PhRvD..90d1502K} and the death of a massive star \citep{1992Natur.357..472U, 1998A&A...333L..87D, 2000ApJ...537..810W, 2004ApJ...611..380T, 2007MNRAS.380.1541B,2011MNRAS.413.2031M}.  These remnants could accrete material and inject energy into the blastwave.  The spinning magnetized NS or the central BH inject energy due to either a magnetic spin-down  \citep{1997A&A...319..122R, 1998A&A...333L..87D, 2000ApJ...537..803D, 2001ApJ...552L..35Z} or accretion \citep[]{2006MNRAS.370L..61P,2005ApJ...635L.133B,  2005ApJ...630L.113K, 2006Sci...311.1127D, 2006ApJ...636L..29P, 2006MNRAS.370L..61P, 2005Sci...309.1833B, 2007ApJ...671.1903C, 2013ApJ...765..125L, 2013ApJ...767L..36W, 2017MNRAS.464.4399D}, respectively.   For instance, \cite{2015MNRAS.450.1777K} summarized a variety of final remnants from a merger of BCOs, which might inject energy into the blastwave; a hyper-massive NS (HMNS) that collapses into a BH in short timescales, a non-spinning BH, and a spinning magnetized NS and BH. In the case of a spinning magnetized NS or BH, the authors reported that the late-time activity via spin-down power or accreting material could be expected up to timescales of years. In this case, the synchrotron light curves from ejected materials would be modified by energy injection into the blastwave.

\vspace{1cm}
\subsection{A spinning magnetized NS with fall-back accretion}\label{subsec:2.1}

Rapid spinning magnetized NSs called ``millisecond magnetars" are potential candidates for long and short GRBs.   The energy reservoir of a millisecond magnetar is the total rotation energy which is given by 
\be\label{Erot}
E_{\rm}=\frac12 I\, \Omega^2\,\approx 2.6 \times 10^{52}\,{\rm erg}\,M^{\frac32}_{\rm ns,1.4}\,P^{-2}_{-3}\,,
\ee
where $P$ is the spin period associated to an angular frequency $\Omega=2\pi/P$ and $I\simeq 1.3\times 10^{45}\,M^{\frac32}_{\rm ns,1.4}\,{\rm g\,cm^2}$ \citep{2005ApJ...629..979L} is the NS moment of inertia  with $M_{\rm ns}=1.4\, M_\odot$ the NS mass. The merger of two NSs or CC-SN usually leave a fraction of the stellar progenitor bound to the NS. This fraction of material will begin to rotate into an accretion disk and to fall-back over a long period \citep{1989ApJ...346..847C, 2007MNRAS.376L..48R, 2012ApJ...752...32W, 2012MNRAS.419L...1Q}.  The fall-back accretion rate can be written as \citep{2018ApJ...857...95M}

\be\label{dotM}
\dot{M}=\frac23\frac{M_{\rm fb}}{t_{\rm fb}}\cases{ 
1  \hspace{1.6cm}{\rm for} \hspace{0.4cm} t< t_{\rm fb}   \cr
\left(\frac{t}{t_{\rm fb}}\right)^{-\frac{5}{3}} \hspace{0.4cm}{\rm for} \hspace{0.3cm} t_{\rm fb}< t\,,\cr
}
\ee

where  $M_{\rm fb}$ is the accreting mass  over a characteristic fall-back time $t_{\rm fb}$.  Once the millisecond magnetar is formed, the NS might be subject to fall-back accretion.  This accretion depends on the dipole magnetic moment ($\mu$), and the Alf\'en ($r_{\rm m}$), the co-rotation ($r_{\rm c}$) and cylinder ($r_{\rm lc}$) radii. The spin evolution of an accreting magnetar is given by \citep{2011ApJ...736..108P}

\be\label{dif_eq}
I\frac{d\Omega}{dt}=-N_{\rm dip}+N_{\rm acc}\,,
\ee

where the spin-down terms due to the torque and the accretion are

\be\label{N_dip}
N_{\rm dip} \simeq  \mu^2\Omega^3 \cases{ 
 \frac{r^2_{\rm lc}}{r^2_{\rm m}}  \hspace{0.8cm} r_{\rm m} \lesssim  r_{\rm lc}\,, \cr
1  \hspace{1.05cm} r_{\rm lc} \lesssim r_{\rm m}\,, \cr
}
\ee
and 
\be\label{Nacc}
N_{\rm acc}=\dot{M}(G_N\,M_{\rm ns}\,r_{\rm m})^\frac12\, \left[ 1-\left( \frac{r_{\rm m}}{r_{\rm c}} \right)^\frac32 \right]\,,
\ee
respectively, with $G_N$ the gravitational constant, $\mu=BR^3_{\rm ns}$ where $R_{\rm ns}= 1.2\times 10^6\, {\rm cm}$ is the NS radius and $B$ is the strength of the dipole magnetic field. For details, see \cite{2018ApJ...857...95M}. In this scenario, the isotropic-equivalent kinetic energy due to injection would be $E_{\rm t}=\int L_{\rm inj}\, dt= \frac{\eta}{f_b}\int L_{\rm sd}\, dt$.

The left-hand panels of Figure \ref{luminosity} show the spin-down luminosity and the synchrotron forward-shock light curves from the millisecond magnetar with an initial spin period $P_0=10^{-3}\,{\rm s}$ \citep{2018ApJ...857...95M}, an accreting mass $M_{\rm fb}=0.8M_{\odot}$ \citep{2018ApJ...857...95M}, a half-openig angle $\theta_j=30^\circ$ \citep{2020ApJ...888...67D} and an efficiency $\eta=0.1$ \citep{2019ApJ...878...62X}.  The spin-down luminosities exhibit a ``plateau" phase for an energy injection index of $q=0$ (as indicated), and different time scales are observed for four different sets of parameters. The black solid curve corresponds to a magnetic field strength of $B=10^{14}\,{\rm G}$ and a characteristic fall-back time of $t_{\rm fb}=10^{3}\,{\rm s}$ \citep{2018ApJ...857...95M}, the gray solid line takes the values $B=10^{16}\,{\rm G}$ and $t_{\rm fb}=10^{3}\,{\rm s}$, the black dashed one takes $B=10^{14}\,{\rm G}$ and $t_{\rm fb}=10^{7}\,{\rm s}$ and the gray dashed curve represents $B=10^{16}\,{\rm G}$ and $t_{\rm fb}=10^{7}\,{\rm s}$. It is worth nothing that the gray dashed curve  displays a spin-down luminosity $\approx 10^{42}\,{\rm erg}\, {\rm s}^{-1}$ during the ``plateau" phase.
%
 Light curves show one or two ``plateau" phases depending on the values of the magnetic field, spin period and the characteristic timescale of fallback. The first ``plateau" related with a precursor is $\sim$ two order of magnitude less than the second one, which is associated with the prompt emission.  For small values of the characteristic timescale of fallback and larger values of magnetic field, the light curves exhibit a `plateau" in a timescale of seconds. Larger values of the characteristic timescale and small magnetic fields lead to a small luminosity during the ``plateau" phase. The first ``plateau" is explained when the Alv\'en radius is larger than the co-rotation one ($r_c\ll r_m$). In this case,  the spin-down luminosity becomes
{\small 
\be\label{precursor}
L_{\rm sd}\simeq \left( \frac{\mu^2}{c\, r^2_{\rm m}}+ \dot{M}\,r^2_{\rm m} \right) \Omega_0 \exp\left(-\frac{t}{t_{\rm sd}}\right)\,,
\ee
}
with {\small $t_{\rm sd}=\frac12 \left( \frac{\mu^2}{c^3 I r^2_{\rm m}}+\frac{\dot{M}\,r^2_{\rm m}}{I} \right)^{-1}$}. For a large value of $t_{\rm sd}$ the spin-down luminosity decays very slowly.   The second ``plateau" can be explained once the equilibrium is reached and  an analytic solution of Eq. \ref{dif_eq} can be derived \citep[see][]{2018ApJ...857...95M,2021ApJ...918...12F}. In this case, the spin-down luminosity becomes
{\small
\be\label{L_sd}
L_{\rm sd}\approx 10^{40.7}\,{\rm erg\, s^{-1}}B^{-\frac67}_{16}\,M^{\frac{12}{7}}_{\rm ns,1.4}\,R^{-\frac{18}{7}}_{\rm ns,6.1}\,\cases{ 
t^0\,  \hspace{0.6cm}{\rm for} \hspace{0.1cm} t\ll t_{\rm fb}   \cr
 t^{-\frac{50}{21}}_{8} \hspace{0.25cm}{\rm for} \hspace{0.1cm} t\gg t_{\rm fb}\,,\cr
}
\ee
}
where the typical values of the accreting mass $M_{\rm fb}=0.8M_{\odot}$ and the characteristic fall-back time $t_{\rm fb}=10^8\,{\rm s}$ are used. The profile of the spin-down luminosity (Eq. \ref{L_sd}) agrees with those profiles shown in the upper left-hand panel in Figure  \ref{luminosity}.

 The black solid curve displays an uninterrupted drop during $10^3\ \mathrm{days}$. The gray solid line presents a very similar behaviour to its black counterpart as it decays with no interruption with the same power law. The main difference between both curves being that the luminosity achieved by this solution is smaller by about two orders of magnitude. The black dashed curve shows an initial ``plateau" phase for approximately $10^{-1}\ \mathrm{days}$. Afterwards it decreases by about two orders of magnitude until it reaches its second plateau phase after a couple of days. This phase happens for roughly one month after which the luminosity drops once again with a power law slightly less steep than the one present in the previously mentioned solid curves. Finally, the gray dashed profile presents the same type of behaviour as its black analogue, however this solution starts off with larger values than the black line at early times and then drops below them during the black curve's first ``plateau" phase. Afterwards, the gray profile remains below the black one.

The lower left-hand panel shows the light curves in this same scenario in three energy bands: radio (3 GHz), optical (R-band) and X-ray (1 keV). All curves assume evolution in a constant density medium $k=0$ with $n=5.88\times 10^{-3}\,{\rm cm^{-3}}$, $\tilde{E}=10^{48}\,{\rm erg}$, $z=0.023$, $p= 2.15$, $\epsilon_{B}=10^{-3}$, $\epsilon_{e}=10^{-1}$,  $\alpha=3.0$ and $\beta=0.3$ \citep{2017LRR....20....3M, 2019ApJ...886L..17H, 2021ApJ...907...78F, 2021arXiv210402070H}. The values of the magnetic field strength and characteristic time scale were chosen to be the same as the ones of the gray dashed line from the upper panel.

All three light curves grow with the same power law during the first $10^{4}$ days. After this point in time only the X-ray profile reaches its maximum and starts to drop, while the other two continue to rise, albeit with a smaller slope. The panel shows that the flux density of a profile increases with the energy of the band, that is, that the X-ray curve has the largest flux density, while the radio one has the smallest value.

\subsection{Fall-back material onto a BH}\label{subsec:2.2}

The Eddington luminosity for pure ionized heavy elements is

\be
L_{\rm Edd}=\frac{4\pi G_N M_{\rm BH}m_p}{\sigma_T}=10^{38.9}\,{\rm erg\,s^{-1}},
\ee

where $M_{\rm BH}=2.3\,M_{\odot}$ is the mass of the remnant BH, $\sigma_T$ is the Thompson cross section and $Y_e\approx 0.4$.\\ 

The fall-back material onto the remmant BH could create an accretion disk, powering a new material via Blandford-Znajek \citep[BZ;][]{1977MNRAS.179..433B},  or neutrino-annihilation mechanism \citep{1999ApJ...518..356P}.  The BZ jet power from a BH with mass $M_{\rm BH}$ and angular momentum $J_{\rm BH}$ can be described as \citep{2000PhR...325...83L}
\begin{equation}\label{L_BZ}
    L_{\rm BZ}\approx 10^{49}{\rm erg\, s^{-1}}\, G(a) \dot{M}_{\rm BH, -5}\,, 
\end{equation}
where $\dot{M}_{\rm BH, -5}=\frac{\dot{M}_{\rm BH}}{10^{-5}\,M_\odot\,s^{-1}}$, $a=\frac{J_{\rm BH}}{G_N M^2_{\rm BH}}$ is the dimensionless spin parameter \citep[e.g., see][]{2008MNRAS.388..551T}, $G(a)=\frac{a^2 F(a)}{(1+\sqrt{1-a^2})^2}$
with $F(a)=[\frac{1+q'^2}{q'^2}][(q'+\frac{1}{q'})\arctan q' - 1]$ and $q'=\frac{a}{1+\sqrt{1-a^2}}$ \citep{2013ApJ...767L..36W}. The term $\dot{M}_{\rm BH}$ is the accretion rate onto the BH given by  \citep{2008Sci...321..376K, 2008MNRAS.388.1729K}
\begin{equation}\label{M_odot}
    \dot{M}_{\rm BH}=\frac{1}{\tau_{\rm vis}}e^{-\frac{t}{\tau_{\rm vis}}}\int^t_{t_0}e^{\frac{t'}{\tau_{\rm vis}}}\dot{M}_{\rm fb}\,dt'\,,
\end{equation}
where $\tau_{\rm vis}$ is the viscous timescale, $t_0$ is the starting time of the accretion
and $\dot{M}_{\rm fb}$ is the accretion rate described by \citep{1989ApJ...346..847C, 1999ApJ...524..262M, 2001ApJ...550..410M, 2008ApJ...679..639Z}

\be\label{dotMfb}
\dot{M}_{\rm fb}=\frac12\,\dot{M}_{\rm p}\cases{ 
\left(\frac{t-t_0}{t_{\rm p} - t_0}\right)^{\frac{1}{2}}  \hspace{1.cm}{\rm for} \hspace{0.4cm} t< t_{\rm p}   \cr
\left(\frac{t-t_0}{t_{\rm p} - t_0}\right)^{-\frac{5}{3}} \hspace{0.75cm}{\rm for} \hspace{0.4cm} t_{\rm p}<t\,,\cr
}
\ee

with $\dot{M}_{\rm p}$ and $t_{\rm p}$ the fallback rate and the time at the peak, respectively. For details, see \cite{2013ApJ...767L..36W}. In this scenario, the isotropic-equivalent kinetic energy due to injection would be $E_{\rm t}=\frac{\eta}{f_b}\int L_{\rm BZ}\, dt$.\\

\begin{figure}
{\centering
\resizebox*{0.9\textwidth}{0.65\textheight}
{\includegraphics{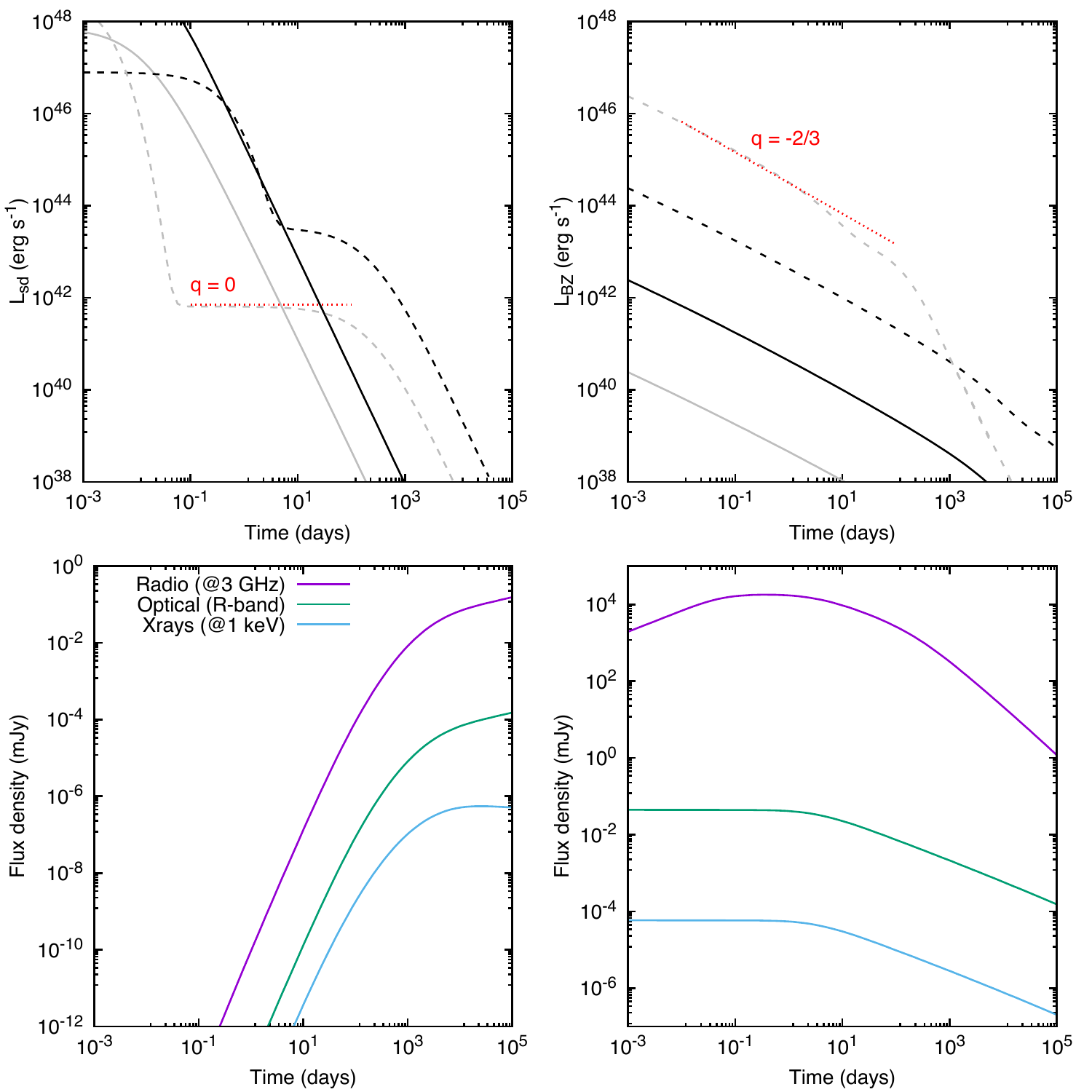}}
}   
\caption{The spin-down and BZ luminosities (panels above) together with synchrotron forward-shock light curves (panels below) from a millisecond magnetar (left) and BH (right) with accretion.   The spin-down luminosities are shown for $P_0=10^{-3}\,{\rm s}$ and the parameters $B=10^{14}\,{\rm G}$ and $t_{\rm fb}=10^3\,{\rm s}$ (black solid line), $10^{16}\,{\rm G}$ and $10^3\,{\rm s}$ (gray solid line), $10^{14}\,{\rm G}$ and $10^7\,{\rm s}$ (black dashed line) and $10^{16}\,{\rm G}$ and $10^7\,{\rm s}$ (gray  dashed line).   The BZ luminosities are shown for $t_0=0\,{\rm s}$, $t_p=10^3\,{\rm s}$, $M_{\rm BH}=2.3\,{\rm M_{\odot}}$ and $a=0.7$, and the parameters  $\tau_{\rm vis}=10^{9}\,{\rm s}$ and  $\dot{M}_{\rm p}=10^{-6}\,{\rm M_\odot\,s^{-1}}$ (black solid line), $10^{6}\,{\rm s}$ and $10^{-5}\,{\rm M_\odot\,s^{-1}}$ (gray solid line) $10^{9}\,{\rm s}$ and $10^{-4}\,{\rm M_\odot\,s^{-1}}$ (black dashed line), and $10^{12}\,{\rm s}$ and $10^{-5}\,{\rm M_\odot\,s^{-1}}$ (gray  dashed line).  The synchrotron forward-shock light curves are shown in radio (3 GHz),  optical (R-band) and X-ray (1 keV) bands with  $\tilde{E}=10^{48}\,{\rm erg}$, $p= 2.15$, $\epsilon_{B}=10^{-3}$, $\epsilon_{e}=10^{-1}$,  $\alpha=3.0$ and $\beta=0.3$, and the parameters of the magnetar ($B=10^{16}\,{\rm G}$ and $t_{\rm fb}=10^7\,{\rm s}$) and the BH ($t_0=0\,{\rm s}$, $t_p=10^3\,{\rm s}$, $M_{\rm BH}=2.3\,{\rm M_{\odot}}$, $a=0.7$,   $\tau_{\rm vis}=10^{9}\,{\rm s}$ and $\dot{M}_{\rm p}=10^{-6}\,{\rm M_\odot\,s^{-1}}$) model. We consider that the synchrotron light curves evolve in a constant-density medium with ${\rm k=0}$ and $n=5.88\times 10^{-3}\,{\rm cm^{-3}}$ (left) and in a stellar-wind medium with ${\rm k=2}$ and $A_{\rm W}=10^{-2}$ (right).}
\label{luminosity}
\end{figure}

The right-hand panels of Figure \ref{luminosity} are analogous to their left-hand counterparts, which were explained in the previous subsection, but now in the context of the fall-back material onto a BH scenario.  The upper panel presents the BZ luminosity with $q=2/3$ (as indicated in the dashed gray line) with different time scales. This behavior corresponds to an accretion rate of $\dot{M}_{\rm fb}\propto t^{\frac12}$.    The change of the slope in the BZ luminosity is associated with the variation of the accretion rate from $\dot{M}_{\rm fb}\propto t^{\frac12}$ to $\propto t^{-\frac53}$ (see Eq. \ref{dotMfb}).  It is worth noting that the BH torus system has no effect on the afterglow evolution at later times, which corresponds to an accretion rate of $\propto t^{-\frac53}$ \citep[e.g.,][]{2001ApJ...550..410M, 2004MNRAS.355..950J, 2006ApJ...642..354Z}.  We consider typical values of starting time scale of accretion $t_0=1\,{\rm s}$, time scale at the peak $t_p=10^3\,{\rm s}$ \citep{2013ApJ...767L..36W}, BH mass $M_{\rm BH}=2.3\,{\rm M_{\odot}}$ and dimensionless spin parameter $a=0.7$ \citep{2013ApJ...767L..36W, 2021ApJ...906...60Z}.   Once again, four different sets of parameters were considered. The black solid curve corresponds to a viscous timescale of $\tau_{\rm vis}=10^{9}\,{\rm s}$ and a fallback rate of $\dot{M}_{\rm p}=10^{-6}\,{\rm M_\odot\,s^{-1}}$ \citep{2021ApJ...906...60Z}, the gray solid line takes the values $10^{6}\,{\rm s}$ and $10^{-5}\,{\rm M_\odot\,s^{-1}}$, the black dashed one takes $10^{9}\,{\rm s}$ and $10^{-4}\,{\rm M_\odot\,s^{-1}}$ and the gray dashed curve represents $10^{12}\,{\rm s}$ and $10^{-5}\,{\rm M_\odot\,s^{-1}}$.   In a similar fashion to the left-hand panel, both of the solid curves have a similar behaviour. They both decrease following the same power law and the black profile is greater than the gray one by about two orders of magnitude. In the case of the dashed lines, the black solution decreases uninterrupted, but it changes its power law at around $10^{4}\,{\rm days}$. The gray curve also decreases with no interruption and follows the same slope as the black one at early times, but it changes its slope at approximately $10^{2}\,{\rm days}$. It is also interesting to note, that similarly to the left panel, the gray dashed profile starts off with a greater luminosity than its black counterpart, but it falls below it at later times, namely in the vicinity of $10^{3}\,{\rm days}$.

The lower right-hand panel shows the synchrotron forward-shock light curves in this scenario in the same energy bands as in the panel to its left. The deceleration was once again assumed to be in a wind-like medium ${\rm k=2}$ with $A_{\rm W}=10^{-2}\,{\rm cm^{-3}}$ and the values of the viscous timescale and fallback rate were chosen to be the same as the ones of the gray dashed line from the upper panel. The rest of the parameters were selected as $\tilde{E}=10^{48}\,{\rm erg}$, $z=0.023$, $p=2.15$, $\epsilon_{B}=10^{-3}$, $\epsilon_{e}=10^{-1}$, $\alpha=3.0$ and $\beta=0.3$
\citep{2017LRR....20....3M, 2019ApJ...886L..17H, 2021ApJ...907...78F, 2021arXiv210402070H}. In all cases the BZ luminosity is much higher than the Eddington luminosity.

The optical and X-ray bands have the same exact behaviour, namely that their light curve reaches its peak very early, it remains constant until $\sim10\ \mathrm{days}$ and from this moment onward its flux density decreases. It is worth to note that both of these curves follow the same power laws; the difference between them being that the optical band lies two orders of magnitude above the X-ray band. The case of the radio light curve is different, as it can be observed that it reaches its maximum later than the previously mentioned profiles. Its constant phase also lasts less time and its behaviour at late times is according to a power law with a steeper slope than the one of the other two energy bands.

Upon comparison of the two upper panels of Figure \ref{luminosity}, it can be observed that, in general, the luminosities in the context of a BH decrease in time less steeply than the ones from a millisecond magnetar. It can also be noted that for a BH, there is no appearance of the so-called ``plateau" phase, which is a characteristic that could discriminate between both scenarios.

In the case of the lower panels of the aforementioned Figure, it can be concluded that the flux density increases in smaller timescales in the BH scenario, as in this case the maximum is reached at $\sim10^{-1}\ \mathrm{days}$, while in the case of the magnetar it is obtained at $\sim10^{3}\ \mathrm{days}$. The same observation can be made for the drop of the flux, as it can be noticed from the Figure that the light curves in the right panel begin to decrease at $\sim10^{1}\ \mathrm{days}$, while on the left panel a drop is not apparent in none of the shown energy bands even at $\sim10^{5}\ \mathrm{days}$.

An analytic solution of Eq. \ref{L_BZ} can be derived for $t_{\rm p}<t$ ($\dot{M}_{\rm fb}\propto t^{-\frac53}$) with $t\ll \tau_{\rm vis}$. In this case, the term  $e^{\pm \frac{t'}{\tau_{\rm vis}}}\approx 1$ and therefore the BZ luminosity becomes

\be\label{L_bz}
L_{\rm BZ}\approx 10^{49}{\rm erg\, s^{-1}}\,\tau^{-1}_{\rm vis,7}\, \dot{M_p}_{-6} t^{-\frac23}_{7}\,,
\ee

with $\dot{M}_{p,-6}=\frac{\dot{M}_{\rm BH}}{10^{-6}\,M_\odot\,s^{-1}}$.  The profile of the BZ luminosity (Eq. \ref{L_bz}) agrees with those profiles shown in the upper right-hand panel in Figure  \ref{luminosity}.\\

It is worth mentioning that the neutrino-annihilation luminosity could not describe late-time activities \cite[e.g., see][]{2021ApJ...906...60Z} and therefore, it cannot be considered in this work.  

\vspace{1cm}
\section{Analysis of the multiwavelength aferglow light curves}\label{sec:analysis}
\subsection{Synchrotron emission}\label{subsec:synchrotron}

\begin{figure}
{\centering
\resizebox*{0.9\textwidth}{0.65\textheight}
{\includegraphics{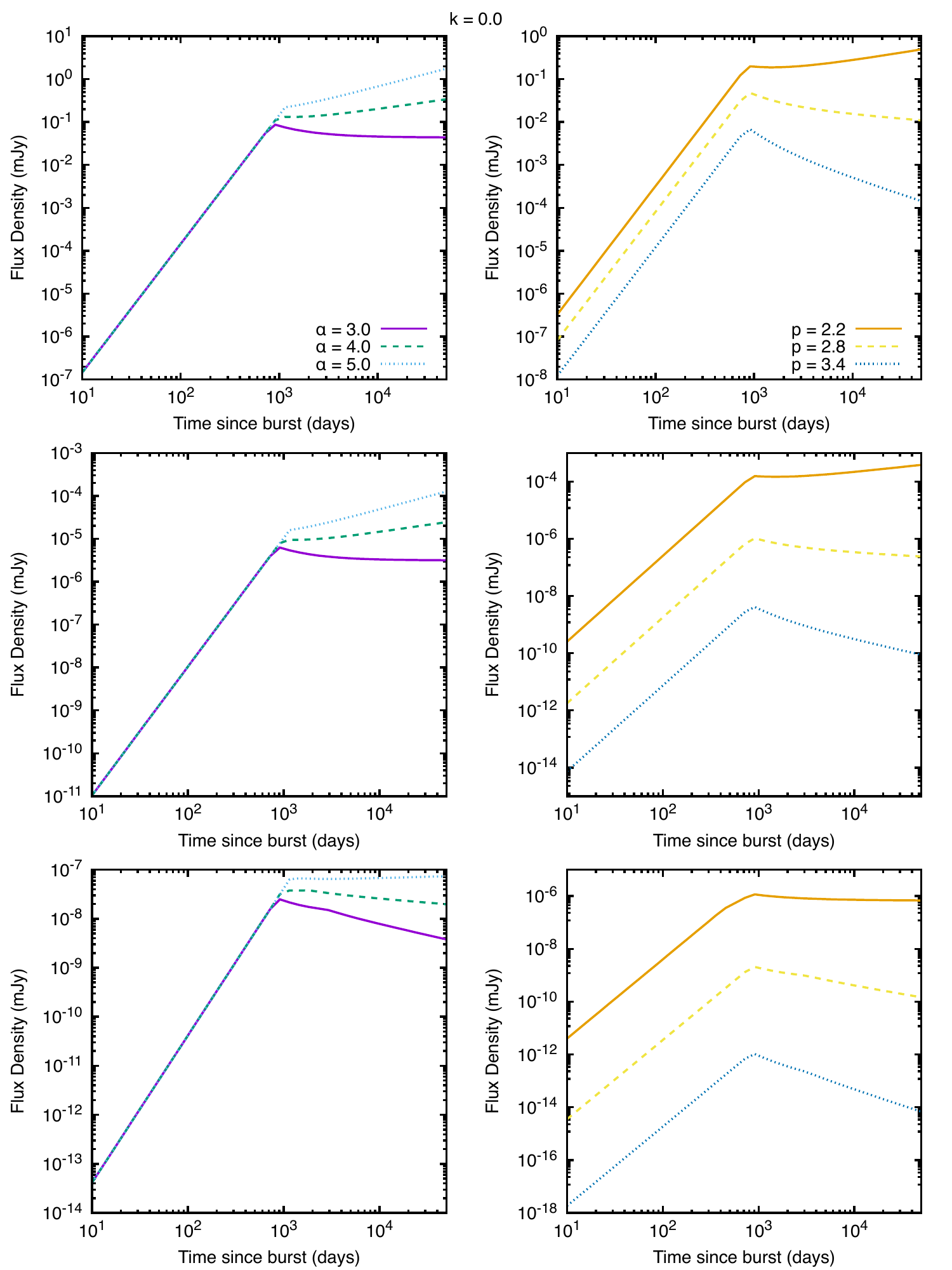}}
}   
\caption{Synchrotron light curves generated by the deceleration of the sub-relativistic ejecta for ${\rm k=0}$. Panels from top to bottom correspond to radio (1.6 GHz),  optical (R-band) and X-ray (1 keV) bands, respectively.  The left-hand panels show the light curves for $p=2.6$ with $\alpha=3$, $4$ and $5$, and the right-hand panels show the light curves for $\alpha=3$ with $p=2.2$, $2.8$  and $3.4$.   The following parameters $P_0=10^{-3}\,{\rm s}$,  $B=10^{14}\,{\rm G}$,  $t_{\rm fb}=10^{3}\,{\rm s}$, $\tilde{E}= 10^{51}\,{\rm erg}$,  $A_0=1\,{\rm cm^{-3}}$, $\epsilon_{\rm B}=10^{-4}$, $\epsilon_{\rm e}=10^{-1}$ and $d_{z}\approx 100\ \mathrm{Mpc}$ ($z=0.023$) are used.}
\label{k_0}
\end{figure}

\begin{figure}
{\centering
\resizebox*{0.9\textwidth}{0.65\textheight}
{\includegraphics{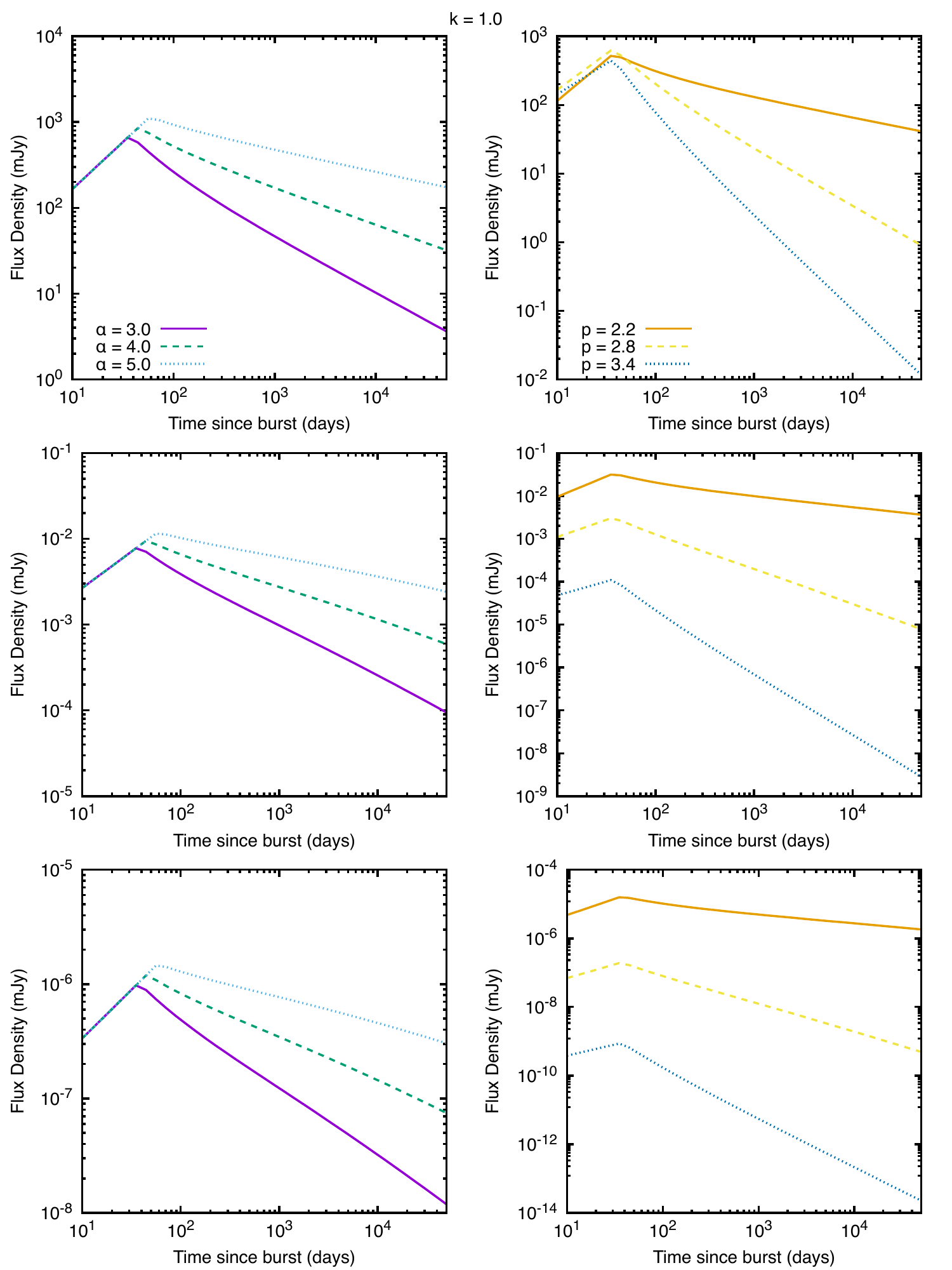}}
}   
\caption{The same as Figure \ref{k_0}, but for ${\rm k=1.0}$ with $A_1=1.5\times 10^{19}\,{\rm cm^{-2}}$ }
\label{k_1}
\end{figure}

\begin{figure}
{\centering
\resizebox*{0.9\textwidth}{0.65\textheight}
{\includegraphics{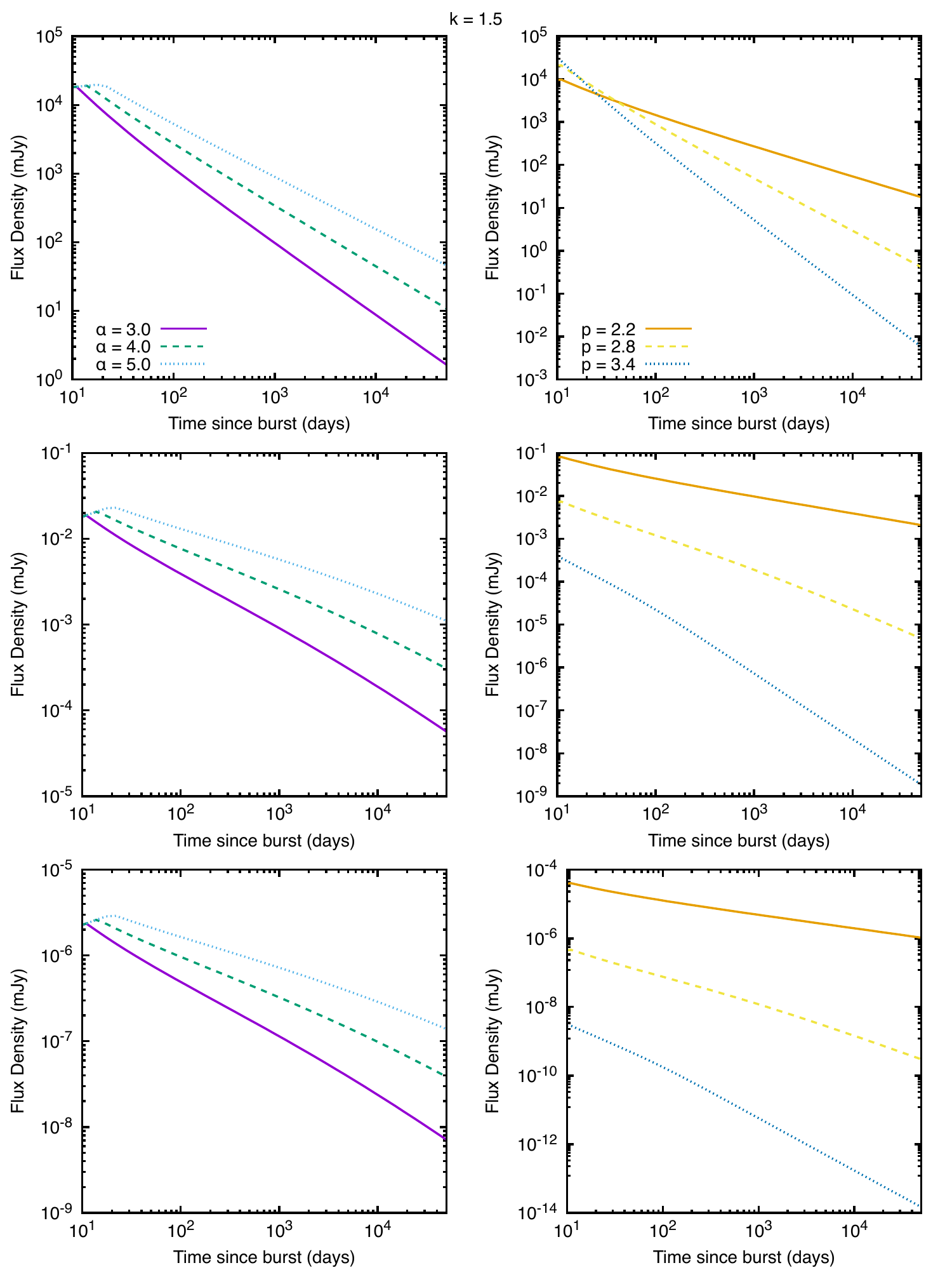}}
}   
\caption{The same as Figure \ref{k_0}, but for ${\rm k=1.5}$ with $A_{1.5}=2.7\times 10^{28}\,{\rm cm^{-\frac32}}$. }
\label{k_1.5}
\end{figure}

\begin{figure}
{\centering
\resizebox*{0.9\textwidth}{0.65\textheight}
{\includegraphics{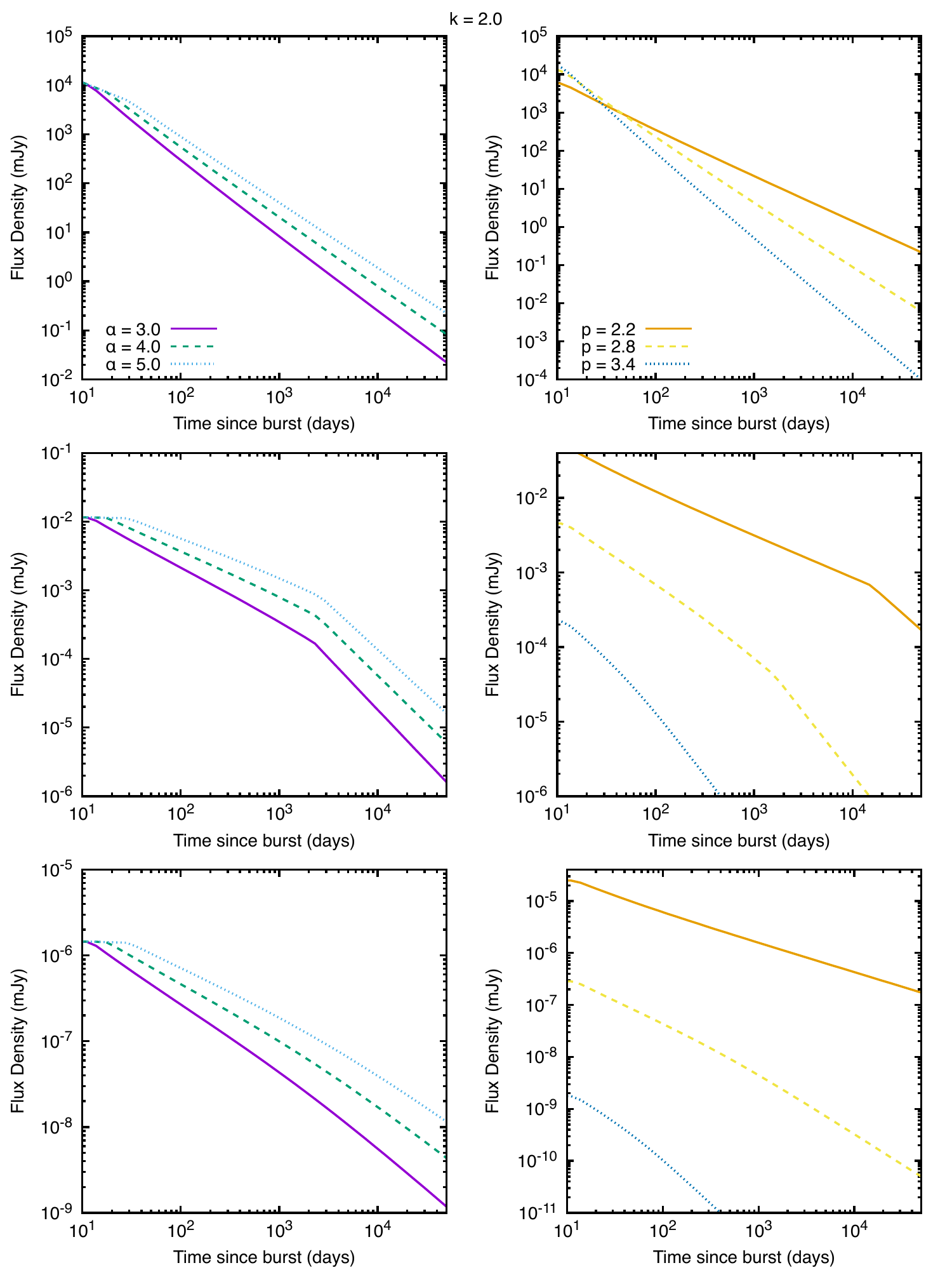}}
}   
\caption{The same as Figure \ref{k_0}, but for ${\rm k=2.0}$ with $A_2=3\times 10^{36}\,{\rm cm^{-1}}$. }
\label{k_2.0}
\end{figure}

\begin{figure}
{\centering
\resizebox*{0.9\textwidth}{0.65\textheight}
{\includegraphics{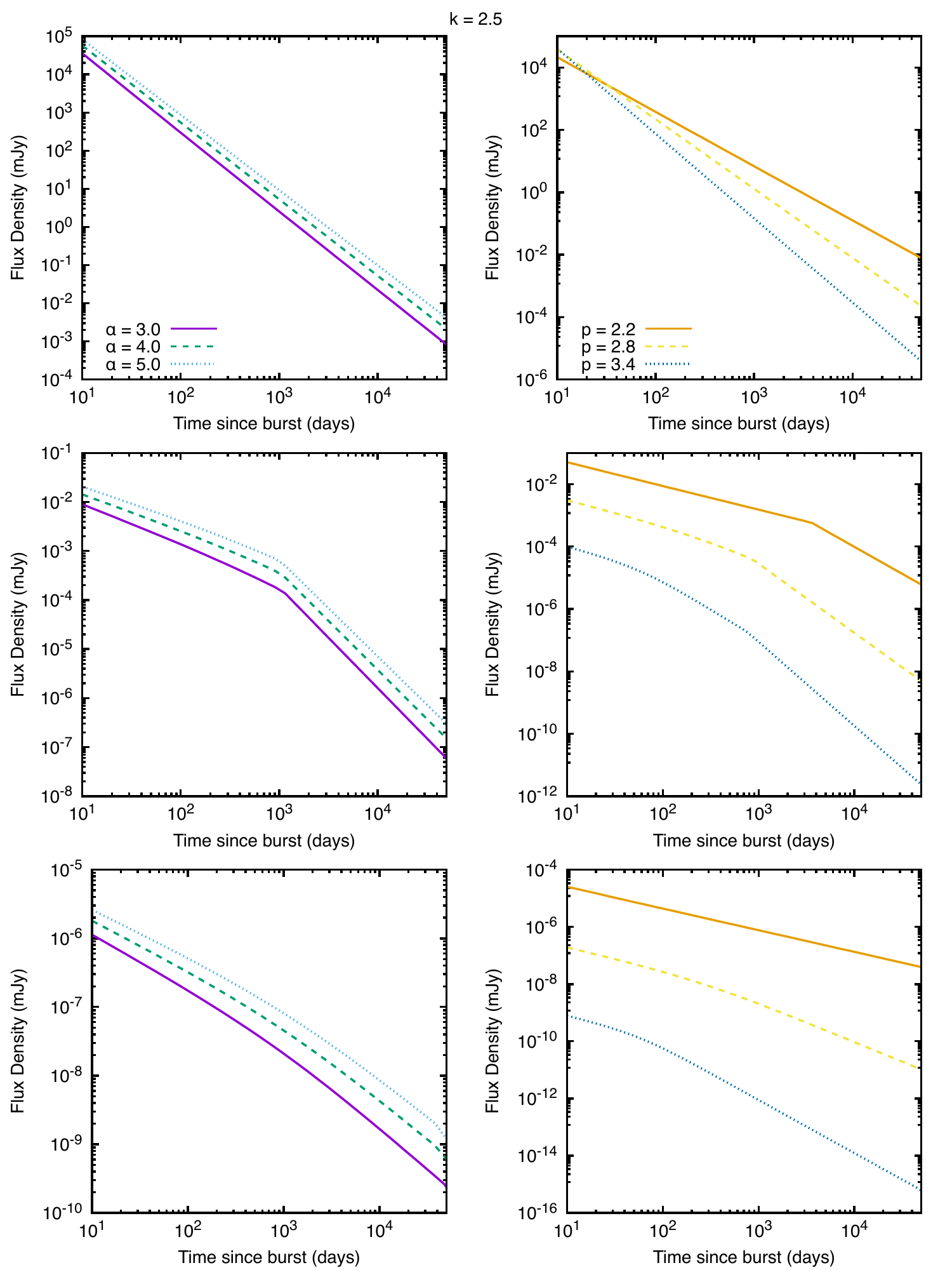}}
}   
\caption{The same as Figure \ref{k_0}, but for ${\rm k=2.5}$ with $A_{2.5}=1.3\times 10^{45}\,{\rm cm^{-\frac12}}$. }
\label{k_2.5}
\end{figure}

Some examples of the light curves in several energy bands are shown in Figures \ref{k_0} - \ref{k_2.5}. Each Figure, in ascending order, presents the radiative behaviour produced by the interaction between the sub-relativistic ejecta and its surrounding medium, which is described by a density profile  $A_{\rm k} r^{-\rm k}$ with ${\rm k}=0$, $1$, $1.5$, $2$ and $2.5$, respectively. Panels from top to bottom correspond to the electromagnetic bands in radio at 1.6 GHz, optical at the R-band and X-rays at 1 keV for $\tilde{E}=10^{49}\,{\rm erg}$, $\epsilon_{\rm B}=10^{-3}$, $\epsilon_{\rm e}=10^{-1}$ and $z=0.023$ \citep{2017LRR....20....3M, 2019ApJ...886L..17H, 2021ApJ...907...78F, 2021arXiv210402070H}.  The left-hand panels show the light curves for $p=2.6$ with $\alpha=3.0$, $4.0$ and $5.0$, and the right-hand panels show the light curves for $\alpha=3.0$ with $p=2.2$, $2.8$  and $3.4$.\\

We present the predicted synchrotron light curves in the previously mentioned energy bands for typical values of GRB afterglows. Most light curves peak on timescales from several months to a few years, which is in agreement with the observations of some SNe, such as SN 2014C \citep{2017ApJ...835..140M} and SN2016aps \citep{2020NatAs.tmp...78N}. There is an outlier, however, present in the flux in the radio band for $p=3.4$ and $\alpha=3.0$, which reaches its maximum in a timescale of a couple of days for stratified media when ${\rm k>1.5}$.  The synchrotron light curves are shown using a stratified medium with density profile $\propto r^{-\rm k}$ with ${\rm k}=0$, $1$, $1.5$, $2$ and $2.5$, which covers both long and short GRB progenitors.  The constant-density medium (${\rm k=0}$) is usually related to short GRBs which stem from the merger of two NSs, while the stratified medium ($1\leq {\rm k} \leq 2.5$) is only associated to long GRBs from dying massive stars with different mass-loss evolution. For instance, in their article, \cite{2013ApJ...776..120Y} investigated the evolution of the emission of forward-reverse shocks propagating in a medium described by the previously mentioned PL distribution. They applied their model to 19 GRBs and found that the density profile index took values of $ 0.4 \leq {\rm k}\leq  1.4$, with a typical value of ${\rm k\sim1}$. This value was also obtained by \cite{2013ApJ...774...13L}, who analyzed a bigger sample of 146 GRBs.\\

All Figures show that, during the early stages of the evolution, there is an increase in the flux. During this epoch, when the sub-relativistic material decelerates in a constant-density medium, the light curve grows steeply. For more stratified media, this growth is not as evident, as the example presented in Figure \ref{k_2.5}, where the early-time behaviour is more gradual. That is to say, the amount of time during which the flux grows depends on the stratification of the surrounding medium. This is exemplified by Figures \ref{k_0} and \ref{k_1}, where the rise in the flux is evident, while in subsequent Figures this growth in not observed, which means that this phase happens much quicker as the stratification is increased. This result implies that if such a flattening or rebrightening at timescales from months to years  in the light curve was observed together with GW detection, then this would be associated with the deceleration of a sub-relativistic material launched during the merger of two NSs.  Otherwise, we show that an observed flux that gradually decreases on timescales from months to years could be associated with the deceleration of a sub-relativistic material launched during the death of a massive star with different mass-loss evolution at the end of its life. It is worth noting that all these results are for on-axis observers, and for an off-axis observer the flux would decrease as the viewing angle between the material and the observer increases. For a relativistic off-axis component in the outflow, the spectral breaks  and the maximum flux are corrected by the Doppler factor ($\delta_D$) as $\nu_{\rm i}=\delta_D/(1+z)\nu'_{\rm i}$ with ${\rm i=a,\,m, c}$ and {\small $F_{\rm \nu, max}=\frac{(1+z)^2\delta^3_D}{4\pi d_z^2}N_eP'_{\nu'_m}$}, respectively, where {\small $P_{\nu_m}=\delta_D/(1+z) P'_{\nu'_m}$} is the radiation power and {\small $N_e=(\Omega/4\pi)\, n(r) \frac{4\pi}{3-k} r^3$} is the total number of emitting electrons with $r=\delta_D/(1+z) \Gamma\beta c t$ the shock radius and the transformation law for the solid angle as $\Omega= \Omega'/\delta^2_D$.  The Doppler factor is defined as $\delta_D=\frac{1}{\Gamma(1-\mu\beta)}$ with $\mu=\cos \Delta \theta$, $\Gamma$ the bulk Lorentz factor and $\Delta \theta=\theta_{\rm obs} - \theta_{\rm j}$ given by the viewing angle ($\theta_{\rm obs}$) and the half-opening angle of the jet ($\theta_{\rm j}$).\\

Table \ref{table:dens} shows the evolution of the density parameter in each cooling condition of  the synchrotron afterglow model. For instance,  the synchrotron light curve in the slow-cooling regime as a function of the density parameter is given by $F_{\nu}:~\propto A_{\rm k}^{\frac{4\alpha+13}{3(\alpha+5-k)}}$ for $\nu_{\rm a,1}<\nu <\nu_{\rm m}$, $\propto A_{\rm k}^{\frac{5\alpha+19+p(\alpha-5)}{4(\alpha+5-k)}}$ for $\nu_{\rm m}< \nu < \nu_{\rm c}$ and $A_{\rm k}^{\frac{p(\alpha-5)+2(\alpha+5)}{4(\alpha+5-k)}}$ for $\nu_{\rm c}< \nu$. Any variation of the density will be better observed in low-energy frequencies, such as radio. Additionally,  it shows that variations of the density profile index are more apparent in the radio light curve when compared to the other fluxes in the other energy bands. This change in the density profile is also more easily appreciated for large values of $\alpha$, namely 4.0 and 5.0. Therefore, a transition between density profiles will be more easily observed in the radio band with high values of the velocity distribution parameter.\\

\subsection{Comparison: with and without energy injection}

\begin{figure}
{\centering
\resizebox*{0.9\textwidth}{0.65\textheight}
{\includegraphics{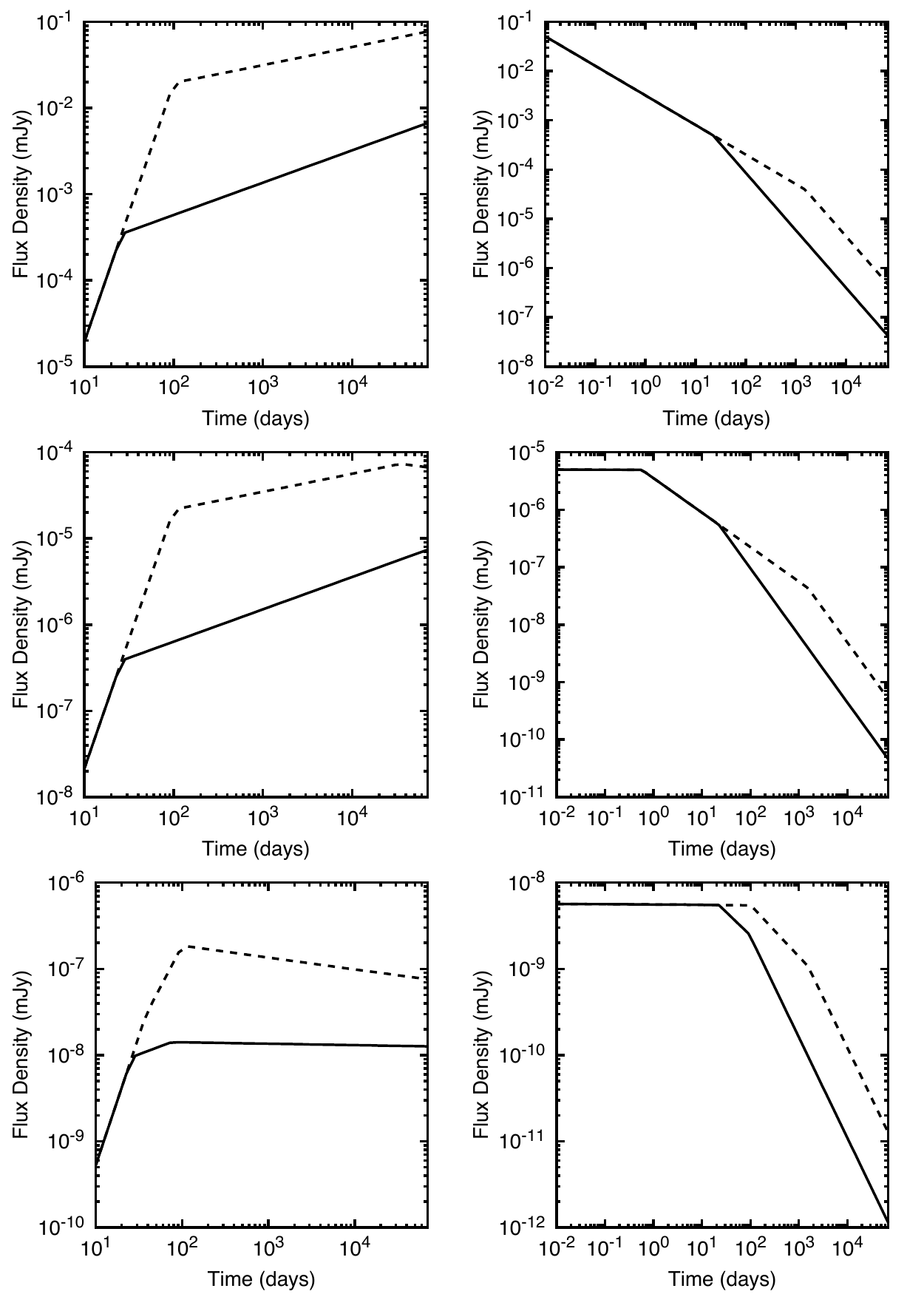}}
} 
\caption{Synchrotron light curves generated by the deceleration of the sub-relativistic ejecta with and without energy injection for a constant-density (left; $k=0$ and $n=1\,{\rm cm^{-3}}$, $q=0$) and stellar-wind (right; $k=2$, and $A_{\rm W}=100$, $q=0.66$) medium.  The millisecond magnetar scenario was required in the left-hand panels with the parameter values of $P_0=10^{-3}\,{\rm s}$,  $B=10^{14}\,{\rm G}$ and $t_{\rm fb}=10^{3}\,{\rm s}$, and the fall-back material onto BH scenario was used in the right-hand panels with the parameter values of $t_0=0\,{\rm s}$, $t_p=10^3\,{\rm s}$, $M_{\rm BH}=2.3\,{\rm M_{\odot}}$, $a=0.7$,   $\tau_{\rm vis}=10^{9}\,{\rm s}$ and $\dot{M}_{\rm p}=10^{-6}\,{\rm M_\odot\,s^{-1}}$.  Panels from top to bottom correspond to radio (1.6 GHz),  optical (R-band) and X-ray (1 keV) bands, respectively. The dashed  line represents the case of energy injection and the solid line without energy injection. The parameter values used in both panels are  $\epsilon_{\rm e}=10^{-1}$, $\epsilon_{\rm B}=10^{-4}$, $\beta=0.4$, $\tilde{E}=10^{51}\,{\rm erg}$ and $\alpha=3.0$.}
\label{compar_w_and_wout}
\end{figure}

Figure \ref{compar_w_and_wout} shows several synchrotron light curves in order to compare the effects of the continuous injection of energy into the blastwave. It is divided into two columns, the one on the left considers a constant-density medium (${\rm k=0}$) with the millisecond magnetar remnant, while the one on the right takes into account a stellar wind (${\rm k=2}$) with the scenario of fall-back material onto a BH. The panels from top to bottom correspond to radio (1.6 GHz), optical (R-band) and X-ray (1 keV), respectively.  Every panel shows two curves, the dashed line represents the evolution of the flux density with energy injection, and the solid line stands for an evolution with no injection of energy.\\

It can be seen in all panels on the left that regardless of injection of energy or not, both solutions are the same at early times. However, all solid curves remain several orders of magnitude below their dashed counterparts after the evolution has reached times of approximately $20$ days. There also seems to be a difference in timescales when the maximum is reached, as the dashed curves reach it within the limits of the plot, while the solid lines continue their upward trend. This is noted by observing that for late times ($>10^{4}\, \rm days$) the rise in the light curve is steeper when there is energy injection, while the solution represented by the dashed lines reaches its peak with a smaller slope and then begins to drop.\\

On the other hand, the panels on the right present contrasting behaviour, as both the case with energy injection and the case without reach the same peak flux, the only difference between them being the decay of the light curves. For a stellar-wind environment, the late-time behaviour is the opposite of the constant medium case, namely that the flux density drops and this decay is less sharp when there is energy injection.\\

Upon comparison between the right and left columns, it can also be observed that the early-time behaviour is different in the optical and X-ray bands. Regardless of whether there is energy injection or not, the flux density in the constant medium increases until its peak. For a stellar-wind environment, however, the light curve remains constant and begins to decay very slowly. This behavior is due to the change in the energy injection. At early times it evolves as $q=2/3$ (up to $10^3\,{\rm days}$), and at later times with $\dot{M}_{\rm fb}\propto t^{-\frac53}$ the BH torus system has no effect on the afterglow evolution \citep[e.g.,][]{2001ApJ...550..410M, 2004MNRAS.355..950J, 2006ApJ...642..354Z}.  It is worth noting that after $10^{3}\,{\rm days}$ both curves decrease with the similar slope, as expected. We emphasize that for $q=1$ the standard synchrotron light curves are recovered \citep[e.g.,][]{2006ApJ...642..354Z}.   

\section{Synchrotron emission from different ejected Materials and Applications}

It is believed that sub-energetic GRBs are quasi-spherical explosions whose dominating components are the sub-relativistic materials, which contribute approximately $99.9\%$ of the explosion's energy. The mildly relativistic materials, on the other hand, correspond to only $\approx0.1\%$ \citep[e.g., see][]{2014ApJ...797..107M, 2020ApJ...892..153M}.  There is wide agreement in the community that the origin of sGRBs and lGRBs is closely related to the merger of BCOs and the death of massive stars leading to KNe and SNe, respectively. In addition to KN and SN materials, other types of materials are launched into the circumstellar medium with different velocities and, as such, will contribute at distinct timescales in distinct energy bands with contrasting intensities. In the following we will give a brief introduction about the values of masses, the isotropic-equivalent kinetic energies and velocities of each decelerated material that is ejected during the merger of two NSs, namely the dynamical ejecta, the shock breakout material, the disk wind and the cocoon material.

\paragraph{Dynamical ejecta}

At the moment of the merger of two NSs, matter is ejected dynamically from their surfaces due to gravitational and hydrodynamical interactions \citep{1994ApJ...431..742D, 1997A&A...319..122R, 1999A&A...341..499R}. Based on numerical simulations, the mass of the material liberated, the kinetic energy, and the velocities lie in the ranges of $10^{-4}\lesssim M_{\rm ej}\lesssim10^{-2}\,{\rm M_{\odot}}$,  $10^{49}\lesssim \tilde{E}\lesssim10^{51}\,{\rm erg}$ and $0.1\lesssim\beta\Gamma\lesssim0.3$, respectively \citep[e.g., see][]{2011ApJ...738L..32G, 2013ApJ...778L..16H,2013ApJ...773...78B, 2013MNRAS.430.2121P, 2014ApJ...789L..39W, 2014MNRAS.439..757G}.

\paragraph{Shock breakout material}  A shock at the interface between the two NSs is formed the moment directly after their coalescence. This shock manages to break out from the NS core to the crust at sub-relativistic velocities \citep[$\beta_{\rm in}\simeq 0.25$   e.g., see][]{2014MNRAS.437L...6K, 2015MNRAS.446.1115M}. When the shocked material reaches half of the escape velocity it converts a fraction of the shock-heated internal energy into kinetic energy and it escapes the merger into a nearly vacuum region \citep[for details see][]{2014MNRAS.437L...6K, 2019ApJ...871..200F}.  The shock breakout material's properties depend on the mass, radius and velocity of the merger remnant. Numerical simulations indicate that the material mass, the kinetic energy and the velocities lie in the ranges of $10^{-6}\lesssim M_{\rm ej}\lesssim10^{-4}\,{\rm M_{\odot}}$,  $10^{47}\lesssim \tilde{E}\lesssim10^{50.5}\,{\rm erg}$ and $\beta\Gamma \gtrsim0.8$, respectively  \citep[e.g., see][]{2014MNRAS.437L...6K, 2015MNRAS.446.1115M}.\\

\paragraph{Disk wind}  The coalescence of the NS binary will end in a tidal disruption and some of the material of the stars will be shed, forming an accretion disk around the central remnant. This component of the sub-relativistic material represents a significant portion of the total material mass and might dominate over other constituents \citep{2017PhRvL.119w1102S}. The mass of the accretion disk will depend on the initial NS spins and is located in the range of $10^{-3}\lesssim M_{\rm ej}\lesssim 0.3\,{\rm M_{\odot}}$
\citep{2006PhRvD..73f4027S, 2013ApJ...778L..16H}. The disk's kinetic energy and the velocities lie in the ranges of  $10^{47}\lesssim \tilde{E}\lesssim10^{50}\,{\rm erg}$ and $0.03 \lesssim \beta\Gamma \lesssim0.1$, respectively  \citep[e.g., see][]{2009ApJ...690.1681D, 2014MNRAS.441.3444M, 2015MNRAS.446..750F}.

\paragraph{Cocoon material}  The GRB jet will deposit energy as it travels through the neutrino-driven or magnetically driven wind (previously expelled during the merger of two NSs). The energy deposited laterally will produce a cocoon with an energy similar to that of the jet's electromagnetic emission. \cite{2014ApJ...788L...8M} looked into the conditions required for cocoon formation as a function of the jet's luminosity. A weak cocoon emission was predicted, regardless of the magnitude of the jet's luminosity. In particular, when \cite{2014ApJ...784L..28N} numerically examined a low-luminosity jet, the authors concluded that a hot cocoon enclosing the jet would form. The cocoon would break free and spread along the axis of the relativistic jet as soon as it would reach the shock-breakout material.
The external pressure would then drop dramatically beyond the breakout material, allowing the cocoon to accelerate and expand relativistically until it became transparent.   The material mass liberated in the cocoon, the kinetic energy and the velocities lie in the ranges of  $10^{-6}\lesssim M_{\rm ej}\lesssim10^{-4}\,{\rm M_{\odot}}$,  $10^{47}\lesssim \tilde{E}\lesssim10^{50.5}\,{\rm erg}$ and $0.2\lesssim\beta\Gamma\lesssim10$, respectively  \cite[e.g., see][]{2014ApJ...784L..28N, 2014ApJ...788L...8M,2017ApJ...848L...6L, 2018PhRvL.120x1103L, 2017ApJ...834...28N,2018MNRAS.473..576G}.\\

Figure \ref{Short_long_GRBs} shows the synchrotron light curves with energy injection by a spinning magnetized NS remnant and generated by materials ejected from the merger of two NSs such as the dynamical ejecta,  the cocoon material, the shock breakout material and the wind ejecta. The synchrotron light curves correspond to the (from top to bottom) radio (1.6 GHz) and X-ray (1 keV) bands, respectively for ${\rm k=0}$.  Taking into account the velocities of the wind and the shock breakout material, we also consider the trans-relativistic (TR; $\beta\sim 0.8$) and Deep-Newtonian (DN; $\beta\sim0.08$) regimes, respectively. The TR and DN timescales during the deceleration phase are given in Appendix.   The light curves are shown  for $n=10^{-2}\,{\rm cm^{-3}}$, $\alpha=3$, $P_0=10^{-3}\,{\rm s}$,  $B=10^{16}\,{\rm G}$, $t_{\rm fb}=5\times10^9\,{\rm s}$, $\epsilon_{\rm B}=10^{-2}$, $\epsilon_{\rm e}=10^{-1}$ and $p=2.2$ and the pair of values  of  $\tilde{E}=10^{50}\,{\rm erg}$ and $\beta=0.2$ for the dynamical ejecta, $10^{48}\,{\rm erg}$ and $0.3$ for the cocoon material, $10^{48.5}\,{\rm erg}$ and $0.8$ for the shock breakout material and $10^{50}\,{\rm erg}$ and $0.07$ for  the wind. Besides, we show for completeness the synchrotron afterglow radiation from an on-axis and off-axis outflow with a viewing angle of $30^\circ$, which are specified in the top panel. The synchrotron light curves from the off-axis outflow are considered as detailed in \cite{2019ApJ...884...71F}.\\

The disk wind, the dynamical ejecta, the cocoon and the shock breakout peak at $1.2\times 10^5\, {\rm days}$, $7150\, {\rm days}$, $522.5\, {\rm days}$ and $56.1\, {\rm days}$, respectively.  The total contribution of synchrotron light curves exhibits a brightening once the off-axis emission decreases at a few years, so that, depending on the parameter values and conditions, the synchrotron emission from the ejected materials could be detected or not \cite[see,][for the cocoon material]{2014ApJ...788L...8M, 2014ApJ...784L..28N}.\\

The top panel shows the evolution of the integrated light curve in the radio band. The early-time behaviour of the radiation due to the jet's emission is emphasized by showing different solutions for distinct viewing angles. This panel shows that, as the viewing angle is increased, the light curve at early times decreases and takes longer to enter our line of sight. Once the jet enters on-axis, the behaviour is independent of the initial opening angle, however, the light curve becomes dominated by the emission from the other emitted materials.\\

This last point is highlighted in the subsequent lower panel, where each component's contribution is explicitly plotted. It can be seen that at timescales of $\approx10^3-10^4\,{\rm days}$, the flux density is dominated by the emission from the dynamical ejecta and the shock breakout, while at late times ($\gtrsim 10^5\,{\rm days}$) both the dynamical and the disk wind ejecta have the upper hand as the most influential constituents of the synchrotron light curve. Contrastingly, the cocoon emission lies a couple of orders of magnitude below the other contributions and is not observed.\\ 

Figure \ref{Short_long_GRBs_f2} shows the synchrotron light curves with energy injection produced by a fall-back material onto a BH and generated when ejected materials and a relativistic outflow decelerate in a stellar wind environment. The synchrotron light curves correspond to the (from top to bottom) radio (1.6 GHz) and X-ray (1 keV) bands, respectively for ${\rm k=2}$,  $A_{\rm 2}=3\times 10^{36}\,{\rm cm^{-1}}$ ($A_{\rm W}=10$), $t_0=1\,{\rm s}$, $t_p=10^3\,{\rm s}$, $M_{\rm BH}=2.3\,{\rm M_{\odot}}$, $a=0.7$,   $\tau_{\rm vis}=10^{9}\,{\rm s}$, $\dot{M}_{\rm p}=10^{-6}\,{\rm M_\odot\,s^{-1}}$, $\alpha=3.0$, $\epsilon_{\rm B}=5\times 10^{-3}$, $\epsilon_{\rm e}=10^{-1}$ and $p=3.2$, and the pair of values $10^{48}\,{\rm erg}$ and $0.3$ for the cocoon material, and $10^{48.5}\,{\rm erg}$ and $0.8$ for the shock breakout material.\\

For a stellar-wind surrounding medium, the jet's contribution on the radio light curve is much shorter than for the case of a constant-density medium, as the top panel shows that it lasts less than a day, while the left-hand panels show that this phase has a duration $\gtrsim10^3$ days. On the other hand, the light curve at 1 keV shows that the flux density is completely dominated by the jet's emission, as the cocoon's and the shock breakout's contributions lie several orders of magnitude below the jet's. Therefore, a high variation in the behaviour of the light curves while comparing the different energy bands could be a hint for an outflow evolving in a medium that is stratified. The synchrotron light curves at the radio bands show that, depending on the parameter values, the afterglow emission from the cocoon and shock breakout materials could be detected on timescales of days. However, if the outflow is chocked, radio fluxes could be detected on timescales of hours.\\

GRBs could be successful or chocked, this is determined by the range of values in the observables such as luminosity, duration and bulk Lorentz factor \citep[e.g., see][]{2001ApJ...550..410M, 2001PhRvL..87q1102M, 2014ApJ...788L...8M, 2014MNRAS.437.2187F, 2014ApJ...784L..28N, 2017MNRAS.472..616S, 2011ApJ...739L..55B}. As an indication of this behavior, successful GRBs might be less frequent than choked ones, only limited by the ratio of SNe (types Ib/c and II) to lGRB rates \citep{2003ApJ...598.1151T, 2005PhRvL..95f1103A}. Some SNe of type Ic-BL not connected with GRBs have been suggested to arise from events such as off-axis GRBs or failed jets \citep[e.g., see][]{2019Natur.565..324I, 2020arXiv200405941I,2020MNRAS.493.3521B}. One exponent of such a possibility is the failed burst GRB 171205A, which besides being associated to SN 2017iuk,  exhibited material with high expansion velocities $\beta \sim 0.4$ interpreted as mildly relativistic cocoon material  \citep{2019Natur.565..324I}. As another example, in the context of off-axis GRBs that enter on-axis after some time, \cite{2020arXiv200405941I}  found that the X-ray observations from the nearby SN 2020bvc were consistent with the afterglow emission generated by an off-axis jet with viewing angle of $23^\circ$ when it decelerated in a circumburst medium  with a density profile with ${\rm k}=1.5$. However, as no prompt emission was detected, the authors implied that this was a hint for the first orphan GRB detected through its associated SN emission.\\

As follows we apply the current model to describe the latest multi-wavelength afterglow observations ($\gtrsim 900$ days) of the GW170817/GRB 170817A event, and using multi-wavelength upper limits associated with i) promising GW events in GWTC-2 and GWTC-3 that could generate electromagnetic emission, ii) short-bursts with the lowest-redshifts ($100\leq {\rm z}\leq 200\, {\rm Mpc}$), and iii) evidence of KNe, we provide constraints on the possible afterglow emission.

\subsection{GW170817/GRB 170817A event}
GW radiation from the merger of two NSs \citep{2017LRR....20....3M} is expected together with a short gamma-ray prompt and an UV-optical-IR KN emission in timescales of $\sim 1\, {\rm s}$ and a few days, respectively \citep{1998ApJ...507L..59L, 2005ApJ...634.1202R, 2010MNRAS.406.2650M, 2013ApJ...774...25K, 2017LRR....20....3M}. A KN classified as ``blue" and ``red" is a transient powered by radioactive decay of unstable heavy nuclei via the rapid neutron capture (r-process) synthesized in merger ejecta.   The ``blue" KN situated at the polar regions has low opacity and fast velocity $\beta \simeq 0.3$ and the ``red" KN positioned at the equatorial plane has high opacity due to the Lanthanide-bearing matter and slower velocity $\beta\simeq 0.1$ \citep{2014MNRAS.441.3444M, 2014MNRAS.443.3134P, 2014ApJ...789L..39W, 2019PhRvD.100b3008M}.\\

As follows, we present the GW170817/GRB 170817A observations, focusing on and describing the latest ones with synchrotron afterglow radiation from a sub-relativistic material, i.e., a KN afterglow emission, and when the spinning magnetized NS remnant is accreting and injecting energy into afterglow. We consider the characteristics of the ``blue" KN as used in  \cite{2021ApJ...907...78F}

\subsubsection{Multi-band observations}
On 2017 August 17 12:41:06, the short GRB 170817A was first detected by the Gamma-ray Burst Monitor (GBM) instrument aboard the Fermi satellite with a reported location of R.A.$=176.8$ and Dec.$=−39.8$ with an error of $11.6\ \mathrm{deg}$ \citep{2017ApJ...848L..14G}.   Approximately two seconds before the GBM trigger, the LIGO Scientific Collaboration and the Virgo Collaboration reported the identification of a GW candidate (GW170817) consistent with the same location of GRB 170817A \citep{2017GCN.21506....1C, PhysRevLett.119.161101, 2041-8205-848-2-L12}. Relevant evidence soon connected the progenitor of GRB 170817A with the merger of two NSs, being the first detection of GWs from this merger \citep{PhysRevLett.119.161101, 2041-8205-848-2-L12}.\\ 

About 10.9 hours after the GW trigger, this event also exhibited a transient and fading optical source.  This optical transient, named Swope Supernova Survey 2017a (SSS17a), coincident  with  the  quiescent  galaxy  NGC 4993  at  a  distance  of $d_z=40.7\pm2.36\,{\rm Mpc}$ \citep{2018ApJ...854L..31C} was associated with KN emission  \citep[AT2017gfo;][]{2017ApJ...848L..16S, 2017Sci...358.1556C, 2017ApJ...848L..17C, 2017ApJ...848L..27T, 2017Natur.551...75S, 2018MNRAS.473..576G}.  This burst was followed up by an enormous observational campaign covering radio, optical and X-ray bands \citep[e.g., see][and references therein]{2041-8205-848-2-L12, 2017ApJ...848L..12A, 2018Natur.554..207M, 2017Natur.547..425T, 2018NatAs...2..751L, 2018ApJ...856L..18M, 2017Natur.551...71T, 2018ApJ...863L..18A, 2018A&A...613L...1D}. The observations of the non-thermal spectrum of GRB 170817A gathered during the first $\approx 900$ days after the initial merger were analyzed by several authors and it was shown that they were consistent with synchrotron radiation from an off-axis structured jet decelerated in a constant-density medium.  This relativistic jet observed from a viewing angle of $15^{\circ}\leq \theta_{\mathrm{obs}}\leq 25^{\circ}$ was described with an opening angle $\theta_{j}\approx5^{\circ}$ \citep{2017Sci...358.1559K, 2017MNRAS.472.4953L,  2018Natur.561..355M, 2019ApJ...871..123F}.\\

\cite{2021arXiv210402070H} analyzed the latest X-ray and radio observations of GRB 170817A collected with the Chandra X-ray Observatory, the Very Large Array (VLA), and the MeerKAT radio interferometer about 3.3 years after the initial merger, and reported evidence of a new X-ray emission component. This new measurement was not in agreement with the synchrotron off-axis afterglow model in constant-density. Given these contrasting properties, the authors offered the solution to explain this phenomena in the framework of either radiation from accretion processes on the compact-object remnant or a KN afterglow.

\subsubsection{Analysis, Description and Discussion}
\begin{figure}
{\centering
\resizebox*{\textwidth}{0.6\textheight}
{\includegraphics{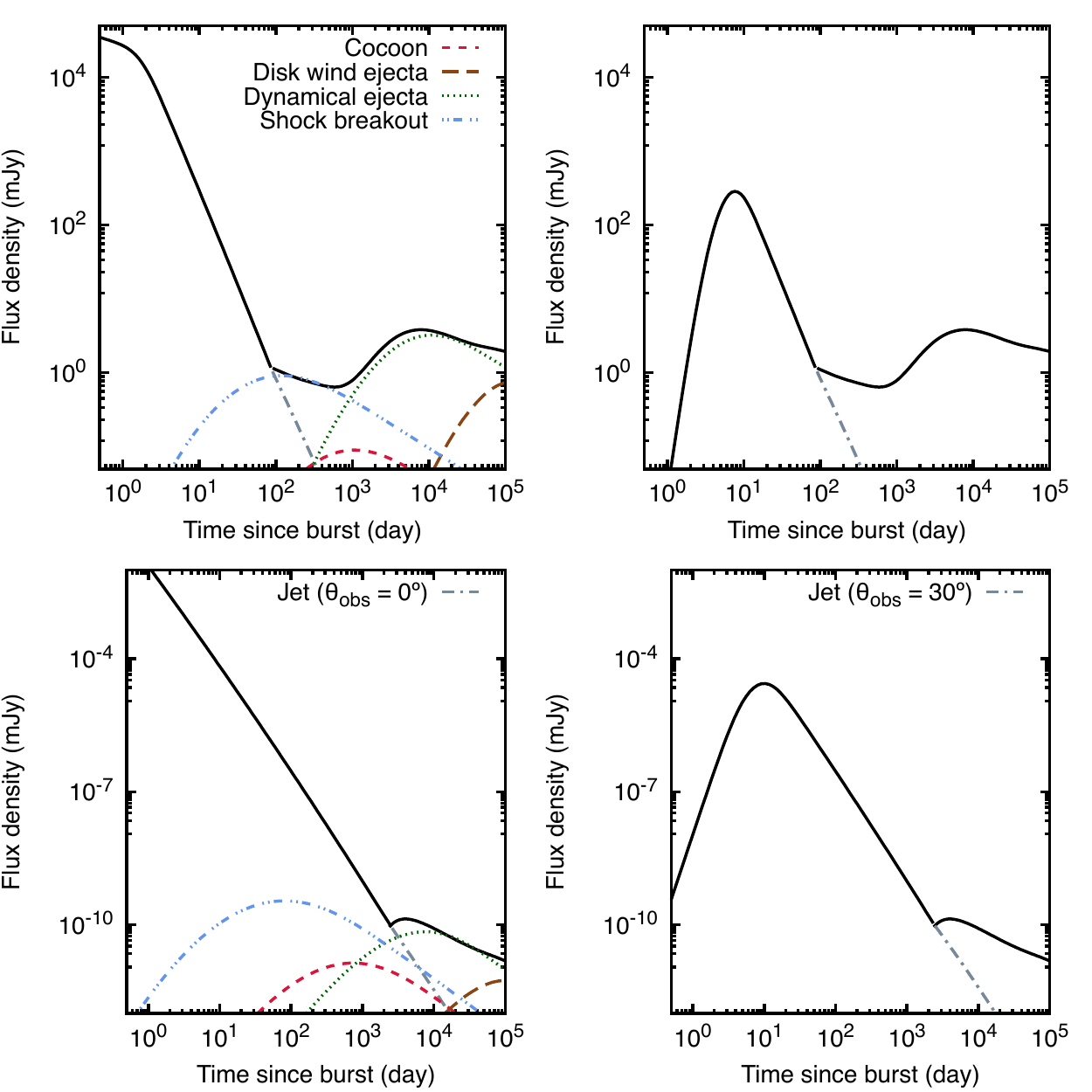}}
} 
\caption{Synchrotron light curves generated by the deceleration of the relativistic jet, cocoon, the shock breakout, the dynamical, and the wind ejecta  in the constant-density medium ($k=0$). The left-hand panels shows the light curves from an on-axis jet and the right-hand ones from an off-axis jet with a viewing angle of $30^\circ$. The spinning magnetized NS scenario with the parameter values $P_0=10^{-3}\,{\rm s}$,  $B=10^{16}\,{\rm G}$ and $t_{\rm fb}=5\times 10^9\,{\rm s}$ was used.  
Upper and lower panels correspond to radio (1.6 GHz) and X-rays (1 keV) bands, respectively.   The gray dash-dotted,  the red dashed, brown long-dashed, the green dotted and the blue double-dash-dotted curves represent the contribution of the relativistic jet, the cocoon, the shock breakout, the dynamical and the wind ejecta.   The values of $\tilde{E}=10^{50}\,{\rm erg}$,  $\epsilon_{\rm B}=10^{-2}$, $\epsilon_{\rm e}=10^{-1}$,  $n=10^{-2}\,{\rm cm^{-3}}$, $p=2.2$  and $d_{z}\approx 100\ \mathrm{Mpc}$ ($z=0.023$) were used \citep{2019ApJ...884...71F}.
}
\label{Short_long_GRBs}
\end{figure}

\begin{figure}
{\centering
\resizebox*{\textwidth}{0.6\textheight}
{\includegraphics{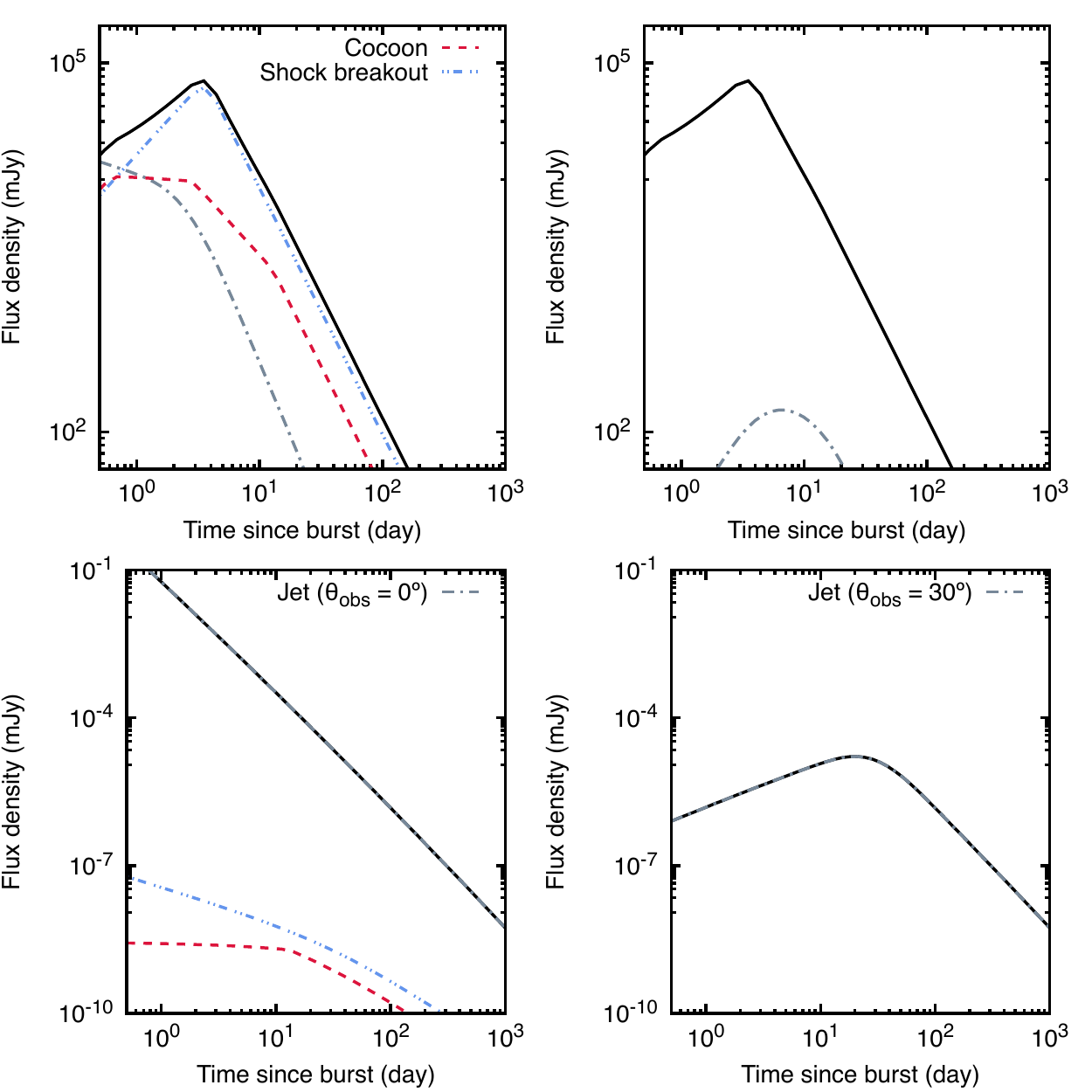}}
} 
\caption{The same as Figure \ref{Short_long_GRBs}, but the deceleration of the cocoon and the shock breakout in the stellar wind medium ($k=2$). The fall-back material onto BH scenario with the parameter values $t_0=1\,{\rm s}$, $t_p=10^3\,{\rm s}$, $M_{\rm BH}=2.3\,{\rm M_{\odot}}$, $a=0.7$,  $\tau_{\rm vis}=10^{9}\,{\rm s}$ and $\dot{M}_{\rm p}=10^{-6}\,{\rm M_\odot\,s^{-1}}$ was used.
}
\label{Short_long_GRBs_f2}
\end{figure}

The observations of the non-thermal spectrum of GRB 170817A gathered during the first $\approx 900$ days after the initial merger have been modelled with synchrotron forward-shock emission generated by the deceleration of a relativistic off-axis jet \citep{troja2017a, 2017Sci...358.1559K, 2017MNRAS.472.4953L, 2018MNRAS.478..733L, 2018ApJ...867...57R, 2017ApJ...848L..20M}, a cocoon \citep{2017ApJ...848L...6L, 2018MNRAS.479..588G, 2019ApJ...884...71F} and a shock breakout \citep{2018MNRAS.479..588G, 2018ApJ...867...95H, 2019ApJ...871..200F} in a constant-density medium. While the syncrotron radiation from an off-axis jet peaked at $\approx 110-130\,{\rm days}$,  the synchrotron radiation from the relativistic cocoon material with a bulk Lorentz factor ($\Gamma_c\gtrsim 4$) increased gradually during the first weeks, after reached a maximum flux at $\sim 15-45\, {\rm days}$ and decreased afterwards \citep{2017ApJ...848L...6L, 2019ApJ...884...71F}. A similar description was performed considering the relativistic shock breakout material \cite[e.g., see][]{2019ApJ...871..200F}. On the other hand, Figure \ref{Short_long_GRBs} shows that disk wind ejecta peaks at time scales as longer as $\approx 10^5\,{\rm s}$ (see eq. \ref{t_dec_DN}). Therefore, we only consider the sub-relativistic decelerated material with the typical parameters of the dynamical ejecta which peaks at timescales of years (see Eq. \ref{t_dec_SubRel}), as shown in Figure \ref{Short_long_GRBs}.\\

\begin{figure}
{\centering
\resizebox*{\textwidth}{0.4\textheight}
{\includegraphics{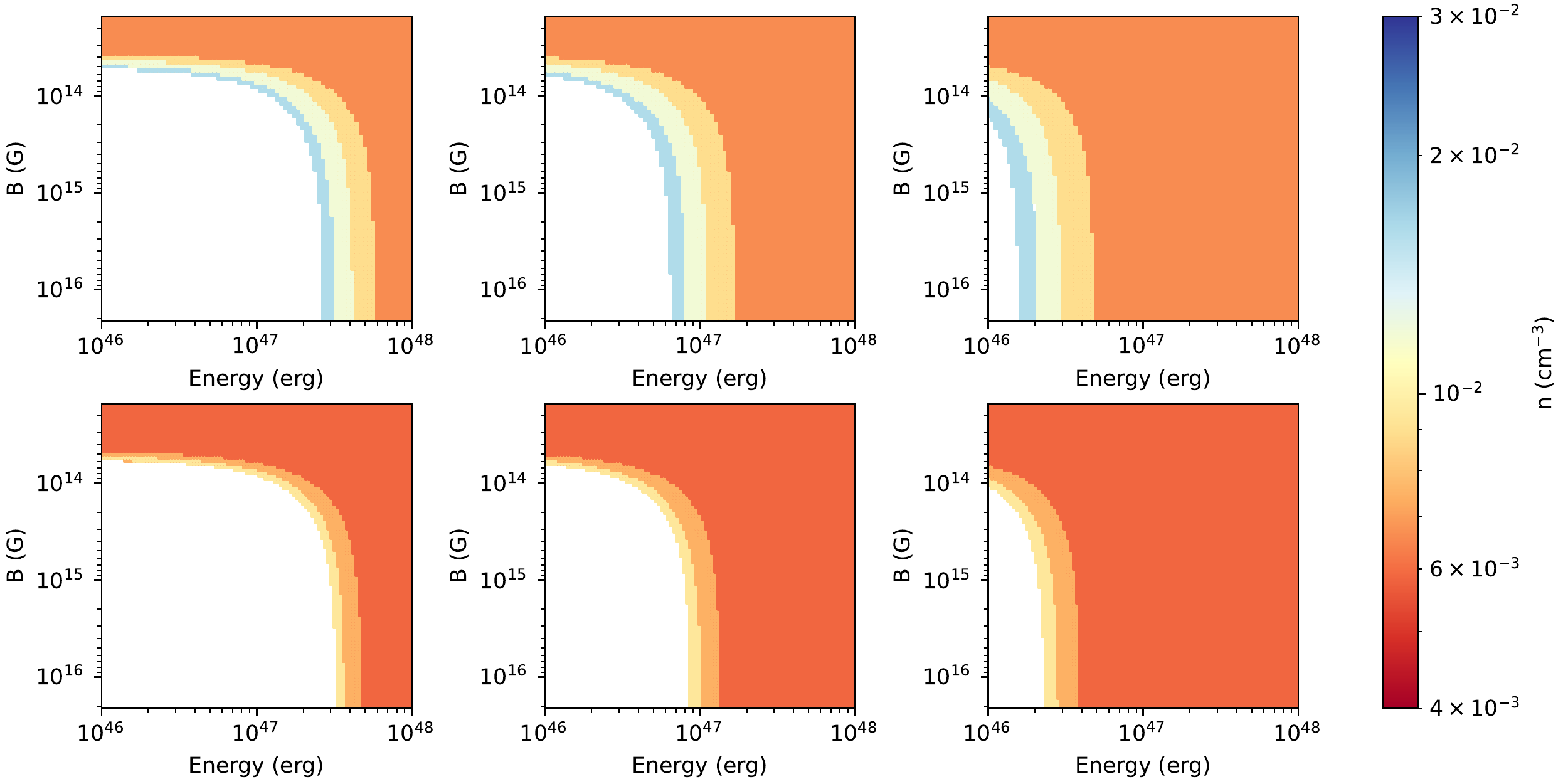}}}
\caption{The {\bf allowed} parameter space of the luminosity injection, isotropic-equivalent kinetic energy and constant-density medium for $\epsilon_{e}=10^{-1}$, $\epsilon_{B}=10^{-3}$, $\beta=0.3$, $p=2.05$ (upper row) and $p=2.15$ (lower row), $\alpha=3$ (left column),  $\alpha=4$ (middle column) and $\alpha=5$ (right column). {\bf The white spaces correspond to the excluded parameters.}} 
\label{par_space_beta0.3}
\end{figure}

\begin{figure}
{\centering
\resizebox*{\textwidth}{0.4\textheight}
{\includegraphics{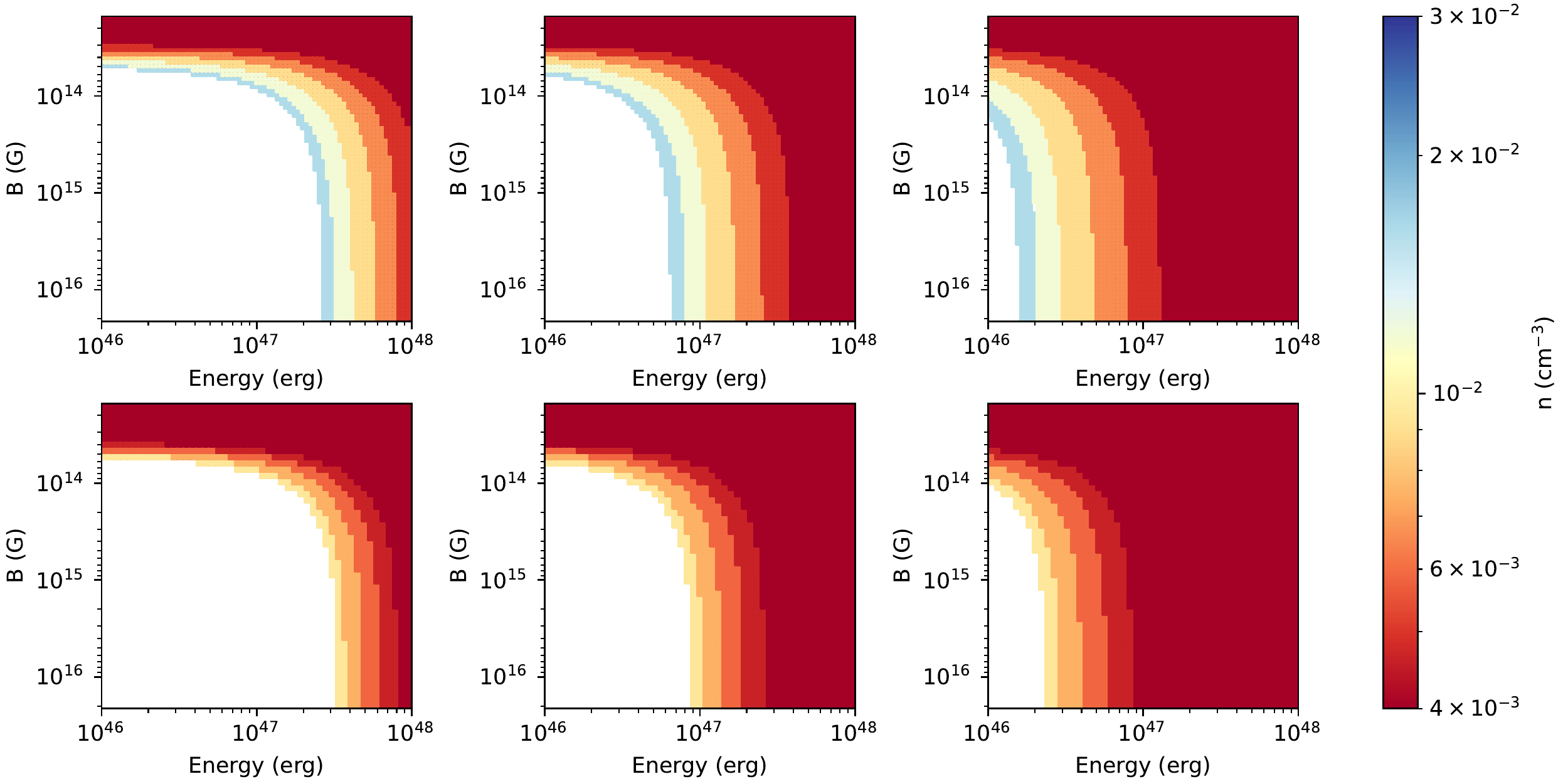}}
} 
\caption{The same as Figure \ref{par_space_beta0.4}, but for $\beta=0.4$.}
\label{par_space_beta0.4}
\end{figure}

We use the X-ray, optical and radio observations of GRB 170817A displayed in \cite{2021ApJ...907...78F} and \cite{2021arXiv210402070H},  together with the best-fit curve found by the off-axis jet with cocoon model, which is introduced in \cite{2019ApJ...884...71F}. To describe the latest multi-band observations through the sub-relativistic decelerated material in a constant-density medium, we constrain the parameter space that reproduces their synchrotron light curves.  Figures \ref{par_space_beta0.3} and \ref{par_space_beta0.4} show the parameter space allowed for $\beta=0.3$ and $0.4$, respectively. We consider a spinning magnetized NS as the remnant of the merger of two NSs which is continuously injecting energy into the blastwave due to magnetic spin-down. For a typical value of efficiency $\eta=0.1$ \citep{2019ApJ...878...62X} and the half-opening angle of the KN AT2017gfo associated with GW170817  \citep[$\theta_j=30^\circ$;][]{2020ApJ...888...67D}, the luminosity injected to the afterglow becomes similar to the spin-down luminosity.   Both figures are displayed as a function of the magnetic field of the NS remnant, the isotropic-equivalent kinetic energy $\tilde{E}$ and the constant-density medium $n=A_{\rm 0}$ that describe the latest multi-band observations for values of the synchrotron afterglow model $\epsilon_{\rm e}=10^{-1}$, $\epsilon_{\rm B}=10^{-3}$, $\alpha=\{3.0,\,4.0, 5.0\}$ and $p=\{2.05,\,2.15\}$ \citep{2013ApJ...773...78B, 2017Natur.551...71T, 2017LRR....20....3M, 2019ApJ...883L...1F, 2018Natur.561..355M, 2019ApJ...871..123F, 2019LRR....23....1M, 2019MNRAS.487.3914K, 2020arXiv200601150T, 2021ApJ...907...78F}. We can see that these parameter spaces are strongly degenerate. \\  

The three columns in Figures \ref{par_space_beta0.3} and \ref{par_space_beta0.4} correspond to the values $\alpha=3.0$, $4.0$ and $5.0$ from left to right, respectively. In a similar manner, the two rows correspond to the values $p=2.05$ and $2.15$ from top to bottom, respectively. Figure \ref{par_space_beta0.3} shows that approximately the same parameter space of the magnetic field and isotropic-equivalent kinetic energy is allowed for both values of the electron energy distribution index $p$. Nevertheless, there are differences in the allowed values of the circumburst density, as Figure \ref{par_space_beta0.3} shows that for $p=2.05$, larger densities ($n\sim10^{-2}\,{\rm cm^{-3}}$) are preferred, while for $p=2.15$, densities of the order $n\sim10^{-3}\,{\rm cm^{-3}}$ are favored. On the other hand, upon increase of $\alpha$, it is shown that the 3D parameter space shrinks, which means that smaller values of $\alpha$ are able to give stronger constraints on the rest of the parameters.

Figure \ref{par_space_beta0.4} is the same as \ref{par_space_beta0.3}, but it considers $\beta=0.4$. Overall, the regions of allowed energy and magnetic field  parameter space are very similar to the ones from Figure \ref{par_space_beta0.3} and the same behaviour as described in the previous paragraph can be observed. The main difference between both figures is that for $\beta=0.4$ lower number densities are preferred, as shown by the deeper red color of the plots.  For both values of velocity ($\beta=0.3$ and $0.4$), $t_{\rm fb}=5\times 10^9\,{\rm s}$ and $10^{13.7}\lesssim B \lesssim 10^{16.3}\,{\rm G}$, the values of $p=2.15$, $n<5\times 10^{-3}\,{\rm cm^{-3}}$ and $\alpha=5$ are preferred, although the value of $\alpha=3.0$ is not discarded.

\begin{figure}
{\centering
\resizebox*{\textwidth}{0.6\textheight}
{\includegraphics{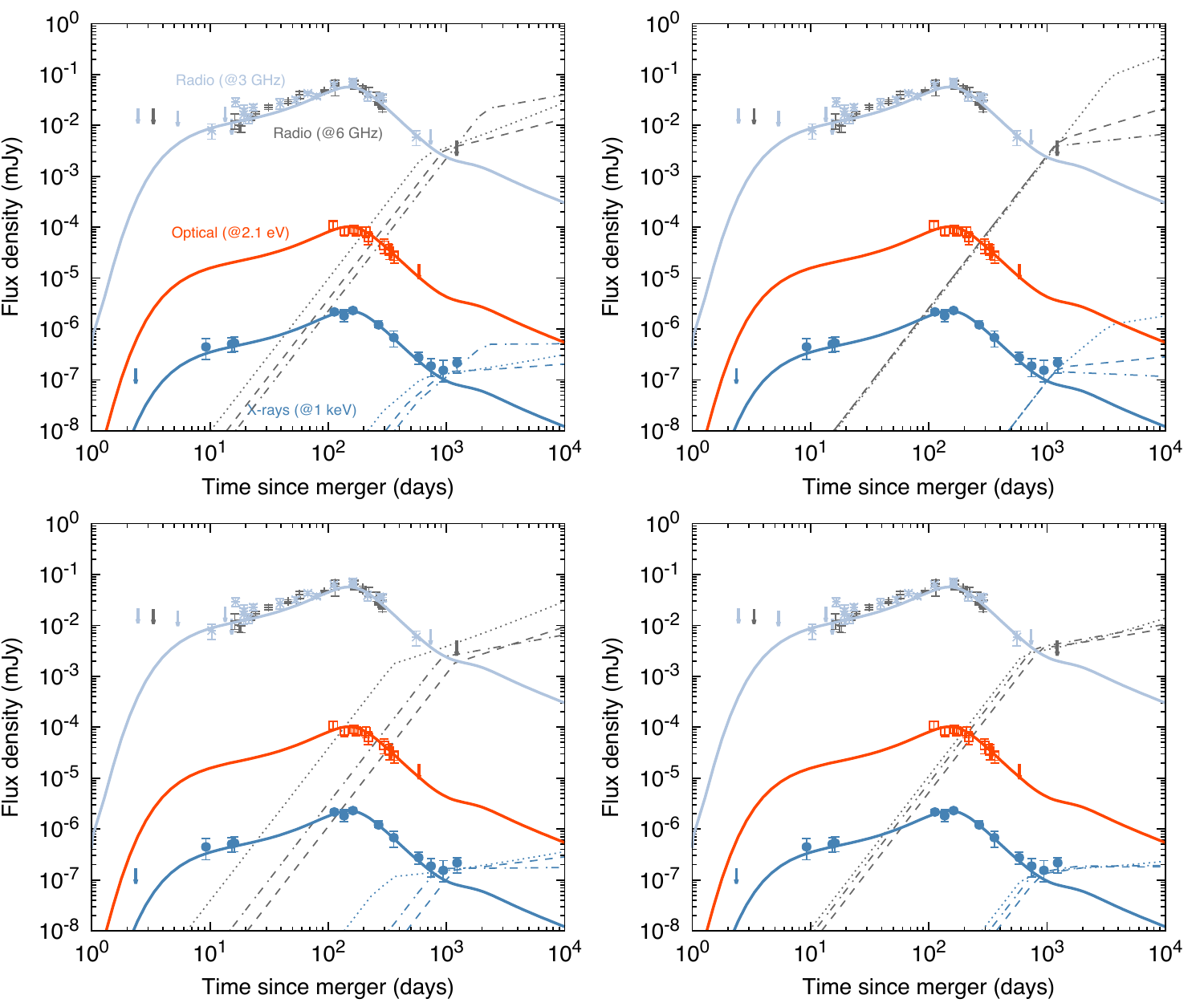}}
} 
\caption{The multi-wavelength afterglow observations of GW170817/GRB 170817A at X-rays, optical bands and radio wavelengths with the best-fit synchrotron light curves from a relativistic structured jet \cite[solid lines;][]{2019ApJ...884...71F} and sub-relativistic material (dotted, dashed and dot-dashed lines).  The latest X-ray data point and the radio upper limit are taken from \cite{2021arXiv210402070H}. The synchrotron light curves with energy injection from the deceleration of sub-relativistic material in a constant-density are shown in radio at 3 GHz and X-rays at 1 keV \citep{2018ApJ...854L..31C}. The millisecond magnetar scenario  with accretion is considered in all curves for $P_0=10^{-3}\,{\rm s}$ and $t_{\rm fb}=10^9\,{\rm s}$.  The parameters used for curves of the above panels are (Left; $\beta=0.3$, $p=2.05$, $\epsilon_{\rm B}=10^{-4}$ and $\epsilon_{\rm e}=10^{-1}$): the dotted lines ($\tilde{E}=10^{46}\,{\rm erg}$, $B=1.5\times 10^{13}\,{\rm G}$, $n=6.61\times 10^{-3}\,{\rm cm^{-3}}$ and $\alpha=3$), the dashed lines ($\tilde{E}=3.93\times 10^{46}\,{\rm erg}$, $B=6.5\times 10^{13}\,{\rm G}$, $n=8.88\times 10^{-3}\,{\rm cm^{-3}}$ and $\alpha=4$) and the dash-dotted lines ($\tilde{E}=10^{46}\,{\rm erg}$, $B=1.3\times 10^{14}\,{\rm G}$,  $n=1.60\times 10^{-2}\,{\rm cm^{-3}}$ and $\alpha=5$), and (Right; $\beta=0.3$ $p=2.15$, $\epsilon_{\rm B}=10^{-3}$, $\epsilon_{\rm e}=10^{-1}$): the dotted lines ($\tilde{E}=10^{46}\,{\rm erg}$,  $B=4.8\times 10^{13}\,{\rm G}$,  $n=5.87\times 10^{-3}\,{\rm cm^{-3}}$ and $\alpha=3$), the dashed lines ($\tilde{E}=1.9\times 10^{47}\,{\rm erg}$, $B=3.2\times 10^{15}\,{\rm G}$,  $n=5.87\times 10^{-3}\,{\rm cm^{-3}}$ and $\alpha=4$) and the dash-dotted lines ($\tilde{E}=10^{48}\,{\rm erg}$, $B=2.7\times 10^{13}\,{\rm G}$,  $n=5.87\times 10^{-3}\,{\rm cm^{-3}}$ and  $\alpha=5$).
%
%
The parameters used for curves of the below panels are (Left; $\beta=0.4$ $p=2.05$, $\epsilon_{\rm B}=10^{-3}$, and $\epsilon_{\rm e}=10^{-1}$): the dotted line ($\tilde{E}=3.1\times 10^{47}\,{\rm erg}$, $B=5.4\times 10^{13}\,{\rm G}$, $n=2.72\times 10^{-3}\,{\rm cm^{-3}}$ and  $\alpha=3$), the dashed lines ($\tilde{E}=4.73\times 10^{47}\,{\rm erg}$, $B=4.2\times 10^{14}\,{\rm G}$,  $n=1.51\times 10^{-3}\,{\rm cm^{-3}}$ and $\alpha=4$) and the dash-dotted line ($\tilde{E}=10^{46}\,{\rm erg}$, $B=1.1\times 10^{14}\,{\rm G}$,  $n=1.61\times 10^{-2}\,{\rm cm^{-3}}$ and $\alpha=5$), and (Right; $\beta=0.4$ $p=2.15$, $\epsilon_{\rm B}=10^{-3}$ and $\epsilon_{\rm e}=10^{-1}$): the dotted line ($\tilde{E}=9.10\times 10^{47}\,{\rm erg}$, $B=4.7\times 10^{15}\,{\rm G}$,  $n=3.66\times 10^{-3}\,{\rm cm^{-3}}$ and  $\alpha=3$), the dashed lines ($\tilde{E}=4.09\times 10^{46}\,{\rm erg}$, $B=4.3\times 10^{13}\,{\rm G}$, $n=2.89\times 10^{-3}\,{\rm cm^{-3}}$ and $\alpha=4$) and  the dash-dotted lines ($\tilde{E}=1.45\times 10^{46}\,{\rm erg}$, $B=5.7\times 10^{13}\,{\rm G}$, $n=4.64\times 10^{-3}\,{\rm cm^{-3}}$ and $\alpha=5$).
}
\label{GRB170817A_modelling}
\end{figure}

Figure \ref{GRB170817A_modelling} shows the multi-band afterglow observations of GRB 170817A, the best-fit curve from a relativistic structured off-axis jet \cite[solid lines;][]{2019ApJ...884...71F} and several allowed curves (dotted, dashed and dash-dotted lines) from the sub-relativistic material shown in this work. The afterglow observations are shown at X-rays, optical bands and radio wavelengths, and the synchrotron light curves are obtained at 1 keV (blue), 2.1 eV (red), 6 GHz (black) and 3 GHz (gray).  In each panel we consider the PL indexes $\alpha=3.0$ (dotted line), $4.0$ (dashed line) and $5.0$ (dash-dotted line) for $\beta=0.3$ (panels above) and $\beta=0.4$ (panels below) with $p=2.05$ (left) $p=2.15$ (right).   We can see that for different sets of parameters; we can obtain similar results about the description of the latest observations. It indicates, as expected, that our results are not unique but are possible solutions because the synchrotron equations are degenerate in these parameters.\\ 

It is relevant to mention that the value of the mean opacity for which the KN ejecta is transparent $k_s \approx \beta^{2}_{\rm -1} M^{-1}_{\rm ej,-1.3}t_{\rm th,3}^2\approx 10^{3.5}\,{\rm g^{-1}\,cm^{2}}$  agrees with the radiation transfer simulations, which is $\gtrsim 10\,{\rm g^{-1}\,cm^{-2}}$ for lanthanide-rich ejecta \citep{2013ApJ...775...18B, 2013ApJ...775..113T, 2019ARNPS..69...41S}. The luminosity would be estimated as $L_{\rm bol}\approx 10^{38}\,{\rm erg\,s^{-1}}\,t^{1.3}_{\rm th,3}M_{\rm ej, -1.3}$ \citep{2019ARNPS..69...41S}, where $t_{\rm th}$ is the timescale when the KN eject enters the thin regime.  It is worth noting that the features of the synchrotron emission of the sub-relativistic material is consistent with the faster ``blue" kilonova afterglow.\\
 
We have considered the spinning magnetized NS scenario and discarded the fall-back material on BH scenario proposed in subsection \ref{subsec:2.2} because the rate of fall-back accretion estimated at early times with the parameters used to describe the latest observations are fully different from  the rate of fall-back accretion used in hydrodynamical simulations \citep[e.g., see][]{2001ApJ...550..410M,
2008ApJ...679..639Z}. It can be demonstrated as follows. At $t=t_{\rm p}=1000\,{\rm days}$, the expectation accretion rate and the BZ luminosity (Eq. \ref{M_odot}) are $\dot{M}_{\rm BH} \approx \dot{M}_p\left( t/t_{\rm p} \right)^{-5/3}$  with $\tau_{\rm vis}\approx t_{\rm p}$ and  $L_{\rm BZ}\approx 10^{39}{\rm erg\, s^{-1}}\, G(a)\,\dot{M}_{\rm p,-2}\left( t/{\rm 1000\,days} \right)^{-5/3}$, respectively.  The previous derivation is similar to that resulted found by \cite{2021arXiv210402070H} after modelling the latest observations at 1000 days.  However, extrapolating the rate of fall-back accretion at early times $\sim 1\, {\rm s}$, it would be $\dot{M}_{\rm fb}\approx 10^{-17}\,{\rm M_\odot\,s^{-1}}$, which is different from the rate of fall-back accretion used in hydrodynamical simulations \citep[e.g., see][]{2001ApJ...550..410M,2008ApJ...679..639Z}.  It is worth noting that in the BH scenario the rate of fall-back accretion at early times is $\propto \left( t/t_{\rm p} \right)^{1/2}$ instead of $\left( t/t_{\rm p} \right)^{-5/3}$, as considered by \cite{2021arXiv210402070H}.

\subsection{Short GRBs with evidence of a KN}

Candidates discussed in the literature with evidence of a KN emission are GRB 050709 \citep{2016NatCo...712898J}, GRB 060614 \citep{2015NatCo...6.7323Y}, GRB 130603B \citep{2013Natur.500..547T, 2013ApJ...774L..23B} and GRB 160821B  \citep{2017ApJ...843L..34K, 2019MNRAS.489.2104T}.  As follows, we present the four claimed KN observations, and then we show the synchrotron light curves with a set of allowed and ruled out parameters, assuming the characteristics of the ``blue" KN.

\subsubsection{Multi-band observations}
\paragraph{GRB 050709}

GRB 050709 was detected on 2005 July 9 at 22:36:37 UT by the Soft X-Ray Camera (SXC), the Wide-Field X-Ray Monitor (WXM) and the French Gamma Telescope (FREGATE) instruments on board the High Energy Transient Explorer 2 satellite (HETE) with a reported location of $\textrm{R.A.}=+23^{\textrm{h}} 01^{\textrm{m}}30^{\textrm{s}}$, $\textrm{Dec}=-38^{\circ}58'33''$ (J2000) \citep{2005Natur.437..855V}. The prompt emission had an approximate duration of 0.5 seconds in the form of a hard spike in the 3–400 keV energy band, which was followed by an extended X-ray emission lasting $\sim$ 130 seconds \citep{2016NatCo...712898J}. The accurate location of the burst led to the first-ever identification of the optical afterglow of a short-hard burst in ground-based experiments and HST \citep{2005Natur.437..859H, 2005Natur.437..845F}. This, in turn, led to the determination of its host galaxy, which lied at redshift $z=0.16$.

\paragraph{GRB 060614}

GRB 060614 was detected on 2006 June 14 at 12:43:48 UT by the Swift-BAT instrument. Its location was found to be at $\textrm{R.A.}=+21^{\textrm{h}} 23^{\textrm{m}}27^{\textrm{s}}$, $\textrm{Dec}=-53^{\circ}02'02''$ (J2000). The event had a duration of 102 seconds in the 15-350 keV energy range, which places this burst in the long GRB category \citep{2006Natur.444.1044G}. However, subsequent observations showed that the event lacked an associated supernova, which is expected for lGRBs and that its temporal lag and peak luminosity were in line with those of short-duration GRBs \citep{2016NatCo...712898J}. This contrasting behaviour led to denoting GRB 060614 as a ‘hybrid GRB'.

\paragraph{GRB 130603B}

GRB 130603B was simultaneously detected on 2013 June 3 at 15:49:14 UT by Swift BAT and by Konus-Wind \citep{2013Natur.500..547T}. Its location was found to be at $\textrm{R.A.}=+21^{\textrm{h}} 23^{\textrm{m}}27^{\textrm{s}}$, $\textrm{Dec}=-53^{\circ}02'02''$ (J2000). According to the BAT instrument, it had a duration of $T_{90}\approx0.18\pm0.02\,{\rm s}$ in the 15-350 keV band \citep{2013GCN.14741....1B}, which places it in the sGRB class. Optical and near-IR observations of the event were performed which demonstrated the presence of excess near-IR emission matching a KN \citep{2013ApJ...774L..23B}.

\paragraph{GRB 160821B}

GRB 160821B was detected on 2016 August 21 at 22:29:13 UT by the Swift BAT instrument \citep{2021ApJ...908...90A}. The proposed host galaxy's location was found to be at $\textrm{R.A.}=+18^{\textrm{h}} 39^{\textrm{m}}53.968^{\textrm{s}}$, $\textrm{Dec}=62^{\circ}23'34.35''$ (J2000) at redshift $z=0.162$, making this GRB one of the lowest redshift burst observed by Swift \citep{2019ApJ...883...48L}. It had a duration of $T_{90}=0.48\pm0.07\,{\rm s}$ in the 15–350 keV energy band \citep{2019MNRAS.489.2104T}, which places it in the sGRB class. Upon analysis of the X-ray light curves, \cite{2019MNRAS.489.2104T}  found that there is evidence for continued energy injection from a long-lived central engine. On the other hand, the optical and near IR observations showed behaviour consistent with a KN.

\subsubsection{Analysis and Description}

\begin{figure}
{\centering
\resizebox*{\textwidth}{0.45\textheight}
{\includegraphics{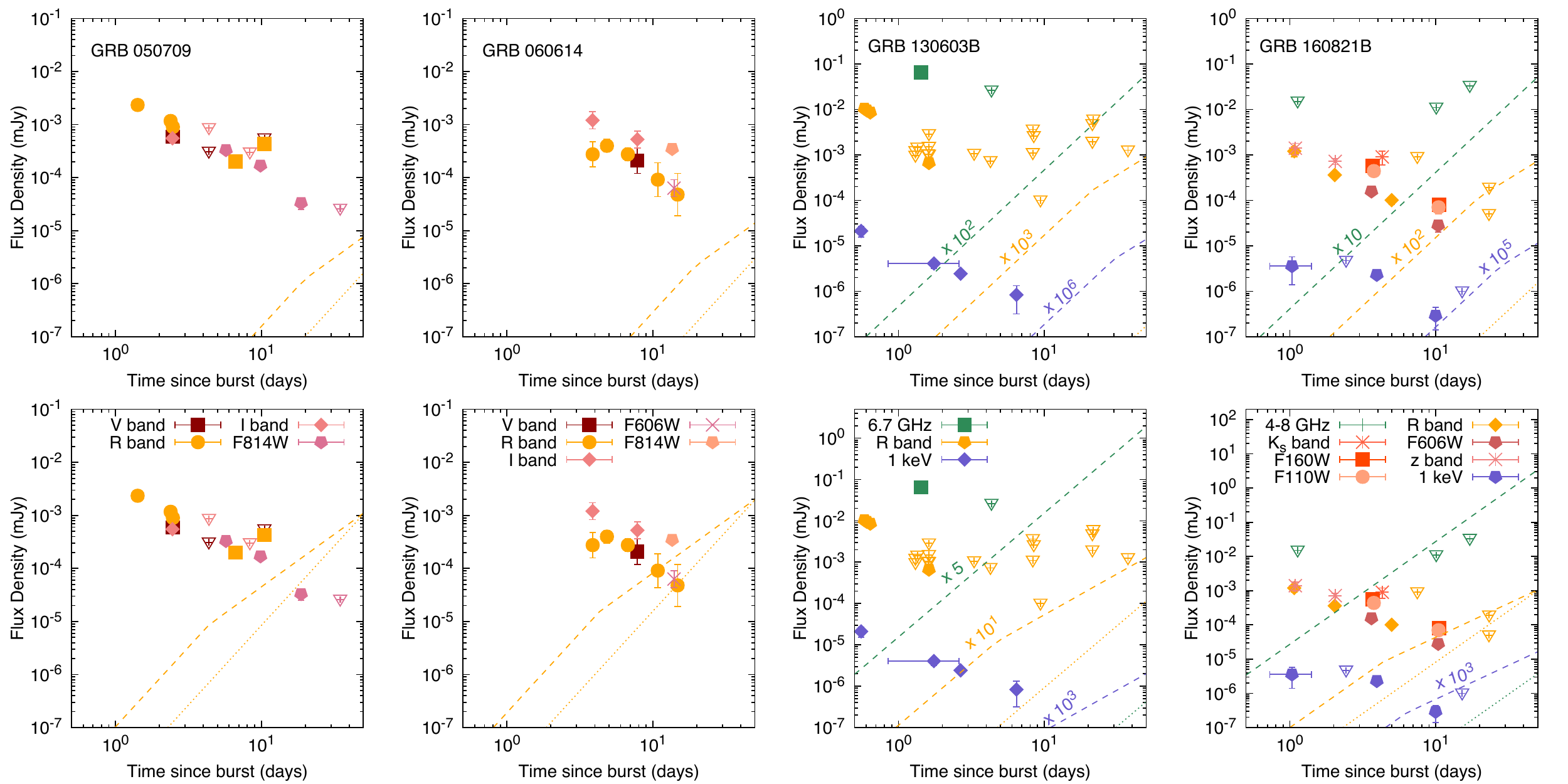}}
} 
\caption{Multi-band afterglow observations of sGRBs with evidence of a KN emission and the synchrotron light curves from the cocoon (upper panels; $\beta=0.3$) and the shock breakout (lower panels; $\beta=0.8$) materials decelerating in a constant-density medium with $n=1\,{\rm cm^{-3}}$ (dashed lines) and $10^{-2}\,{\rm cm^{-3}}$ (dotted lines). The synchrotron light curves are shown in X-ray (1 keV; blue), optical (R-band; gold) and radio (5 GHz; green) bands.  The parameter values used are $P_0=10^{-3}\,{\rm s}$, $B=7\times 10^{14}\,{\rm G}$, $t_{\rm fb}=5\times 10^5\,{\rm s}$, $\tilde{E}=1.29\times 10^{51}\,{\rm erg}$, $\epsilon_{\rm e}=0.3$, $\epsilon_{\rm B}=0.1$, $p=2.05$ and $\alpha=3.0$.}
\label{kn_candidates}
\end{figure}

Figure \ref{kn_candidates} presents four columns, where each one corresponds to a different short GRB with evidence of KN emission.  Each panel shows the multi-band afterglow observations of bursts with evidence of a KN emission and the synchrotron light curves from the cocoon (upper) with $\beta=0.3$ and the shock breakout (lower) with $\beta=0.8$ decelerating in a constant-density medium with $n=1\,{\rm cm^{-3}}$ (dashed lines) and $10^{-2}\,{\rm cm^{-3}}$ (dotted lines).  The synchrotron light curves are presented at 1 keV (blue), R-band (gold) and 5 GHz (green).   The disk wind and dynamical ejecta  are not displayed because these decelerated materials peak at timescales longer than $\approx 10^3\,{\rm days}$, and the multi-band afterglow observations with the respective upper limits are reported at timescales from days to $\sim$ one month.    We consider a spinning magnetized NS with accretion as remnant of merger of two NSs.  The parameter values used are $P_0=10^{-3}\,{\rm s}$, $B=7\times 10^{14}\,{\rm G}$, $t_{\rm fb}=5\times 10^5\,{\rm s}$, $\tilde{E}=2.1\times 10^{49}\,{\rm erg}$, $\epsilon_{\rm e}=0.3$, $\epsilon_{\rm B}=0.1$, $p=2.05$ and $\alpha=3.0$.  For GRB 050709,  the synchrotron emission at the F814W band from the shock breakout material with a velocity $\beta=0.8$ is ruled out for both $n=1\,{\rm cm^{-3}}$ and $n=10^{-2}\,{\rm cm^{-3}}$, but not for the cocoon material. For GRB 060614,  the synchrotron emission at the R-band from the shock breakout material is ruled out for both $n=1\,{\rm cm^{-3}}$ and $n=10^{-2}\,{\rm cm^{-3}}$, but not for the cocoon material. For GRB 130603B, the synchrotron curves from the shock breakout and cocoon are allowed at all bands. For GRB 160821B, the synchrotron curves at 5 GHz and at the R-band from the cocoon material are ruled out for a density of $n=1\,{\rm cm^{-3}}$, but not for $n=10^{-2}\,{\rm cm^{-3}}$. The synchrotron curves at 1 keV from the shock breakout and cocoon are allowed are allowed.  Similarly,  all synchrotron curves from the cocoon  are allowed.  The value of the uniform-density medium with $n=1\,{\rm cm^{-3}}$ is ruled out in our model for GRB 050709, GRB 060614 and GRB 160821B, but nor for GRB 130603B. This result is consistent with the mean value reported for sGRBs \citep[e.g., see][]{2014ARA&A..52...43B}.\\

Continuous energy injection by the progenitor on the afterglow could generate a long-lived reverse shock at very long timescales, and shocked-accelerated electrons in this region would radiate modifying the forward-shock light curves, as found in some GRB afterglows \cite[e.g., see][]{2006ApJ...651..381C, 2007RSPTA.365.1241V, 2014MNRAS.444.3151V, 2018ApJ...862...94L}. In a forthcoming manuscript, we will present a detailed analysis of this long-lived reverse shock scenario. 
\section{Summary}\label{sec:Summary}
We have extended the synchrotron model presented in \cite{2021ApJ...907...78F} and presented the dynamics of deceleration of a sub-relativistic material when the central engine (a remnant of either a spinning magnetized NS or a fall-back material onto BH) injects energy into the blastwave and the external medium is stratified with a density profile $A_{\rm k} r^{\rm -k}$ with $0 \leq {\rm k}\leq 2.5$. We have considered different profiles of the energy injection and also GRB progenitors. The energy injection index $q=0$ is connected with a spinning magnetized NS and $q$, in general, with fall-back material onto a central black hole.  The total isotropic-equivalent kinetic energy was introduced as the superposition of the energy distribution $E_{\rm \beta}$  and the energy injection $E_{\rm t}$. When the condition $E_{\rm t}\ll E_{\rm \beta}$ is satisfied, the synchrotron light curve mimics the ones without energy injection. Otherwise,  for $E_{\rm t}\ll E_{\rm \beta}$ the continuous energy injection into the afterglow dominates and many differences are observed.  The constant-density medium (${\rm k=0}$) is associated with the death of massive stars and the merger of two NSs, and the stratified medium ($1\leq {\rm k} \leq 2.5$) is only expected  with the death of massive stars.   We have presented the synchrotron light curves in radio at 1.6 GHz, optical at the R-band  and X-rays at 1 keV with typical values of GRB afterglows and the scenarios of spinning magnetized NS and fall-back material onto BH with different characteristic timescales for a generic remnant located at $d_z=100\,{\rm Mpc}$. 
The synchrotron light curves exhibit the maximum flux on timescales from days to years, although if the remnant injects large amounts of energy, the maximum flux could be expected on timescales of hours. These light curves exhibit that, during the early stages, there is an increase in the flux. During this epoch, when the sub-relativistic ejecta decelerates in a constant-density medium, the light curve grows steeply. For more stratified media, this growth is not as evident. For instance, in the light curves with a density profile with $k=2.5$,  the early-time behaviour is more gradual. This result implies that if such a flattening or rebrightening at  timescales from months to years  in the light curve was observed together with GW detection, then this would be associated with the deceleration of a sub-relativistic material launched during the merger of two NSs. Otherwise, we showed that an observed flux that gradually decreases on timescales from months to years could be associated with the deceleration of a sub-relativistic material launched during the death of a massive star with different mass-loss evolution at the end of its life.  We have shown that variations of the stratification parameter are more apparent in the radio light curve when compared to the other fluxes in the other energy bands. This change in the density profile is also more easily appreciated for large values of $\alpha$, (i.e., $\alpha=4.0$ and $5.0$). Therefore, a transition between density profiles will be more easily observed in the radio band with high values of $\alpha$.

The two NS merger ejects sub-relativistic material with distinct velocities which are decelerated by the external medium, generating the synchrotron light curves at different frequencies peaking at timescales from days to years. It is important to mention that before the ejected materials are in the sub-relativistic regime, these might begin momentarily in the TR regime (i.e., with $\beta\sim 0.8$). Similarly,  they could finally with the passage of time ($\sim 10^3\,{\rm years}$) lie in the DN regime with a velocity (i.e., $\beta\sim 0.08$). Therefore, we have introduced timescales in both regime.  For instance, with the values of parameters used we have shown that the dynamical ejecta, the cocoon, the shock breakout and the wind peak at $7150\ \rm days$, $522.5\ \mathrm{days}$, $56.1\ \mathrm{days}$ and $321.8\ \mathrm{years}$, respectively.  The total contribution of synchrotron light curves exhibits a brightening once the off-axis emission decreases at a few years, so that, depending on the parameter values and conditions, the synchrotron emission from the sub-relativistic materials could be detected or not. The synchrotron light curves could be associated with GW detections.  We showed that, in the case of a failed or an off-axis GRB, the non-thermal emission generated by the deceleration of sub-relativistic materials could be detected at early times.   In the case of an on-axis GRB, the afterglow emission originated from deceleration of the relativistic jet would have to decrease substantially so that  the afterglow emission from the sub-relativistic ejecta could be observed.    In addition, we gave an important tool  to distinguish the afterglow emission among the sub-relativistic materials from  the relativistic jet through the evolution of the synchrotron flux.

We have applied our model to describe the latest multi-wavelength afterglow observations ($> 900$ days) of GW170817/GRB 170817A, and 
constraints on the afterglow emission of some short-bursts with evidence of KNe.  Regarding to the GW170817/GRB 170817A event,  we have constrained the parameter space that reproduces the synchrotron light curves evolving in a constant-density medium from the faster ``blue" KN afterglow. We have considered a spinning magnetized NS as remnant of the merger of two NSs which is continuously injecting energy due to spin-down luminosity into the blastwave. We plot the parameter space that describes the X-ray observations and is below the radio upper limits as a function of $\epsilon_{\rm e}=10^{-1}$, $\epsilon_{\rm B}=10^{-3}$, $\alpha=\{3.0,\,4.0, 5.0\}$ and $p=\{2.05,\,2.15\}$.  We have shown that for both values of velocity ($\beta=0.3$ and $0.4$), $t_{\rm fb}=5\times 10^9\,{\rm s}$ and $10^{13.7}\lesssim B \lesssim 10^{16.3}\,{\rm G}$, the values of $p=2.15$, $n<5\times 10^{-3}\,{\rm cm^{-3}}$ and $\alpha=5$ are preferred, although the value of $\alpha=3.0$ cannot be discarded.   The value allowed of $\alpha=3.0$ in our theoretical model agrees with the values found in numerical simulations \citep[e.g., see][]{2013ApJ...773...78B} and used for describing the KN emission \citep{2017LRR....20....3M, 2019LRR....23....1M}, and $\alpha=5.0$ is more consistent with those reported in \cite{2019MNRAS.487.3914K, 2021arXiv210402070H}. It is worth noting that the features of the synchrotron emission of the sub-relativistic material is consistent with the faster ``blue" KN afterglow.

\acknowledgements

We would like to mention our deeply gratitude to the anonymous referee for his or her careful reading of the paper and useful recommendations that helped improve the clarity of the manuscript.   We thank Tanmoy Laskar, Paz Beniamini, Bin-bin Zhang and Bing Zhang for useful discussions. NF acknowledges financial  support  from UNAM-DGAPA-PAPIIT  through  grant IN106521. RLB acknowledges support from CONACyT postdoctoral fellowships and the support from the DGAPA/UNAM IG100820 and IN105921.

\newpage

\bibliography{Bib_kilonova}

\begin{thebibliography}{}
\expandafter\ifx\csname natexlab\endcsname\relax\def\natexlab#1{#1}\fi
\providecommand{\url}[1]{\href{#1}{#1}}
\providecommand{\dodoi}[1]{doi:~\href{http://doi.org/#1}{\nolinkurl{#1}}}
\providecommand{\doeprint}[1]{\href{http://ascl.net/#1}{\nolinkurl{http://ascl.net/#1}}}
\providecommand{\doarXiv}[1]{\href{https://arxiv.org/abs/#1}{\nolinkurl{https://arxiv.org/abs/#1}}}

\bibitem[{Abbott {et~al.}(2017{\natexlab{a}})Abbott, Abbott, Abbott, \&
  et~al.}]{PhysRevLett.119.161101}
Abbott, B.~P., Abbott, R., Abbott, T.~D., \& et~al. 2017{\natexlab{a}}, Phys.
  Rev. Lett., 119, 161101, \dodoi{10.1103/PhysRevLett.119.161101}

\bibitem[{Abbott {et~al.}(2017{\natexlab{b}})Abbott, Abbott, Abbott, \&
  et~al.}]{2041-8205-848-2-L12}
---. 2017{\natexlab{b}}, The Astrophysical Journal Letters, 848, L12.
\newblock \url{http://stacks.iop.org/2041-8205/848/i=2/a=L12}

\bibitem[{{Abbott} {et~al.}(2017){Abbott}, {Abbott}, {Abbott}, {Acernese},
  {Ackley}, {Adams}, {Adams}, {Addesso}, {Adhikari}, {Adya}, {Affeldt},
  {Afrough}, {Agarwal}, {Agathos}, {Agatsuma}, {Aggarwal}, {Aguiar}, {Aiello},
  {Ain}, {Ajith}, {Allen}, {Allen}, {Allocca}, {Altin}, {Amato}, {Ananyeva},
  {Anderson}, {Anderson}, {Angelova}, {Antier}, {Appert}, {Arai}, {Araya},
  {Areeda}, {Arnaud}, {Arun}, {Ascenzi}, {Ashton}, {Ast}, {Aston}, {Astone},
  {Atallah}, {Aufmuth}, {Aulbert}, {AultONeal}, {Austin}, {Avila-Alvarez},
  {Babak}, {Bacon}, {Bader}, {Bae}, {Baker}, {Baldaccini}, {Ballardin},
  {Ballmer}, {Banagiri}, {Barayoga}, {Barclay}, {Barish}, {Barker}, {Barkett},
  {Barone}, {Barr}, {Barsotti}, {Barsuglia}, {Barta}, {Barthelmy}, {Bartlett},
  {Bartos}, {Bassiri}, {Basti}, {Batch}, {Bawaj}, {Bayley}, {Bazzan},
  {B{\'e}csy}, {Beer}, {Bejger}, {Belahcene}, {Bell}, {Berger}, {Bergmann},
  {Bero}, {Berry}, {Bersanetti}, {Bertolini}, {Betzwieser}, {Bhagwat},
  {Bhandare}, {Bilenko}, {Billingsley}, {Billman}, {Birch}, {Birney},
  {Birnholtz}, {Biscans}, {Biscoveanu}, {Bisht}, {Bitossi}, {Biwer},
  {Bizouard}, {Blackburn}, {Blackman}, {Blair}, {Blair}, {Blair}, {Bloemen},
  {Bock}, {Bode}, {Boer}, {Bogaert}, {Bohe}, {Bondu}, {Bonilla}, {Bonnand},
  {Boom}, {Bork}, {Boschi}, {Bose}, {Bossie}, {Bouffanais}, {Bozzi},
  {Bradaschia}, {Brady}, {Branchesi}, {Brau}, {Briant}, {Brillet}, {Brinkmann},
  {Brisson}, {Brockill}, {Broida}, {Brooks}, {Brown}, {Brown}, {Brunett},
  {Buchanan}, {Buikema}, {Bulik}, {Bulten}, {Buonanno}, {Buskulic}, {Buy},
  {Byer}, {Cabero}, {Cadonati}, {Cagnoli}, {Cahillane}, {Calder{\'o}n
  Bustillo}, {Callister}, {Calloni}, {Camp}, {Canepa}, {Canizares}, {Cannon},
  {Cao}, {Cao}, {Capano}, {Capocasa}, {Carbognani}, {Caride}, {Carney},
  {Casanueva Diaz}, {Casentini}, {Caudill}, {Cavagli{\`a}}, {Cavalier},
  {Cavalieri}, {Cella}, {Cepeda}, {Cerd{\'a}-Dur{\'a}n}, {Cerretani},
  {Cesarini}, {Chamberlin}, {Chan}, {Chao}, {Charlton}, {Chase},
  {Chassande-Mottin}, {Chatterjee}, {Chatziioannou}, {Cheeseboro}, {Chen},
  {Chen}, {Chen}, {Cheng}, {Chia}, {Chincarini}, {Chiummo}, {Chmiel}, {Cho},
  {Cho}, {Chow}, {Christensen}, {Chu}, {Chua}, {Chua}, {Chung}, {Chung},
  {Ciani}, {Ciolfi}, {Cirelli}, {Cirone}, {Clara}, {Clark}, {Clearwater},
  {Cleva}, {Cocchieri}, {Coccia}, {Cohadon}, {Cohen}, {Colla}, {Collette},
  {Cominsky}, {Constancio}, {Conti}, {Cooper}, {Corban}, {Corbitt},
  {Cordero-Carri{\'o}n}, {Corley}, {Cornish}, {Corsi}, {Cortese}, {Costa},
  {Coughlin}, {Coughlin}, {Coulon}, {Countryman}, {Couvares}, {Covas}, {Cowan},
  {Coward}, {Cowart}, {Coyne}, {Coyne}, {Creighton}, {Creighton}, {Cripe},
  {Crowder}, {Cullen}, {Cumming}, {Cunningham}, {Cuoco}, {Dal Canton},
  {D{\'a}lya}, {Danilishin}, {D'Antonio}, {Danzmann}, {Dasgupta}, {Da Silva
  Costa}, {Dattilo}, {Dave}, {Davier}, {Davis}, {Daw}, {Day}, {De}, {DeBra},
  {Degallaix}, {De Laurentis}, {Del{\'e}glise}, {Del Pozzo}, {Demos}, {Denker},
  {Dent}, {De Pietri}, {Dergachev}, {De Rosa}, {DeRosa}, {De Rossi}, {DeSalvo},
  {de Varona}, {Devenson}, {Dhurandhar}, {D{\'\i}az}, {Di Fiore}, {Di
  Giovanni}, {Di Girolamo}, {Di Lieto}, {Di Pace}, {Di Palma}, {Di Renzo},
  {Doctor}, {Dolique}, {Donovan}, {Dooley}, {Doravari}, {Dorrington},
  {Douglas}, {Dovale {\'A}lvarez}, {Downes}, {Drago}, {Dreissigacker},
  {Driggers}, {Du}, {Ducrot}, {Dupej}, {Dwyer}, {Edo}, {Edwards}, {Effler},
  {Ehrens}, {Eichholz}, {Eikenberry}, {Eisenstein}, {Essick}, {Estevez},
  {Etienne}, {Etzel}, {Evans}, {Evans}, {Factourovich}, {Fafone}, {Fair},
  {Fairhurst}, {Fan}, {Farinon}, {Farr}, {Farr}, {Fauchon-Jones}, {Favata},
  {Fays}, {Fee}, {Fehrmann}, {Feicht}, {Fejer}, {Fernandez-Galiana},
  {Ferrante}, {Ferreira}, {Ferrini}, {Fidecaro}, {Finstad}, {Fiori},
  {Fiorucci}, {Fishbach}, {Fisher}, {Fitz-Axen}, {Flaminio}, {Fletcher},
  {Fong}, {Font}, {Forsyth}, {Forsyth}, {Fournier}, {Frasca}, {Frasconi},
  {Frei}, {Freise}, {Frey}, {Frey}, {Fries}, {Fritschel}, {Frolov}, {Fulda},
  {Fyffe}, {Gabbard}, {Gadre}, {Gaebel}, {Gair}, {Gammaitoni}, {Ganija},
  {Gaonkar}, {Garcia-Quiros}, {Garufi}, {Gateley}, {Gaudio}, {Gaur},
  {Gayathri}, {Gehrels}, {Gemme}, {Genin}, {Gennai}, {George}, {George},
  {Gergely}, {Germain}, {Ghonge}, {Ghosh}, {Ghosh}, {Ghosh}, {Giaime},
  {Giardina}, {Giazotto}, {Gill}, {Glover}, {Goetz}, {Goetz}, {Gomes},
  {Goncharov}, {Gonz{\'a}lez}, {Gonzalez Castro}, {Gopakumar}, {Gorodetsky},
  {Gossan}, {Gosselin}, {Gouaty}, {Grado}, {Graef}, {Granata}, {Grant}, {Gras},
  {Gray}, {Greco}, {Green}, {Gretarsson}, {Griswold}, {Groot}, {Grote},
  {Grunewald}, {Gruning}, {Guidi}, {Guo}, {Gupta}, {Gupta}, {Gushwa},
  {Gustafson}, {Gustafson}, {Halim}, {Hall}, {Hall}, {Hamilton}, {Hammond},
  {Haney}, {Hanke}, {Hanks}, {Hanna}, {Hannam}, {Hannuksela}, {Hanson},
  {Hardwick}, {Harms}, {Harry}, {Harry}, {Hart}, {Haster}, {Haughian}, {Healy},
  {Heidmann}, {Heintze}, {Heitmann}, {Hello}, {Hemming}, {Hendry}, {Heng},
  {Hennig}, {Heptonstall}, {Heurs}, {Hild}, {Hinderer}, {Hoak}, {Hofman},
  {Holt}, {Holz}, {Hopkins}, {Horst}, {Hough}, {Houston}, {Howell}, {Hreibi},
  {Hu}, {Huerta}, {Huet}, {Hughey}, {Husa}, {Huttner}, {Huynh-Dinh}, {Indik},
  {Inta}, {Intini}, {Isa}, {Isac}, {Isi}, {Iyer}, {Izumi}, {Jacqmin}, {Jani},
  {Jaranowski}, {Jawahar}, {Jim{\'e}nez-Forteza}, {Johnson}, {Jones}, {Jones},
  {Jonker}, {Ju}, {Junker}, {Kalaghatgi}, {Kalogera}, {Kamai}, {Kandhasamy},
  {Kang}, {Kanner}, {Kapadia}, {Karki}, {Karvinen}, {Kasprzack}, {Katolik},
  {Katsavounidis}, {Katzman}, {Kaufer}, {Kawabe}, {K{\'e}f{\'e}lian}, {Keitel},
  {Kemball}, {Kennedy}, {Kent}, {Key}, {Khalili}, {Khan}, {Khan}, {Khan},
  {Khazanov}, {Kijbunchoo}, {Kim}, {Kim}, {Kim}, {Kim}, {Kim}, {Kim},
  {Kimbrell}, {King}, {King}, {Kinley-Hanlon}, {Kirchhoff}, {Kissel},
  {Kleybolte}, {Klimenko}, {Knowles}, {Koch}, {Koehlenbeck}, {Koley},
  {Kondrashov}, {Kontos}, {Korobko}, {Korth}, {Kowalska}, {Kozak},
  {Kr{\"a}mer}, {Kringel}, {Krishnan}, {Kr{\'o}lak}, {Kuehn}, {Kumar}, {Kumar},
  {Kumar}, {Kuo}, {Kutynia}, {Kwang}, {Lackey}, {Lai}, {Landry}, {Lang},
  {Lange}, {Lantz}, {Lanza}, {Larson}, {Lartaux-Vollard}, {Lasky}, {Laxen},
  {Lazzarini}, {Lazzaro}, {Leaci}, {Leavey}, {Lee}, {Lee}, {Lee}, {Lee}, {Lee},
  {Lehmann}, {Lenon}, {Leonardi}, {Leroy}, {Letendre}, {Levin}, {Li}, {Linker},
  {Littenberg}, {Liu}, {Lo}, {Lockerbie}, {London}, {Lord}, {Lorenzini},
  {Loriette}, {Lormand}, {Losurdo}, {Lough}, {Lousto}, {Lovelace}, {L{\"u}ck},
  {Lumaca}, {Lundgren}, {Lynch}, {Ma}, {Macas}, {Macfoy}, {Machenschalk},
  {MacInnis}, {Macleod}, {Maga{\~n}a Hernandez}, {Maga{\~n}a-Sandoval},
  {Maga{\~n}a Zertuche}, {Magee}, {Majorana}, {Maksimovic}, {Man}, {Mandic},
  {Mangano}, {Mansell}, {Manske}, {Mantovani}, {Marchesoni}, {Marion},
  {M{\'a}rka}, {M{\'a}rka}, {Markakis}, {Markosyan}, {Markowitz}, {Maros},
  {Marquina}, {Marsh}, {Martelli}, {Martellini}, {Martin}, {Martin},
  {Martynov}, {Mason}, {Massera}, {Masserot}, {Massinger}, {Masso-Reid},
  {Mastrogiovanni}, {Matas}, {Matichard}, {Matone}, {Mavalvala}, {Mazumder},
  {McCarthy}, {McClelland}, {McCormick}, {McCuller}, {McGuire}, {McIntyre},
  {McIver}, {McManus}, {McNeill}, {McRae}, {McWilliams}, {Meacher}, {Meadors},
  {Mehmet}, {Meidam}, {Mejuto-Villa}, {Melatos}, {Mendell}, {Mercer}, {Merilh},
  {Merzougui}, {Meshkov}, {Messenger}, {Messick}, {Metzdorff}, {Meyers},
  {Miao}, {Michel}, {Middleton}, {Mikhailov}, {Milano}, {Miller}, {Miller},
  {Miller}, {Millhouse}, {Milovich-Goff}, {Minazzoli}, {Minenkov}, {Ming},
  {Mishra}, {Mitra}, {Mitrofanov}, {Mitselmakher}, {Mittleman}, {Moffa},
  {Moggi}, {Mogushi}, {Mohan}, {Mohapatra}, {Montani}, {Moore}, {Moraru},
  {Moreno}, {Morriss}, {Mours}, {Mow-Lowry}, {Mueller}, {Muir}, {Mukherjee},
  {Mukherjee}, {Mukherjee}, {Mukund}, {Mullavey}, {Munch}, {Mu{\~n}iz},
  {Muratore}, {Murray}, {Napier}, {Nardecchia}, {Naticchioni}, {Nayak},
  {Neilson}, {Nelemans}, {Nelson}, {Nery}, {Neunzert}, {Nevin}, {Newport},
  {Newton}, {Ng}, {Nguyen}, {Nguyen}, {Nichols}, {Nielsen}, {Nissanke}, {Nitz},
  {Noack}, {Nocera}, {Nolting}, {North}, {Nuttall}, {Oberling}, {O'Dea},
  {Ogin}, {Oh}, {Oh}, {Ohme}, {Okada}, {Oliver}, {Oppermann}, {Oram},
  {O'Reilly}, {Ormiston}, {Ortega}, {O'Shaughnessy}, {Ossokine}, {Ottaway},
  {Overmier}, {Owen}, {Pace}, {Page}, {Page}, {Pai}, {Pai}, {Palamos},
  {Palashov}, {Palomba}, {Pal-Singh}, {Pan}, {Pan}, {Pang}, {Pang}, {Pankow},
  {Pannarale}, {Pant}, {Paoletti}, {Paoli}, {Papa}, {Parida}, {Parker},
  {Pascucci}, {Pasqualetti}, {Passaquieti}, {Passuello}, {Patil}, {Patricelli},
  {Pearlstone}, {Pedraza}, {Pedurand}, {Pekowsky}, {Pele}, {Penn}, {Perez},
  {Perreca}, {Perri}, {Pfeiffer}, {Phelps}, {Piccinni}, {Pichot},
  {Piergiovanni}, {Pierro}, {Pillant}, {Pinard}, {Pinto}, {Pirello}, {Pitkin},
  {Poe}, {Poggiani}, {Popolizio}, {Porter}, {Post}, {Powell}, {Prasad},
  {Pratt}, {Pratten}, {Predoi}, {Prestegard}, {Price}, {Prijatelj}, {Principe},
  {Privitera}, {Prodi}, {Prokhorov}, {Puncken}, {Punturo}, {Puppo},
  {P{\"u}rrer}, {Qi}, {Quetschke}, {Quintero}, {Quitzow-James}, {Raab},
  {Rabeling}, {Radkins}, {Raffai}, {Raja}, {Rajan}, {Rajbhandari}, {Rakhmanov},
  {Ramirez}, {Ramos-Buades}, {Rapagnani}, {Raymond}, {Razzano}, {Read},
  {Regimbau}, {Rei}, {Reid}, {Reitze}, {Ren}, {Reyes}, {Ricci}, {Ricker},
  {Rieger}, {Riles}, {Rizzo}, {Robertson}, {Robie}, {Robinet}, {Rocchi},
  {Rolland}, {Rollins}, {Roma}, {Romano}, {Romel}, {Romie}, {Rosi{\'n}ska},
  {Ross}, {Rowan}, {R{\"u}diger}, {Ruggi}, {Rutins}, {Ryan}, {Sachdev},
  {Sadecki}, {Sadeghian}, {Sakellariadou}, {Salconi}, {Saleem}, {Salemi},
  {Samajdar}, {Sammut}, {Sampson}, {Sanchez}, {Sanchez}, {Sanchis-Gual},
  {Sandberg}, {Sanders}, {Sassolas}, {Sathyaprakash}, {Saulson}, {Sauter},
  {Savage}, {Sawadsky}, {Schale}, {Scheel}, {Scheuer}, {Schmidt}, {Schmidt},
  {Schnabel}, {Schofield}, {Sch{\"o}nbeck}, {Schreiber}, {Schuette}, {Schulte},
  {Schutz}, {Schwalbe}, {Scott}, {Scott}, {Seidel}, {Sellers}, {Sengupta},
  {Sentenac}, {Sequino}, {Sergeev}, {Shaddock}, {Shaffer}, {Shah}, {Shahriar},
  {Shaner}, {Shao}, {Shapiro}, {Shawhan}, {Sheperd}, {Shoemaker}, {Shoemaker},
  {Siellez}, {Siemens}, {Sieniawska}, {Sigg}, {Silva}, {Singer}, {Singh},
  {Singhal}, {Sintes}, {Slagmolen}, {Smith}, {Smith}, {Smith}, {Somala}, {Son},
  {Sonnenberg}, {Sorazu}, {Sorrentino}, {Souradeep}, {Spencer}, {Srivastava},
  {Staats}, {Staley}, {Steinke}, {Steinlechner}, {Steinlechner}, {Steinmeyer},
  {Stevenson}, {Stone}, {Stops}, {Strain}, {Stratta}, {Strigin}, {Strunk},
  {Sturani}, {Stuver}, {Summerscales}, {Sun}, {Sunil}, {Suresh}, {Sutton},
  {Swinkels}, {Szczepa{\'n}czyk}, {Tacca}, {Tait}, {Talbot}, {Talukder},
  {Tanner}, {T{\'a}pai}, {Taracchini}, {Tasson}, {Taylor}, {Taylor}, {Tewari},
  {Theeg}, {Thies}, {Thomas}, {Thomas}, {Thomas}, {Thorne}, {Thorne}, {Thrane},
  {Tiwari}, {Tiwari}, {Tokmakov}, {Toland}, {Tonelli}, {Tornasi},
  {Torres-Forn{\'e}}, {Torrie}, {T{\"o}yr{\"a}}, {Travasso}, {Traylor},
  {Trinastic}, {Tringali}, {Trozzo}, {Tsang}, {Tse}, {Tso}, {Tsukada}, {Tsuna},
  {Tuyenbayev}, {Ueno}, {Ugolini}, {Unnikrishnan}, {Urban}, {Usman},
  {Vahlbruch}, {Vajente}, {Valdes}, {van Bakel}, {van Beuzekom}, {van den
  Brand}, {Van Den Broeck}, {Vander-Hyde}, {van der Schaaf}, {van Heijningen},
  {van Veggel}, {Vardaro}, {Varma}, {Vass}, {Vas{\'u}th}, {Vecchio},
  {Vedovato}, {Veitch}, {Veitch}, {Venkateswara}, {Venugopalan}, {Verkindt},
  {Vetrano}, {Vicer{\'e}}, {Viets}, {Vinciguerra}, {Vine}, {Vinet}, {Vitale},
  {Vo}, {Vocca}, {Vorvick}, {Vyatchanin}, {Wade}, {Wade}, {Wade}, {Walet},
  {Walker}, {Wallace}, {Walsh}, {Wang}, {Wang}, {Wang}, {Wang}, {Wang}, {Ward},
  {Warner}, {Was}, {Watchi}, {Weaver}, {Wei}, {Weinert}, {Weinstein}, {Weiss},
  {Wen}, {Wessel}, {Wessels}, {Westerweck}, {Westphal}, {Wette}, {Whelan},
  {Whitcomb}, {Whiting}, {Whittle}, {Wilken}, {Williams}, {Williams},
  {Williamson}, {Willis}, {Willke}, {Wimmer}, {Winkler}, {Wipf}, {Wittel},
  {Woan}, {Woehler}, {Wofford}, {Wong}, {Worden}, {Wright}, {Wu}, {Wysocki},
  {Xiao}, {Yamamoto}, {Yancey}, {Yang}, {Yap}, {Yazback}, {Yu}, {Yu}, {Yvert},
  {Zadro{\.z}ny}, {Zanolin}, {Zelenova}, {Zendri}, {Zevin}, {Zhang}, {Zhang},
  {Zhang}, {Zhang}, {Zhao}, {Zhou}, {Zhou}, {Zhu}, {Zhu}, {Zimmerman},
  {Zucker}, {Zweizig}, {LIGO Scientific Collaboration}, {Virgo Collaboration},
  {Wilson-Hodge}, {Bissaldi}, {Blackburn}, {Briggs}, {Burns}, {Cleveland},
  {Connaughton}, {Gibby}, {Giles}, {Goldstein}, {Hamburg}, {Jenke}, {Hui},
  {Kippen}, {Kocevski}, {McBreen}, {Meegan}, {Paciesas}, {Poolakkil}, {Preece},
  {Racusin}, {Roberts}, {Stanbro}, {Veres}, {von Kienlin}, {GBM}, {Savchenko},
  {Ferrigno}, {Kuulkers}, {Bazzano}, {Bozzo}, {Brandt}, {Chenevez},
  {Courvoisier}, {Diehl}, {Domingo}, {Hanlon}, {Jourdain}, {Laurent}, {Lebrun},
  {Lutovinov}, {Martin-Carrillo}, {Mereghetti}, {Natalucci}, {Rodi}, {Roques},
  {Sunyaev}, {Ubertini}, {INTEGRAL}, {Aartsen}, {Ackermann}, {Adams},
  {Aguilar}, {Ahlers}, {Ahrens}, {Samarai}, {Altmann}, {Andeen}, {Anderson},
  {Ansseau}, {Anton}, {Arg{\"u}elles}, {Auffenberg}, {Axani}, {Bagherpour},
  {Bai}, {Barron}, {Barwick}, {Baum}, {Bay}, {Beatty}, {Becker Tjus},
  {Bernardini}, {Besson}, {Binder}, {Bindig}, {Blaufuss}, {Blot}, {Bohm},
  {B{\"o}rner}, {Bos}, {Bose}, {B{\"o}ser}, {Botner}, {Bourbeau}, {Bourbeau},
  {Bradascio}, {Braun}, {Brayeur}, {Brenzke}, {Bretz}, {Bron},
  {Brostean-Kaiser}, {Burgman}, {Carver}, {Casey}, {Casier}, {Cheung},
  {Chirkin}, {Christov}, {Clark}, {Classen}, {Coenders}, {Collin}, {Conrad},
  {Cowen}, {Cross}, {Day}, {de Andr{\'e}}, {De Clercq}, {DeLaunay},
  {Dembinski}, {De Ridder}, {Desiati}, {de Vries}, {de Wasseige}, {de With},
  {DeYoung}, {D{\'\i}az-V{\'e}lez}, {di Lorenzo}, {Dujmovic}, {Dumm},
  {Dunkman}, {Dvorak}, {Eberhardt}, {Ehrhardt}, {Eichmann}, {Eller}, {Evenson},
  {Fahey}, {Fazely}, {Felde}, {Filimonov}, {Finley}, {Flis}, {Franckowiak},
  {Friedman}, {Fuchs}, {Gaisser}, {Gallagher}, {Gerhardt}, {Ghorbani}, {Giang},
  {Glauch}, {Gl{\"u}senkamp}, {Goldschmidt}, {Gonzalez}, {Grant}, {Griffith},
  {Haack}, {Hallgren}, {Halzen}, {Hanson}, {Hebecker}, {Heereman}, {Helbing},
  {Hellauer}, {Hickford}, {Hignight}, {Hill}, {Hoffman}, {Hoffmann},
  {Hokanson-Fasig}, {Hoshina}, {Huang}, {Huber}, {Hultqvist}, {H{\"u}nnefeld},
  {In}, {Ishihara}, {Jacobi}, {Japaridze}, {Jeong}, {Jero}, {Jones},
  {Kalaczynski}, {Kang}, {Kappes}, {Karg}, {Karle}, {Kauer}, {Keivani},
  {Kelley}, {Kheirandish}, {Kim}, {Kim}, {Kintscher}, {Kiryluk}, {Kittler},
  {Klein}, {Kohnen}, {Koirala}, {Kolanoski}, {K{\"o}pke}, {Kopper}, {Kopper},
  {Koschinsky}, {Koskinen}, {Kowalski}, {Krings}, {Kroll}, {Kr{\"u}ckl},
  {Kunnen}, {Kunwar}, {Kurahashi}, {Kuwabara}, {Kyriacou}, {Labare},
  {Lanfranchi}, {Larson}, {Lauber}, {Lesiak-Bzdak}, {Leuermann}, {Liu}, {Lu},
  {L{\"u}nemann}, {Luszczak}, {Madsen}, {Maggi}, {Mahn}, {Mancina}, {Maruyama},
  {Mase}, {Maunu}, {McNally}, {Meagher}, {Medici}, {Meier}, {Menne}, {Merino},
  {Meures}, {Miarecki}, {Micallef}, {Moment{\'e}}, {Montaruli}, {Moore},
  {Moulai}, {Nahnhauer}, {Nakarmi}, {Naumann}, {Neer}, {Niederhausen},
  {Nowicki}, {Nygren}, {Obertacke Pollmann}, {Olivas}, {O'Murchadha},
  {Palczewski}, {Pandya}, {Pankova}, {Peiffer}, {Pepper}, {P{\'e}rez de los
  Heros}, {Pieloth}, {Pinat}, {Price}, {Przybylski}, {Raab}, {R{\"a}del},
  {Rameez}, {Rawlins}, {Rea}, {Reimann}, {Relethford}, {Relich}, {Resconi},
  {Rhode}, {Richman}, {Robertson}, {Rongen}, {Rott}, {Ruhe}, {Ryckbosch},
  {Rysewyk}, {S{\"a}lzer}, {Sanchez Herrera}, {Sandrock}, {Sandroos},
  {Santander}, {Sarkar}, {Sarkar}, {Satalecka}, {Schlunder}, {Schmidt},
  {Schneider}, {Schoenen}, {Sch{\"o}neberg}, {Schumacher}, {Seckel},
  {Seunarine}, {Soedingrekso}, {Soldin}, {Song}, {Spiczak}, {Spiering},
  {Stachurska}, {Stamatikos}, {Stanev}, {Stasik}, {Stettner}, {Steuer},
  {Stezelberger}, {Stokstad}, {St{\"o}ssl}, {Strotjohann}, {Stuttard},
  {Sullivan}, {Sutherland}, {Taboada}, {Tatar}, {Tenholt}, {Ter-Antonyan},
  {Terliuk}, {Te{\v{s}}i{\'c}}, {Tilav}, {Toale}, {Tobin}, {Toscano}, {Tosi},
  {Tselengidou}, {Tung}, {Turcati}, {Turley}, {Ty}, {Unger}, {Usner},
  {Vandenbroucke}, {Van Driessche}, {van Eijndhoven}, {Vanheule}, {van Santen},
  {Vehring}, {Vogel}, {Vraeghe}, {Walck}, {Wallace}, {Wallraff}, {Wandler},
  {Wandkowsky}, {Waza}, {Weaver}, {Weiss}, {Wendt}, {Werthebach}, {Whelan},
  {Wiebe}, {Wiebusch}, {Wille}, {Williams}, {Wills}, {Wolf}, {Wood}, {Woolsey},
  {Woschnagg}, {Xu}, {Xu}, {Xu}, {Yanez}, {Yodh}, {Yoshida}, {Yuan}, {Zoll},
  {IceCube Collaboration}, {Balasubramanian}, {Mate}, {Bhalerao},
  {Bhattacharya}, {Vibhute}, {Dewangan}, {Rao}, {Vadawale}, {AstroSat Cadmium
  Zinc Telluride Imager Team}, {Svinkin}, {Hurley}, {Aptekar}, {Frederiks},
  {Golenetskii}, {Kozlova}, {Lysenko}, {Oleynik}, {Tsvetkova}, {Ulanov},
  {Cline}, {IPN Collaboration}, {Li}, {Xiong}, {Zhang}, {Lu}, {Song}, {Cao},
  {Chang}, {Chen}, {Chen}, {Chen}, {Chen}, {Chen}, {Chen}, {Cui}, {Cui},
  {Deng}, {Dong}, {Du}, {Fu}, {Gao}, {Gao}, {Gao}, {Ge}, {Gu}, {Guan}, {Guo},
  {Han}, {Hu}, {Huang}, {Huo}, {Jia}, {Jiang}, {Jiang}, {Jin}, {Jin}, {Li},
  {Li}, {Li}, {Li}, {Li}, {Li}, {Li}, {Li}, {Li}, {Li}, {Li}, {Liang}, {Liao},
  {Liu}, {Liu}, {Liu}, {Liu}, {Liu}, {Liu}, {Liu}, {Lu}, {Lu}, {Luo}, {Ma},
  {Meng}, {Nang}, {Nie}, {Ou}, {Qu}, {Sai}, {Sun}, {Tan}, {Tao}, {Tao}, {Tuo},
  {Wang}, {Wang}, {Wang}, {Wang}, {Wang}, {Wen}, {Wu}, {Wu}, {Xiao}, {Xu},
  {Xu}, {Yan}, {Yang}, {Yang}, {Yang}, {Zhang}, {Zhang}, {Zhang}, {Zhang},
  {Zhang}, {Zhang}, {Zhang}, {Zhang}, {Zhang}, {Zhang}, {Zhang}, {Zhang},
  {Zhang}, {Zhang}, {Zhang}, {Zhang}, {Zhang}, {Zhang}, {Zhao}, {Zhao}, {Zhao},
  {Zheng}, {Zhu}, {Zhu}, {Zou}, {Insight-HXMT Collaboration}, {Albert},
  {Andr{\'e}}, {Anghinolfi}, {Ardid}, {Aubert}, {Aublin}, {Avgitas}, {Baret},
  {Barrios-Mart{\'\i}}, {Basa}, {Belhorma}, {Bertin}, {Biagi}, {Bormuth},
  {Bourret}, {Bouwhuis}, {Br{\^a}nza{\c{s}}}, {Bruijn}, {Brunner}, {Busto},
  {Capone}, {Caramete}, {Carr}, {Celli}, {Cherkaoui El Moursli}, {Chiarusi},
  {Circella}, {Coelho}, {Coleiro}, {Coniglione}, {Costantini}, {Coyle},
  {Creusot}, {D{\'\i}az}, {Deschamps}, {De Bonis}, {Distefano}, {Di Palma},
  {Domi}, {Donzaud}, {Dornic}, {Drouhin}, {Eberl}, {El Bojaddaini}, {El
  Khayati}, {Els{\"a}sser}, {Enzenh{\"o}fer}, {Ettahiri}, {Fassi}, {Felis},
  {Fusco}, {Gay}, {Giordano}, {Glotin}, {Gr{\'e}goire}, {Ruiz}, {Graf},
  {Hallmann}, {van Haren}, {Heijboer}, {Hello}, {Hern{\'a}ndez-Rey},
  {H{\"o}ssl}, {Hofest{\"a}dt}, {Hugon}, {Illuminati}, {James}, {de Jong},
  {Jongen}, {Kadler}, {Kalekin}, {Katz}, {Kiessling}, {Kouchner}, {Kreter},
  {Kreykenbohm}, {Kulikovskiy}, {Lachaud}, {Lahmann}, {Lef{\`e}vre}, {Leonora},
  {Lotze}, {Loucatos}, {Marcelin}, {Margiotta}, {Marinelli},
  {Mart{\'\i}nez-Mora}, {Mele}, {Melis}, {Michael}, {Migliozzi}, {Moussa},
  {Navas}, {Nezri}, {Organokov}, {P{\u{a}}v{\u{a}}la{\c{s}}}, {Pellegrino},
  {Perrina}, {Piattelli}, {Popa}, {Pradier}, {Quinn}, {Racca}, {Riccobene},
  {S{\'a}nchez-Losa}, {Salda{\~n}a}, {Salvadori}, {Samtleben}, {Sanguineti},
  {Sapienza}, {Sieger}, {Spurio}, {Stolarczyk}, {Taiuti}, {Tayalati},
  {Trovato}, {Turpin}, {T{\"o}nnis}, {Vallage}, {Van Elewyck}, {Versari},
  {Vivolo}, {Vizzoca}, {Wilms}, {Zornoza}, {Z{\'u}{\~n}iga}, {ANTARES
  Collaboration}, {Beardmore}, {Breeveld}, {Burrows}, {Cenko}, {Cusumano},
  {D'A{\`\i}}, {de Pasquale}, {Emery}, {Evans}, {Giommi}, {Gronwall}, {Kennea},
  {Krimm}, {Kuin}, {Lien}, {Marshall}, {Melandri}, {Nousek}, {Oates},
  {Osborne}, {Pagani}, {Page}, {Palmer}, {Perri}, {Siegel}, {Sbarufatti},
  {Tagliaferri}, {Tohuvavohu}, {Swift Collaboration}, {Tavani}, {Verrecchia},
  {Bulgarelli}, {Evangelista}, {Pacciani}, {Feroci}, {Pittori}, {Giuliani},
  {Del Monte}, {Donnarumma}, {Argan}, {Trois}, {Ursi}, {Cardillo}, {Piano},
  {Longo}, {Lucarelli}, {Munar-Adrover}, {Fuschino}, {Labanti}, {Marisaldi},
  {Minervini}, {Fioretti}, {Parmiggiani}, {Gianotti}, {Trifoglio}, {Di Persio},
  {Antonelli}, {Barbiellini}, {Caraveo}, {Cattaneo}, {Costa}, {Colafrancesco},
  {D'Amico}, {Ferrari}, {Morselli}, {Paoletti}, {Picozza}, {Pilia}, {Rappoldi},
  {Soffitta}, {Vercellone}, {AGILE Team}, {Foley}, {Coulter}, {Kilpatrick},
  {Drout}, {Piro}, {Shappee}, {Siebert}, {Simon}, {Ulloa}, {Kasen}, {Madore},
  {Murguia-Berthier}, {Pan}, {Prochaska}, {Ramirez-Ruiz}, {Rest},
  {Rojas-Bravo}, {1M2H Team}, {Berger}, {Soares-Santos}, {Annis}, {Alexander},
  {Allam}, {Balbinot}, {Blanchard}, {Brout}, {Butler}, {Chornock}, {Cook},
  {Cowperthwaite}, {Diehl}, {Drlica-Wagner}, {Drout}, {Durret}, {Eftekhari},
  {Finley}, {Fong}, {Frieman}, {Fryer}, {Garc{\'\i}a-Bellido}, {Gruendl},
  {Hartley}, {Herner}, {Kessler}, {Lin}, {Lopes}, {Louren{\c{c}}o}, {Margutti},
  {Marshall}, {Matheson}, {Medina}, {Metzger}, {Mu{\~n}oz}, {Muir}, {Nicholl},
  {Nugent}, {Palmese}, {Paz-Chinch{\'o}n}, {Quataert}, {Sako}, {Sauseda},
  {Schlegel}, {Scolnic}, {Secco}, {Smith}, {Sobreira}, {Villar}, {Vivas},
  {Wester}, {Williams}, {Yanny}, {Zenteno}, {Zhang}, {Abbott}, {Banerji},
  {Bechtol}, {Benoit-L{\'e}vy}, {Bertin}, {Brooks}, {Buckley-Geer}, {Burke},
  {Capozzi}, {Carnero Rosell}, {Carrasco Kind}, {Castander}, {Crocce}, {Cunha},
  {D'Andrea}, {da Costa}, {Davis}, {DePoy}, {Desai}, {Dietrich}, {Eifler},
  {Fernandez}, {Flaugher}, {Fosalba}, {Gaztanaga}, {Gerdes}, {Giannantonio},
  {Goldstein}, {Gruen}, {Gschwend}, {Gutierrez}, {Honscheid}, {James},
  {Jeltema}, {Johnson}, {Johnson}, {Kent}, {Krause}, {Kron}, {Kuehn}, {Lahav},
  {Lima}, {Maia}, {March}, {Martini}, {McMahon}, {Menanteau}, {Miller},
  {Miquel}, {Mohr}, {Nichol}, {Ogando}, {Plazas}, {Romer}, {Roodman}, {Rykoff},
  {Sanchez}, {Scarpine}, {Schindler}, {Schubnell}, {Sevilla-Noarbe}, {Sheldon},
  {Smith}, {Smith}, {Stebbins}, {Suchyta}, {Swanson}, {Tarle}, {Thomas},
  {Troxel}, {Tucker}, {Vikram}, {Walker}, {Wechsler}, {Weller}, {Carlin},
  {Gill}, {Li}, {Marriner}, {Neilsen}, {Dark Energy Camera GW-EM
  Collaboration}, {DES Collaboration}, {Haislip}, {Kouprianov}, {Reichart},
  {Sand}, {Tartaglia}, {Valenti}, {Yang}, {DLT40 Collaboration}, {Benetti},
  {Brocato}, {Campana}, {Cappellaro}, {Covino}, {D'Avanzo}, {D'Elia}, {Getman},
  {Ghirlanda}, {Ghisellini}, {Limatola}, {Nicastro}, {Palazzi}, {Pian},
  {Piranomonte}, {Possenti}, {Rossi}, {Salafia}, {Tomasella}, {Amati},
  {Antonelli}, {Bernardini}, {Bufano}, {Capaccioli}, {Casella}, {Dadina}, {De
  Cesare}, {Di Paola}, {Giuffrida}, {Giunta}, {Israel}, {Lisi}, {Maiorano},
  {Mapelli}, {Masetti}, {Pescalli}, {Pulone}, {Salvaterra}, {Schipani},
  {Spera}, {Stamerra}, {Stella}, {Testa}, {Turatto}, {Vergani}, {Aresu},
  {Bachetti}, {Buffa}, {Burgay}, {Buttu}, {Caria}, {Carretti}, {Casasola},
  {Castangia}, {Carboni}, {Casu}, {Concu}, {Corongiu}, {Deiana}, {Egron},
  {Fara}, {Gaudiomonte}, {Gusai}, {Ladu}, {Loru}, {Leurini}, {Marongiu},
  {Melis}, {Melis}, {Migoni}, {Milia}, {Navarrini}, {Orlati}, {Ortu}, {Palmas},
  {Pellizzoni}, {Perrodin}, {Pisanu}, {Poppi}, {Righini}, {Saba}, {Serra},
  {Serrau}, {Stagni}, {Surcis}, {Vacca}, {Vargiu}, {Hunt}, {Jin}, {Klose},
  {Kouveliotou}, {Mazzali}, {M{\o}ller}, {Nava}, {Piran}, {Selsing}, {Vergani},
  {Wiersema}, {Toma}, {Higgins}, {Mundell}, {di Serego Alighieri}, {G{\'o}tz},
  {Gao}, {Gomboc}, {Kaper}, {Kobayashi}, {Kopac}, {Mao}, {Starling}, {Steele},
  {van der Horst}, {GRAWITA: GRAvitational Wave Inaf TeAm}, {Acero}, {Atwood},
  {Baldini}, {Barbiellini}, {Bastieri}, {Berenji}, {Bellazzini}, {Bissaldi},
  {Blandford}, {Bloom}, {Bonino}, {Bottacini}, {Bregeon}, {Buehler}, {Buson},
  {Cameron}, {Caputo}, {Caraveo}, {Cavazzuti}, {Chekhtman}, {Cheung}, {Chiang},
  {Ciprini}, {Cohen-Tanugi}, {Cominsky}, {Costantin}, {Cuoco}, {D'Ammando}, {de
  Palma}, {Digel}, {Di Lalla}, {Di Mauro}, {Di Venere}, {Dubois}, {Fegan},
  {Focke}, {Franckowiak}, {Fukazawa}, {Funk}, {Fusco}, {Gargano}, {Gasparrini},
  {Giglietto}, {Giordano}, {Giroletti}, {Glanzman}, {Green}, {Grondin},
  {Guillemot}, {Guiriec}, {Harding}, {Horan}, {J{\'o}hannesson}, {Kamae},
  {Kensei}, {Kuss}, {La Mura}, {Latronico}, {Lemoine-Goumard}, {Longo},
  {Loparco}, {Lovellette}, {Lubrano}, {Magill}, {Maldera}, {Manfreda},
  {Mazziotta}, {McEnery}, {Meyer}, {Michelson}, {Mirabal}, {Monzani},
  {Moretti}, {Morselli}, {Moskalenko}, {Negro}, {Nuss}, {Ojha}, {Omodei},
  {Orienti}, {Orlando}, {Palatiello}, {Paliya}, {Paneque}, {Pesce-Rollins},
  {Piron}, {Porter}, {Principe}, {Rain{\`o}}, {Rando}, {Razzano}, {Razzaque},
  {Reimer}, {Reimer}, {Reposeur}, {Rochester}, {Saz Parkinson}, {Sgr{\`o}},
  {Siskind}, {Spada}, {Spandre}, {Suson}, {Takahashi}, {Tanaka}, {Thayer},
  {Thayer}, {Thompson}, {Tibaldo}, {Torres}, {Torresi}, {Troja}, {Venters},
  {Vianello}, {Zaharijas}, {Fermi Large Area Telescope Collaboration},
  {Allison}, {Bannister}, {Dobie}, {Kaplan}, {Lenc}, {Lynch}, {Murphy},
  {Sadler}, {Australia Telescope Compact Array}, {Hotan}, {James}, {Oslowski},
  {Raja}, {Shannon}, {Whiting}, {Australian SKA Pathfinder}, {Arcavi},
  {Howell}, {McCully}, {Hosseinzadeh}, {Hiramatsu}, {Poznanski}, {Barnes},
  {Zaltzman}, {Vasylyev}, {Maoz}, {Las Cumbres Observatory Group}, {Cooke},
  {Bailes}, {Wolf}, {Deller}, {Lidman}, {Wang}, {Gendre}, {Andreoni}, {Ackley},
  {Pritchard}, {Bessell}, {Chang}, {M{\"o}ller}, {Onken}, {Scalzo},
  {Ridden-Harper}, {Sharp}, {Tucker}, {Farrell}, {Elmer}, {Johnston},
  {Venkatraman Krishnan}, {Keane}, {Green}, {Jameson}, {Hu}, {Ma}, {Sun}, {Wu},
  {Wang}, {Shang}, {Hu}, {Ashley}, {Yuan}, {Li}, {Tao}, {Zhu}, {Zhang},
  {Suntzeff}, {Zhou}, {Yang}, {Orange}, {Morris}, {Cucchiara}, {Giblin},
  {Klotz}, {Staff}, {Thierry}, {Schmidt}, {OzGrav}, {(Deeper}, {Wider},
  {program}, {AST3}, {CAASTRO Collaborations}, {Tanvir}, {Levan}, {Cano}, {de
  Ugarte-Postigo}, {Gonz{\'a}lez-Fern{\'a}ndez}, {Greiner}, {Hjorth}, {Irwin},
  {Kr{\"u}hler}, {Mandel}, {Milvang-Jensen}, {O'Brien}, {Rol}, {Rosetti},
  {Rosswog}, {Rowlinson}, {Steeghs}, {Th{\"o}ne}, {Ulaczyk}, {Watson}, {Bruun},
  {Cutter}, {Figuera Jaimes}, {Fujii}, {Fruchter}, {Gompertz}, {Jakobsson},
  {Hodosan}, {J{\`e}rgensen}, {Kangas}, {Kann}, {Rabus}, {Schr{\o}der},
  {Stanway}, {Wijers}, {VINROUGE Collaboration}, {Lipunov}, {Gorbovskoy},
  {Kornilov}, {Tyurina}, {Balanutsa}, {Kuznetsov}, {Vlasenko}, {Podesta},
  {Lopez}, {Podesta}, {Levato}, {Saffe}, {Mallamaci}, {Budnev}, {Gress},
  {Kuvshinov}, {Gorbunov}, {Vladimirov}, {Zimnukhov}, {Gabovich}, {Yurkov},
  {Sergienko}, {Rebolo}, {Serra-Ricart}, {Tlatov}, {Ishmuhametova}, {MASTER
  Collaboration}, {Abe}, {Aoki}, {Aoki}, {Asakura}, {Baar}, {Barway}, {Bond},
  {Doi}, {Finet}, {Fujiyoshi}, {Furusawa}, {Honda}, {Itoh}, {Kanda},
  {Kawabata}, {Kawabata}, {Kim}, {Koshida}, {Kuroda}, {Lee}, {Liu},
  {Matsubayashi}, {Miyazaki}, {Morihana}, {Morokuma}, {Motohara}, {Murata},
  {Nagai}, {Nagashima}, {Nagayama}, {Nakaoka}, {Nakata}, {Ohsawa}, {Ohshima},
  {Ohta}, {Okita}, {Saito}, {Saito}, {Sako}, {Sekiguchi}, {Sumi}, {Tajitsu},
  {Takahashi}, {Takayama}, {Tamura}, {Tanaka}, {Tanaka}, {Terai}, {Tominaga},
  {Tristram}, {Uemura}, {Utsumi}, {Yamaguchi}, {Yasuda}, {Yoshida}, {Zenko},
  {J-GEM}, {Adams}, {Anupama}, {Bally}, {Barway}, {Bellm}, {Blagorodnova},
  {Cannella}, {Chandra}, {Chatterjee}, {Clarke}, {Cobb}, {Cook}, {Copperwheat},
  {De}, {Emery}, {Feindt}, {Foster}, {Fox}, {Frail}, {Fremling}, {Frohmaier},
  {Garcia}, {Ghosh}, {Giacintucci}, {Goobar}, {Gottlieb}, {Grefenstette},
  {Hallinan}, {Harrison}, {Heida}, {Helou}, {Ho}, {Horesh}, {Hotokezaka}, {Ip},
  {Itoh}, {Jacobs}, {Jencson}, {Kasen}, {Kasliwal}, {Kassim}, {Kim}, {Kiran},
  {Kuin}, {Kulkarni}, {Kupfer}, {Lau}, {Madsen}, {Mazzali}, {Miller},
  {Miyasaka}, {Mooley}, {Myers}, {Nakar}, {Ngeow}, {Nugent}, {Ofek},
  {Palliyaguru}, {Pavana}, {Perley}, {Peters}, {Pike}, {Piran}, {Qi}, {Quimby},
  {Rana}, {Rosswog}, {Rusu}, {Sadler}, {Van Sistine}, {Sollerman}, {Xu}, {Yan},
  {Yatsu}, {Yu}, {Zhang}, {Zhao}, {GROWTH}, {JAGWAR}, {Caltech-NRAO},
  {TTU-NRAO}, {NuSTAR Collaborations}, {Chambers}, {Huber}, {Schultz},
  {Bulger}, {Flewelling}, {Magnier}, {Lowe}, {Wainscoat}, {Waters}, {Willman},
  {Pan-STARRS}, {Ebisawa}, {Hanyu}, {Harita}, {Hashimoto}, {Hidaka}, {Hori},
  {Ishikawa}, {Isobe}, {Iwakiri}, {Kawai}, {Kawai}, {Kawamuro}, {Kawase},
  {Kitaoka}, {Makishima}, {Matsuoka}, {Mihara}, {Morita}, {Morita}, {Nakahira},
  {Nakajima}, {Nakamura}, {Negoro}, {Oda}, {Sakamaki}, {Sasaki}, {Serino},
  {Shidatsu}, {Shimomukai}, {Sugawara}, {Sugita}, {Sugizaki}, {Tachibana},
  {Takao}, {Tanimoto}, {Tomida}, {Tsuboi}, {Tsunemi}, {Ueda}, {Ueno}, {Yamada},
  {Yamaoka}, {Yamauchi}, {Yatabe}, {Yoneyama}, {Yoshii}, {MAXI Team}, {Coward},
  {Crisp}, {Macpherson}, {Andreoni}, {Laugier}, {Noysena}, {Klotz}, {Gendre},
  {Thierry}, {Turpin}, {Consortium}, {Im}, {Choi}, {Kim}, {Yoon}, {Lim}, {Lee},
  {Lee}, {Kim}, {Ko}, {Joe}, {Kwon}, {Kim}, {Lim}, {Choi}, {KU Collaboration},
  {Fynbo}, {Malesani}, {Xu}, {Optical Telescope}, {Smartt}, {Jerkstrand},
  {Kankare}, {Sim}, {Fraser}, {Inserra}, {Maguire}, {Leloudas}, {Magee},
  {Shingles}, {Smith}, {Young}, {Kotak}, {Gal-Yam}, {Lyman}, {Homan},
  {Agliozzo}, {Anderson}, {Angus}, {Ashall}, {Barbarino}, {Bauer}, {Berton},
  {Botticella}, {Bulla}, {Cannizzaro}, {Cartier}, {Cikota}, {Clark}, {De Cia},
  {Della Valle}, {Dennefeld}, {Dessart}, {Dimitriadis}, {Elias-Rosa}, {Firth},
  {Fl{\"o}rs}, {Frohmaier}, {Galbany}, {Gonz{\'a}lez-Gait{\'a}n}, {Gromadzki},
  {Guti{\'e}rrez}, {Hamanowicz}, {Harmanen}, {Heintz}, {Hernandez}, {Hodgkin},
  {Hook}, {Izzo}, {James}, {Jonker}, {Kerzendorf}, {Kostrzewa-Rutkowska},
  {Kromer}, {Kuncarayakti}, {Lawrence}, {Manulis}, {Mattila}, {McBrien},
  {M{\"u}ller}, {Nordin}, {O'Neill}, {Onori}, {Palmerio}, {Pastorello},
  {Patat}, {Pignata}, {Podsiadlowski}, {Razza}, {Reynolds}, {Roy}, {Ruiter},
  {Rybicki}, {Salmon}, {Pumo}, {Prentice}, {Seitenzahl}, {Smith}, {Sollerman},
  {Sullivan}, {Szegedi}, {Taddia}, {Taubenberger}, {Terreran}, {Van Soelen},
  {Vos}, {Walton}, {Wright}, {Wyrzykowski}, {Yaron}, {pre=''(''>ePESSTO},
  {Chen}, {Kr{\"u}hler}, {Schady}, {Wiseman}, {Greiner}, {Rau}, {Schweyer},
  {Klose}, {Nicuesa Guelbenzu}, {GROND}, {Palliyaguru}, {Tech University},
  {Shara}, {Williams}, {Vaisanen}, {Potter}, {Romero Colmenero}, {Crawford},
  {Buckley}, {Mao}, {SALT Group}, {D{\'\i}az}, {Macri}, {Garc{\'\i}a Lambas},
  {Mendes de Oliveira}, {Nilo Castell{\'o}n}, {Ribeiro}, {S{\'a}nchez},
  {Schoenell}, {Abramo}, {Akras}, {Alcaniz}, {Artola}, {Beroiz}, {Bonoli},
  {Cabral}, {Camuccio}, {Chavushyan}, {Coelho}, {Colazo}, {Costa-Duarte},
  {Cuevas Larenas}, {Dom{\'\i}nguez Romero}, {Dultzin}, {Fern{\'a}ndez},
  {Garc{\'\i}a}, {Girardini}, {Gon{\c{c}}alves}, {Gon{\c{c}}alves}, {Gurovich},
  {Jim{\'e}nez-Teja}, {Kanaan}, {Lares}, {Lopes de Oliveira}, {L{\'o}pez-Cruz},
  {Melia}, {Molino}, {Padilla}, {Pe{\~n}uela}, {Placco}, {Qui{\~n}ones},
  {Ram{\'\i}rez Rivera}, {Renzi}, {Riguccini}, {R{\'\i}os-L{\'o}pez},
  {Rodriguez}, {Sampedro}, {Schneiter}, {Sodr{\'e}}, {Starck}, {Torres-Flores},
  {Tornatore}, {Zadro{\.z}ny}, {Castillo}, {TOROS: Transient Robotic
  Observatory of South Collaboration}, {Castro-Tirado}, {Tello}, {Hu}, {Zhang},
  {Cunniffe}, {Castell{\'o}n}, {Hiriart}, {Caballero-Garc{\'\i}a},
  {Jel{\'\i}nek}, {Kub{\'a}nek}, {P{\'e}rez del Pulgar}, {Park}, {Jeong},
  {Castro Cer{\'o}n}, {Pandey}, {Yock}, {Querel}, {Fan}, {Wang}, {BOOTES
  Collaboration}, {Beardsley}, {Brown}, {Crosse}, {Emrich}, {Franzen},
  {Gaensler}, {Horsley}, {Johnston-Hollitt}, {Kenney}, {Morales}, {Pallot},
  {Sokolowski}, {Steele}, {Tingay}, {Trott}, {Walker}, {Wayth}, {Williams},
  {Wu}, {Murchison Widefield Array}, {Yoshida}, {Sakamoto}, {Kawakubo},
  {Yamaoka}, {Takahashi}, {Asaoka}, {Ozawa}, {Torii}, {Shimizu}, {Tamura},
  {Ishizaki}, {Cherry}, {Ricciarini}, {Penacchioni}, {Marrocchesi}, {CALET
  Collaboration}, {Pozanenko}, {Volnova}, {Mazaeva}, {Minaev}, {Krugov},
  {Kusakin}, {Reva}, {Moskvitin}, {Rumyantsev}, {Inasaridze}, {Klunko},
  {Tungalag}, {Schmalz}, {Burhonov}, {IKI-GW Follow-up Collaboration},
  {Abdalla}, {Abramowski}, {Aharonian}, {Ait Benkhali}, {Ang{\"u}ner},
  {Arakawa}, {Arrieta}, {Aubert}, {Backes}, {Balzer}, {Barnard}, {Becherini},
  {Becker Tjus}, {Berge}, {Bernhard}, {Bernl{\"o}hr}, {Blackwell},
  {B{\"o}ttcher}, {Boisson}, {Bolmont}, {Bonnefoy}, {Bordas}, {Bregeon},
  {Brun}, {Brun}, {Bryan}, {B{\"u}chele}, {Bulik}, {Capasso}, {Caroff},
  {Carosi}, {Casanova}, {Cerruti}, {Chakraborty}, {Chaves}, {Chen},
  {Chevalier}, {Colafrancesco}, {Condon}, {Conrad}, {Davids}, {Decock}, {Deil},
  {Devin}, {deWilt}, {Dirson}, {Djannati-Ata{\"\i}}, {Donath}, {O'C. Drury},
  {Dutson}, {Dyks}, {Edwards}, {Egberts}, {Emery}, {Ernenwein}, {Eschbach},
  {Farnier}, {Fegan}, {Fernandes}, {Fiasson}, {Fontaine}, {Funk},
  {F{\"u}ssling}, {Gabici}, {Gallant}, {Garrigoux}, {Gat{\'e}}, {Giavitto},
  {Giebels}, {Glawion}, {Glicenstein}, {Gottschall}, {Grondin}, {Hahn},
  {Haupt}, {Hawkes}, {Heinzelmann}, {Henri}, {Hermann}, {Hinton}, {Hofmann},
  {Hoischen}, {Holch}, {Holler}, {Horns}, {Ivascenko}, {Iwasaki},
  {Jacholkowska}, {Jamrozy}, {Jankowsky}, {Jankowsky}, {Jingo}, {Jouvin},
  {Jung-Richardt}, {Kastendieck}, {Katarzy{\'n}ski}, {Katsuragawa},
  {Kerszberg}, {Khangulyan}, {Kh{\'e}lifi}, {King}, {Klepser}, {Klochkov},
  {Klu{\'z}niak}, {Komin}, {Kosack}, {Krakau}, {Kraus}, {Kr{\"u}ger}, {Laffon},
  {Lamanna}, {Lau}, {Lees}, {Lefaucheur}, {Lemi{\`e}re}, {Lemoine-Goumard},
  {Lenain}, {Leser}, {Lohse}, {Lorentz}, {Liu}, {Lypova}, {Malyshev},
  {Marandon}, {Marcowith}, {Mariaud}, {Marx}, {Maurin}, {Maxted}, {Mayer},
  {Meintjes}, {Meyer}, {Mitchell}, {Moderski}, {Mohamed}, {Mohrmann},
  {Mor{\r{a}}}, {Moulin}, {Murach}, {Nakashima}, {de Naurois}, {Ndiyavala},
  {Niederwanger}, {Niemiec}, {Oakes}, {O'Brien}, {Odaka}, {Ohm}, {Ostrowski},
  {Oya}, {Padovani}, {Panter}, {Parsons}, {Pekeur}, {Pelletier}, {Perennes},
  {Petrucci}, {Peyaud}, {Piel}, {Pita}, {Poireau}, {Poon}, {Prokhorov},
  {Prokoph}, {P{\"u}hlhofer}, {Punch}, {Quirrenbach}, {Raab}, {Rauth},
  {Reimer}, {Reimer}, {Renaud}, {de los Reyes}, {Rieger}, {Rinchiuso},
  {Romoli}, {Rowell}, {Rudak}, {Rulten}, {Sahakian}, {Saito}, {Sanchez},
  {Santangelo}, {Sasaki}, {Schlickeiser}, {Sch{\"u}ssler}, {Schulz},
  {Schwanke}, {Schwemmer}, {Seglar-Arroyo}, {Settimo}, {Seyffert}, {Shafi},
  {Shilon}, {Shiningayamwe}, {Simoni}, {Sol}, {Spanier}, {Spir-Jacob},
  {Stawarz}, {Steenkamp}, {Stegmann}, {Steppa}, {Sushch}, {Takahashi},
  {Tavernet}, {Tavernier}, {Taylor}, {Terrier}, {Tibaldo}, {Tiziani},
  {Tluczykont}, {Trichard}, {Tsirou}, {Tsuji}, {Tuffs}, {Uchiyama}, {van der
  Walt}, {van Eldik}, {van Rensburg}, {van Soelen}, {Vasileiadis}, {Veh},
  {Venter}, {Viana}, {Vincent}, {Vink}, {Voisin}, {V{\"o}lk}, {Vuillaume},
  {Wadiasingh}, {Wagner}, {Wagner}, {Wagner}, {White}, {Wierzcholska},
  {Willmann}, {W{\"o}rnlein}, {Wouters}, {Yang}, {Zaborov}, {Zacharias},
  {Zanin}, {Zdziarski}, {Zech}, {Zefi}, {Ziegler}, {Zorn}, {{\.Z}ywucka},
  {H.~E.~S.~S. Collaboration}, {Fender}, {Broderick}, {Rowlinson}, {Wijers},
  {Stewart}, {ter Veen}, {Shulevski}, {LOFAR Collaboration}, {Kavic},
  {Simonetti}, {League}, {Tsai}, {Obenberger}, {Nathaniel}, {Taylor}, {Dowell},
  {Liebling}, {Estes}, {Lippert}, {Sharma}, {Vincent}, {Farella}, {Wavelength
  Array}, {Abeysekara}, {Albert}, {Alfaro}, {Alvarez}, {Arceo},
  {Arteaga-Vel{\'a}zquez}, {Avila Rojas}, {Ayala Solares}, {Barber}, {Becerra
  Gonzalez}, {Becerril}, {Belmont-Moreno}, {BenZvi}, {Berley}, {Bernal},
  {Braun}, {Brisbois}, {Caballero-Mora}, {Capistr{\'a}n}, {Carrami{\~n}ana},
  {Casanova}, {Castillo}, {Cotti}, {Cotzomi}, {Couti{\~n}o de Le{\'o}n}, {De
  Le{\'o}n}, {De la Fuente}, {Diaz Hernandez}, {Dichiara}, {Dingus},
  {DuVernois}, {D{\'\i}az-V{\'e}lez}, {Ellsworth}, {Engel},
  {Enr{\'\i}quez-Rivera}, {Fiorino}, {Fleischhack}, {Fraija},
  {Garc{\'\i}a-Gonz{\'a}lez}, {Garfias}, {Gerhardt}, {Gonz{\~o}lez Mu{\~n}oz},
  {Gonz{\'a}lez}, {Goodman}, {Hampel-Arias}, {Harding}, {Hernandez},
  {Hernandez-Almada}, {Hona}, {H{\"u}ntemeyer}, {Iriarte}, {Jardin-Blicq},
  {Joshi}, {Kaufmann}, {Kieda}, {Lara}, {Lauer}, {Lennarz}, {Le{\'o}n Vargas},
  {Linnemann}, {Longinotti}, {Raya}, {Luna-Garc{\'\i}a}, {L{\'o}pez-Coto},
  {Malone}, {Marinelli}, {Martinez}, {Martinez-Castellanos},
  {Mart{\'\i}nez-Castro}, {Mart{\'\i}nez-Huerta}, {Matthews},
  {Miranda-Romagnoli}, {Moreno}, {Mostaf{\'a}}, {Nellen}, {Newbold}, {Nisa},
  {Noriega-Papaqui}, {Pelayo}, {Pretz}, {P{\'e}rez-P{\'e}rez}, {Ren}, {Rho},
  {Rivi{\`e}re}, {Rosa-Gonz{\'a}lez}, {Rosenberg}, {Ruiz-Velasco}, {Salazar},
  {Salesa Greus}, {Sandoval}, {Schneider}, {Schoorlemmer}, {Sinnis}, {Smith},
  {Springer}, {Surajbali}, {Tibolla}, {Tollefson}, {Torres}, {Ukwatta},
  {Weisgarber}, {Westerhoff}, {Wisher}, {Wood}, {Yapici}, {Yodh}, {Younk},
  {Zhou}, {{\'A}lvarez}, {HAWC Collaboration}, {Aab}, {Abreu}, {Aglietta},
  {Albuquerque}, {Albury}, {Allekotte}, {Almela}, {Alvarez Castillo},
  {Alvarez-Mu{\~n}iz}, {Anastasi}, {Anchordoqui}, {Andrada}, {Andringa},
  {Aramo}, {Arsene}, {Asorey}, {Assis}, {Avila}, {Badescu}, {Balaceanu},
  {Barbato}, {Barreira Luz}, {Becker}, {Bellido}, {Berat}, {Bertaina},
  {Bertou}, {Biermann}, {Biteau}, {Blaess}, {Blanco}, {Blazek}, {Bleve},
  {Boh{\'a}{\v{c}}ov{\'a}}, {Bonifazi}, {Borodai}, {Botti}, {Brack}, {Brancus},
  {Bretz}, {Bridgeman}, {Briechle}, {Buchholz}, {Bueno}, {Buitink}, {Buscemi},
  {Caballero-Mora}, {Caccianiga}, {Cancio}, {Canfora}, {Caruso}, {Castellina},
  {Catalani}, {Cataldi}, {Cazon}, {Chavez}, {Chinellato}, {Chudoba}, {Clay},
  {Cobos Cerutti}, {Colalillo}, {Coleman}, {Collica}, {Coluccia},
  {Concei{\c{c}}{\~a}o}, {Consolati}, {Contreras}, {Cooper}, {Coutu},
  {Covault}, {Cronin}, {D'Amico}, {Daniel}, {Dasso}, {Daumiller}, {Dawson},
  {Day}, {de Almeida}, {de Jong}, {De Mauro}, {de Mello Neto}, {De Mitri}, {de
  Oliveira}, {de Souza}, {Debatin}, {Deligny}, {D{\'\i}az Castro}, {Diogo},
  {Dobrigkeit}, {D'Olivo}, {Dorosti}, {Dos Anjos}, {Dova}, {Dundovic}, {Ebr},
  {Engel}, {Erdmann}, {Erfani}, {Escobar}, {Espadanal}, {Etchegoyen}, {Falcke},
  {Farmer}, {Farrar}, {Fauth}, {Fazzini}, {Feldbusch}, {Fenu}, {Fick},
  {Figueira}, {Filip{\v{c}}i{\v{c}}}, {Freire}, {Fujii}, {Fuster},
  {Ga{\"\i}or}, {Garc{\'\i}a}, {Gat{\'e}}, {Gemmeke}, {Gherghel-Lascu}, {Ghia},
  {Giaccari}, {Giammarchi}, {Giller}, {G{\l}as}, {Glaser}, {Golup}, {G{\'o}mez
  Berisso}, {G{\'o}mez Vitale}, {Gonz{\'a}lez}, {Gorgi}, {Gottowik}, {Grillo},
  {Grubb}, {Guarino}, {Guedes}, {Halliday}, {Hampel}, {Hansen}, {Harari},
  {Harrison}, {Harvey}, {Haungs}, {Hebbeker}, {Heck}, {Heimann}, {Herve},
  {Hill}, {Hojvat}, {Holt}, {Homola}, {H{\"o}randel}, {Horvath},
  {Hrabovsk{\'y}}, {Huege}, {Hulsman}, {Insolia}, {Isar}, {Jandt}, {Johnsen},
  {Josebachuili}, {Jurysek}, {K{\"a}{\"a}p{\"a}}, {Kampert}, {Keilhauer},
  {Kemmerich}, {Kemp}, {Kieckhafer}, {Klages}, {Kleifges}, {Kleinfeller},
  {Krause}, {Krohm}, {Kuempel}, {Kukec Mezek}, {Kunka}, {Kuotb Awad}, {Lago},
  {LaHurd}, {Lang}, {Lauscher}, {Legumina}, {Leigui de Oliveira},
  {Letessier-Selvon}, {Lhenry-Yvon}, {Link}, {Lo Presti}, {Lopes}, {L{\'o}pez},
  {L{\'o}pez Casado}, {Lorek}, {Luce}, {Lucero}, {Malacari}, {Mallamaci},
  {Mandat}, {Mantsch}, {Mariazzi}, {Maris}, {Marsella}, {Martello}, {Martinez},
  {Mart{\'\i}nez Bravo}, {Mas{\'\i}as Meza}, {Mathes}, {Mathys}, {Matthews},
  {Matthiae}, {Mayotte}, {Mazur}, {Medina}, {Medina-Tanco}, {Melo},
  {Menshikov}, {Merenda}, {Michal}, {Micheletti}, {Middendorf}, {Miramonti},
  {Mitrica}, {Mockler}, {Mollerach}, {Montanet}, {Morello}, {Morlino},
  {M{\"u}ller}, {M{\"u}ller}, {Muller}, {M{\"u}ller}, {Mussa}, {Naranjo},
  {Nguyen}, {Niculescu-Oglinzanu}, {Niechciol}, {Niemietz}, {Niggemann},
  {Nitz}, {Nosek}, {Novotny}, {No{\v{z}}ka}, {N{\'u}{\~n}ez}, {Oikonomou},
  {Olinto}, {Palatka}, {Pallotta}, {Papenbreer}, {Parente}, {Parra}, {Paul},
  {Pech}, {Pedreira}, {P{\c{e}}kala}, {Pe{\~n}a-Rodriguez}, {Pereira},
  {Perlin}, {Perrone}, {Peters}, {Petrera}, {Phuntsok}, {Pierog}, {Pimenta},
  {Pirronello}, {Platino}, {Plum}, {Poh}, {Porowski}, {Prado}, {Privitera},
  {Prouza}, {Quel}, {Querchfeld}, {Quinn}, {Ramos-Pollan}, {Rautenberg},
  {Ravignani}, {Ridky}, {Riehn}, {Risse}, {Ristori}, {Rizi}, {Rodrigues de
  Carvalho}, {Rodriguez Fernandez}, {Rodriguez Rojo}, {Roncoroni}, {Roth},
  {Roulet}, {Rovero}, {Ruehl}, {Saffi}, {Saftoiu}, {Salamida}, {Salazar},
  {Saleh}, {Salina}, {S{\'a}nchez}, {Sanchez-Lucas}, {Santos}, {Santos},
  {Sarazin}, {Sarmento}, {Sarmiento-Cano}, {Sato}, {Schauer}, {Scherini},
  {Schieler}, {Schimp}, {Schmidt}, {Scholten}, {Schov{\'a}nek}, {Schr{\"o}der},
  {Schr{\"o}der}, {Schulz}, {Schumacher}, {Sciutto}, {Segreto}, {Shadkam},
  {Shellard}, {Sigl}, {Silli}, {{\v{S}}m{\'\i}da}, {Snow}, {Sommers},
  {Sonntag}, {Soriano}, {Squartini}, {Stanca}, {Stani{\v{c}}}, {Stasielak},
  {Stassi}, {Stolpovskiy}, {Strafella}, {Streich}, {Suarez},
  {Suarez-Dur{\'a}n}, {Sudholz}, {Suomij{\"a}rvi}, {Supanitsky},
  {{\v{S}}up{\'\i}k}, {Swain}, {Szadkowski}, {Taboada}, {Taborda},
  {Timmermans}, {Todero Peixoto}, {Tomankova}, {Tom{\'e}}, {Torralba Elipe},
  {Travnicek}, {Trini}, {Tueros}, {Ulrich}, {Unger}, {Urban}, {Vald{\'e}s
  Galicia}, {Vali{\~n}o}, {Valore}, {van Aar}, {van Bodegom}, {van den Berg},
  {van Vliet}, {Varela}, {Vargas C{\'a}rdenas}, {V{\'a}zquez}, {Veberi{\v{c}}},
  {Ventura}, {Vergara Quispe}, {Verzi}, {Vicha}, {Villase{\~n}or}, {Vorobiov},
  {Wahlberg}, {Wainberg}, {Walz}, {Watson}, {Weber}, {Weindl}, {Wiede{\'n}ski},
  {Wiencke}, {Wilczy{\'n}ski}, {Wirtz}, {Wittkowski}, {Wundheiler}, {Yang},
  {Yushkov}, {Zas}, {Zavrtanik}, {Zavrtanik}, {Zepeda}, {Zimmermann},
  {Ziolkowski}, {Zong}, {Zuccarello}, {Pierre Auger Collaboration}, {Kim},
  {Schulze}, {Bauer}, {Corral-Santana}, {de Gregorio-Monsalvo},
  {Gonz{\'a}lez-L{\'o}pez}, {Hartmann}, {Ishwara-Chandra}, {Mart{\'\i}n},
  {Mehner}, {Misra}, {Micha{\l}owski}, {Resmi}, {ALMA Collaboration}, {Paragi},
  {Agudo}, {An}, {Beswick}, {Casadio}, {Frey}, {Jonker}, {Kettenis}, {Marcote},
  {Moldon}, {Szomoru}, {van Langevelde}, {Yang}, {Euro VLBI Team}, {Cwiek},
  {Cwiok}, {Czyrkowski}, {Dabrowski}, {Kasprowicz}, {Mankiewicz}, {Nawrocki},
  {Opiela}, {Piotrowski}, {Wrochna}, {Zaremba}, {{\.Z}arnecki}, {Pi of Sky
  Collaboration}, {Haggard}, {Nynka}, {Ruan}, {Chandra Team at McGill
  University}, {Bland}, {Booler}, {Devillepoix}, {de Gois}, {Hancock}, {Howie},
  {Paxman}, {Sansom}, {Towner}, {Desert Fireball Network}, {Tonry}, {Coughlin},
  {Stubbs}, {Denneau}, {Heinze}, {Stalder}, {Weiland}, {ATLAS}, {Eatough},
  {Kramer}, {Kraus}, {Time Resolution Universe Survey}, {Troja}, {Piro},
  {Becerra Gonz{\'a}lez}, {Butler}, {Fox}, {Khandrika}, {Kutyrev}, {Lee},
  {Ricci}, {Ryan}, {S{\'a}nchez-Ram{\'\i}rez}, {Veilleux}, {Watson},
  {Wieringa}, {Burgess}, {van Eerten}, {Fontes}, {Fryer}, {Korobkin},
  {Wollaeger}, {RIMAS}, {RATIR}, {Camilo}, {Foley}, {Goedhart}, {Makhathini},
  {Oozeer}, {Smirnov}, {Fender}, {Woudt}, \& {South
  Africa/MeerKAT}}]{2017ApJ...848L..12A}
{Abbott}, B.~P., {Abbott}, R., {Abbott}, T.~D., {et~al.} 2017, \apjl, 848, L12,
  \dodoi{10.3847/2041-8213/aa91c9}

\bibitem[{{Acciari} {et~al.}(2021){Acciari}, {Ansoldi}, {Antonelli}, {Arbet
  Engels}, {Asano}, {Baack}, {Babi{\'c}}, {Baquero}, {Barres de Almeida},
  {Barrio}, {Becerra Gonz{\'a}lez}, {Bednarek}, {Bellizzi}, {Bernardini},
  {Bernardos}, {Berti}, {Besenrieder}, {Bhattacharyya}, {Bigongiari}, {Biland},
  {Blanch}, {Bonnoli}, {Bo{\v{s}}njak}, {Busetto}, {Carosi}, {Ceribella},
  {Cerruti}, {Chai}, {Chilingarian}, {Cikota}, {Colak}, {Colombo}, {Contreras},
  {Cortina}, {Covino}, {D'Amico}, {D'Elia}, {da Vela}, {Dazzi}, {de Angelis},
  {de Lotto}, {Delfino}, {Delgado}, {Delgado Mendez}, {Depaoli}, {di Pierro},
  {di Venere}, {Souto Espi{\~n}eira}, {Dominis Prester}, {Donini}, {Dorner},
  {Doro}, {Elsaesser}, {Fallah Ramazani}, {Fattorini}, {Ferrara}, {Foffano},
  {Fonseca}, {Font}, {Fruck}, {Fukami}, {Garc{\'\i}a L{\'o}pez}, {Garczarczyk},
  {Gasparyan}, {Gaug}, {Giglietto}, {Giordano}, {Gliwny}, {Godinovi{\'c}},
  {Green}, {Green}, {Hadasch}, {Hahn}, {Heckmann}, {Herrera}, {Hoang},
  {Hrupec}, {H{\"u}tten}, {Inada}, {Inoue}, {Ishio}, {Iwamura}, {Jormanainen},
  {Jouvin}, {Kajiwara}, {Karjalainen}, {Kerszberg}, {Kobayashi}, {Kubo},
  {Kushida}, {Lamastra}, {Lelas}, {Leone}, {Lindfors}, {Lombardi}, {Longo},
  {L{\'o}pez-Coto}, {L{\'o}pez-Moya}, {L{\'o}pez-Oramas}, {Loporchio}, {Machado
  de Oliveira Fraga}, {Maggio}, {Majumdar}, {Makariev}, {Mallamaci}, {Maneva},
  {Manganaro}, {Mannheim}, {Maraschi}, {Mariotti}, {Mart{\'\i}nez}, {Mazin},
  {Mender}, {Mi{\'c}anovi{\'c}}, {Miceli}, {Miener}, {Minev}, {Miranda},
  {Mirzoyan}, {Molina}, {Moralejo}, {Morcuende}, {Moreno}, {Moretti},
  {Neustroev}, {Nigro}, {Nilsson}, {Ninci}, {Nishijima}, {Noda}, {Nozaki},
  {Ohtani}, {Oka}, {Otero-Santos}, {Paiano}, {Palatiello}, {Paneque},
  {Paoletti}, {Paredes}, {Pavleti{\'c}}, {Pe{\~n}il}, {Perennes}, {Persic},
  {Prada Moroni}, {Prandini}, {Priyadarshi}, {Puljak}, {Rhode}, {Rib{\'o}},
  {Rico}, {Righi}, {Rugliancich}, {Saha}, {Sahakyan}, {Saito}, {Sakurai},
  {Satalecka}, {Saturni}, {Schleicher}, {Schmidt}, {Schweizer}, {Sitarek},
  {{\v{S}}nidari{\'c}}, {Sobczynska}, {Spolon}, {Stamerra}, {Strom}, {Strzys},
  {Suda}, {Suri{\'c}}, {Takahashi}, {Tavecchio}, {Temnikov}, {Terzi{\'c}},
  {Teshima}, {Torres-Alb{\`a}}, {Tosti}, {Truzzi}, {Tutone}, {van
  Scherpenberg}, {Vanzo}, {Vazquez Acosta}, {Ventura}, {Verguilov}, {Vigorito},
  {Vitale}, {Vovk}, {Will}, {Zari{\'c}}, {MAGIC Collaboration}, \&
  {Nava}}]{2021ApJ...908...90A}
{Acciari}, V.~A., {Ansoldi}, S., {Antonelli}, L.~A., {et~al.} 2021, \apj, 908,
  90, \dodoi{10.3847/1538-4357/abd249}

\bibitem[{{Alexander} {et~al.}(2018){Alexander}, {Margutti}, {Blanchard},
  {Fong}, {Berger}, {Hajela}, \& {et.}}]{2018ApJ...863L..18A}
{Alexander}, K.~D., {Margutti}, R., {Blanchard}, P.~K., {et~al.} 2018, \apjl,
  863, L18, \dodoi{10.3847/2041-8213/aad637}

\bibitem[{{Ando} \& {Beacom}(2005)}]{2005PhRvL..95f1103A}
{Ando}, S., \& {Beacom}, J.~F. 2005, \prl, 95, 061103,
  \dodoi{10.1103/PhysRevLett.95.061103}

\bibitem[{{Arcavi} {et~al.}(2017){Arcavi}, {Hosseinzadeh}, {Howell}, {McCully},
  {Poznanski}, {Kasen}, {Barnes}, {Zaltzman}, {Vasylyev}, {Maoz}, \&
  {Valenti}}]{2017Natur.551...64A}
{Arcavi}, I., {Hosseinzadeh}, G., {Howell}, D.~A., {et~al.} 2017, \nat, 551,
  64, \dodoi{10.1038/nature24291}

\bibitem[{{Barnes} \& {Kasen}(2013)}]{2013ApJ...775...18B}
{Barnes}, J., \& {Kasen}, D. 2013, \apj, 775, 18,
  \dodoi{10.1088/0004-637X/775/1/18}

\bibitem[{{Barniol Duran} \& {Giannios}(2015)}]{2015MNRAS.454.1711B}
{Barniol Duran}, R., \& {Giannios}, D. 2015, \mnras, 454, 1711,
  \dodoi{10.1093/mnras/stv2004}

\bibitem[{{Barthelmy} {et~al.}(2005){Barthelmy}, {Cannizzo}, {Gehrels},
  {Cusumano}, {Mangano}, {O'Brien}, \& {et al.}}]{2005ApJ...635L.133B}
{Barthelmy}, S.~D., {Cannizzo}, J.~K., {Gehrels}, N., {et~al.} 2005, \apjl,
  635, L133, \dodoi{10.1086/499432}

\bibitem[{{Barthelmy} {et~al.}(2013){Barthelmy}, {Baumgartner}, {Cummings},
  {Fenimore}, {Gehrels}, {Krimm}, {Lien}, {Markwardt}, {Melandri}, {Palmer},
  {Sakamoto}, {Sato}, {Stamatikos}, {Tueller}, \&
  {Ukwatta}}]{2013GCN.14741....1B}
{Barthelmy}, S.~D., {Baumgartner}, W.~H., {Cummings}, J.~R., {et~al.} 2013, GRB
  Coordinates Network, 14741, 1

\bibitem[{{Bauswein} {et~al.}(2013){Bauswein}, {Goriely}, \&
  {Janka}}]{2013ApJ...773...78B}
{Bauswein}, A., {Goriely}, S., \& {Janka}, H.~T. 2013, \apj, 773, 78,
  \dodoi{10.1088/0004-637X/773/1/78}

\bibitem[{{Becerra} {et~al.}(2019{\natexlab{a}}){Becerra}, {Watson}, {Fraija},
  {Butler}, {Lee}, {Troja}, {Rom{\'a}n-Z{\'u}{\~n}iga}, {Kutyrev}, {{\'A}lvarez
  Nu{\~n}ez}, {{\'A}ngeles}, {Chapa}, {Cuevas}, {Farah},
  {Fuentes-Fern{\'a}ndez}, {Figueroa}, {Langarica}, {Quir{\'o}s},
  {Ru{\'{\i}}z-D{\'{\i}}az-Soto}, {Tejada}, \& {Tinoco}}]{2019ApJ...872..118B}
{Becerra}, R.~L., {Watson}, A.~M., {Fraija}, N., {et~al.} 2019{\natexlab{a}},
  \apj, 872, 118, \dodoi{10.3847/1538-4357/ab0026}

\bibitem[{{Becerra} {et~al.}(2019{\natexlab{b}}){Becerra}, {De Colle},
  {Watson}, {Fraija}, {Butler}, {Lee}, {Rom{\'a}n-Z{\'u}{\~n}iga}, {Bloom},
  {Gonz{\'a}lez}, {Kutyrev}, {Prochaska}, {Ramirez-Ruiz}, {Richer}, \&
  {Troja}}]{2019ApJ...887..254B}
{Becerra}, R.~L., {De Colle}, F., {Watson}, A.~M., {et~al.} 2019{\natexlab{b}},
  \apj, 887, 254, \dodoi{10.3847/1538-4357/ab5859}

\bibitem[{{Beniamini} {et~al.}(2020){Beniamini}, {Granot}, \&
  {Gill}}]{2020MNRAS.493.3521B}
{Beniamini}, P., {Granot}, J., \& {Gill}, R. 2020, \mnras, 493, 3521,
  \dodoi{10.1093/mnras/staa538}

\bibitem[{{Berger}(2014)}]{2014ARA&A..52...43B}
{Berger}, E. 2014, \araa, 52, 43, \dodoi{10.1146/annurev-astro-081913-035926}

\bibitem[{{Berger} {et~al.}(2013){Berger}, {Fong}, \&
  {Chornock}}]{2013ApJ...774L..23B}
{Berger}, E., {Fong}, W., \& {Chornock}, R. 2013, \apjl, 774, L23,
  \dodoi{10.1088/2041-8205/774/2/L23}

\bibitem[{{Blandford} \& {McKee}(1976)}]{1976PhFl...19.1130B}
{Blandford}, R.~D., \& {McKee}, C.~F. 1976, Physics of Fluids, 19, 1130,
  \dodoi{10.1063/1.861619}

\bibitem[{{Blandford} \& {Znajek}(1977)}]{1977MNRAS.179..433B}
{Blandford}, R.~D., \& {Znajek}, R.~L. 1977, \mnras, 179, 433,
  \dodoi{10.1093/mnras/179.3.433}

\bibitem[{{Bloom} {et~al.}(1999){Bloom}, {Kulkarni}, {Djorgovski},
  {Eichelberger}, {C{\^o}t{\'e}}, \& {et al.}}]{1999Natur.401..453B}
{Bloom}, J.~S., {Kulkarni}, S.~R., {Djorgovski}, S.~G., {et~al.} 1999, \nat,
  401, 453, \dodoi{10.1038/46744}

\bibitem[{{Bromberg} {et~al.}(2011){Bromberg}, {Nakar}, \&
  {Piran}}]{2011ApJ...739L..55B}
{Bromberg}, O., {Nakar}, E., \& {Piran}, T. 2011, \apjl, 739, L55,
  \dodoi{10.1088/2041-8205/739/2/L55}

\bibitem[{{Bucciantini} {et~al.}(2007){Bucciantini}, {Quataert}, {Arons},
  {Metzger}, \& {Thompson}}]{2007MNRAS.380.1541B}
{Bucciantini}, N., {Quataert}, E., {Arons}, J., {Metzger}, B.~D., \&
  {Thompson}, T.~A. 2007, \mnras, 380, 1541,
  \dodoi{10.1111/j.1365-2966.2007.12164.x}

\bibitem[{{Burrows} {et~al.}(2005){Burrows}, {Romano}, {Falcone}, {Kobayashi},
  {Zhang}, \& {et al.}}]{2005Sci...309.1833B}
{Burrows}, D.~N., {Romano}, P., {Falcone}, A., {et~al.} 2005, Science, 309,
  1833, \dodoi{10.1126/science.1116168}

\bibitem[{{Cantiello} {et~al.}(2018){Cantiello}, {Jensen}, {Blakeslee},
  {Berger}, {Levan}, {Tanvir}, {Raimondo}, {Brocato}, {Alexander}, {Blanchard},
  {Branchesi}, {Cano}, {Chornock}, {Covino}, {Cowperthwaite}, {D'Avanzo},
  {Eftekhari}, {Fong}, {Fruchter}, {Grado}, {Hjorth}, {Holz}, {Lyman},
  {Mandel}, {Margutti}, {Nicholl}, {Villar}, \&
  {Williams}}]{2018ApJ...854L..31C}
{Cantiello}, M., {Jensen}, J.~B., {Blakeslee}, J.~P., {et~al.} 2018, \apjl,
  854, L31, \dodoi{10.3847/2041-8213/aaad64}

\bibitem[{{Chevalier}(1989)}]{1989ApJ...346..847C}
{Chevalier}, R.~A. 1989, \apj, 346, 847, \dodoi{10.1086/168066}

\bibitem[{{Chevalier} \& {Fransson}(2006)}]{2006ApJ...651..381C}
{Chevalier}, R.~A., \& {Fransson}, C. 2006, \apj, 651, 381,
  \dodoi{10.1086/507606}

\bibitem[{{Chincarini} {et~al.}(2007){Chincarini}, {Moretti}, {Romano},
  {Falcone}, {Morris}, \& {et al.}}]{2007ApJ...671.1903C}
{Chincarini}, G., {Moretti}, A., {Romano}, P., {et~al.} 2007, \apj, 671, 1903,
  \dodoi{10.1086/521591}

\bibitem[{{Connaughton} {et~al.}(2017){Connaughton}, {GBM-LIGO Group},
  {Blackburn}, {Briggs}, {Broida}, {Burns}, {Camp}, {Dal Canton},
  {Christensen}, {Goldstein}, {Hamburg}, {Hui}, {Jenke}, {Kocevski}, {Leroy},
  {Littenberg}, {McEnery}, {Preece}, {Racusin}, {Shawhan}, {Siellez}, {Singer},
  {Veitch}, {Veres}, \& {Wilson-Hodge}}]{2017GCN.21506....1C}
{Connaughton}, V., {GBM-LIGO Group}, {Blackburn}, L., {et~al.} 2017, GRB
  Coordinates Network, 21506, 1

\bibitem[{{Coulter} {et~al.}(2017){Coulter}, {Foley}, {Kilpatrick}, {Drout},
  {Piro}, {Shappee}, {Siebert}, {Simon}, {Ulloa}, {Kasen}, {Madore},
  {Murguia-Berthier}, {Pan}, {Prochaska}, {Ramirez-Ruiz}, {Rest}, \&
  {Rojas-Bravo}}]{2017Sci...358.1556C}
{Coulter}, D.~A., {Foley}, R.~J., {Kilpatrick}, C.~D., {et~al.} 2017, Science,
  358, 1556, \dodoi{10.1126/science.aap9811}

\bibitem[{{Cowperthwaite} {et~al.}(2017){Cowperthwaite}, {Berger}, {Villar},
  {Metzger}, {Nicholl}, {Chornock}, \& {et al.}}]{2017ApJ...848L..17C}
{Cowperthwaite}, P.~S., {Berger}, E., {Villar}, V.~A., {et~al.} 2017, \apjl,
  848, L17, \dodoi{10.3847/2041-8213/aa8fc7}

\bibitem[{{Dai} \& {Lu}(1998)}]{1998A&A...333L..87D}
{Dai}, Z.~G., \& {Lu}, T. 1998, \aap, 333, L87.
\newblock \doarXiv{astro-ph/9810402}

\bibitem[{{Dai} \& {Lu}(1999)}]{1999ApJ...519L.155D}
---. 1999, \apjl, 519, L155, \dodoi{10.1086/312127}

\bibitem[{{Dai} \& {Lu}(2000)}]{2000ApJ...537..803D}
---. 2000, \apj, 537, 803, \dodoi{10.1086/309044}

\bibitem[{{Dai} {et~al.}(2006){Dai}, {Wang}, {Wu}, \&
  {Zhang}}]{2006Sci...311.1127D}
{Dai}, Z.~G., {Wang}, X.~Y., {Wu}, X.~F., \& {Zhang}, B. 2006, Science, 311,
  1127, \dodoi{10.1126/science.1123606}

\bibitem[{{Dall'Osso} {et~al.}(2017){Dall'Osso}, {Perna}, {Tanaka}, \&
  {Margutti}}]{2017MNRAS.464.4399D}
{Dall'Osso}, S., {Perna}, R., {Tanaka}, T.~L., \& {Margutti}, R. 2017, \mnras,
  464, 4399, \dodoi{10.1093/mnras/stw2695}

\bibitem[{{D'Avanzo} {et~al.}(2018{\natexlab{a}}){D'Avanzo}, {Campana},
  {Salafia}, {Ghirlanda}, {Ghisellini}, {Melandri}, {Bernardini}, {Branchesi},
  {Chassande-Mottin}, {Covino}, {D'Elia}, {Nava}, {Salvaterra}, {Tagliaferri},
  \& {Vergani}}]{2018arXiv180106164D}
{D'Avanzo}, P., {Campana}, S., {Salafia}, O.~S., {et~al.} 2018{\natexlab{a}},
  ArXiv e-prints.
\newblock \doarXiv{1801.06164}

\bibitem[{{D'Avanzo} {et~al.}(2018{\natexlab{b}}){D'Avanzo}, {Campana},
  {Salafia}, {Ghirland a}, {Ghisellini}, {Melandri}, {Bernardini}, {Branchesi},
  {Chassande-Mottin}, {Covino}, {D'Elia}, {Nava}, {Salvaterra}, {Tagliaferri},
  \& {Vergani}}]{2018A&A...613L...1D}
---. 2018{\natexlab{b}}, \aap, 613, L1, \dodoi{10.1051/0004-6361/201832664}

\bibitem[{{Davies} {et~al.}(1994){Davies}, {Benz}, {Piran}, \&
  {Thielemann}}]{1994ApJ...431..742D}
{Davies}, M.~B., {Benz}, W., {Piran}, T., \& {Thielemann}, F.~K. 1994, \apj,
  431, 742, \dodoi{10.1086/174525}

\bibitem[{{Dessart} {et~al.}(2009){Dessart}, {Ott}, {Burrows}, {Rosswog}, \&
  {Livne}}]{2009ApJ...690.1681D}
{Dessart}, L., {Ott}, C.~D., {Burrows}, A., {Rosswog}, S., \& {Livne}, E. 2009,
  \apj, 690, 1681, \dodoi{10.1088/0004-637X/690/2/1681}

\bibitem[{{Dhawan} {et~al.}(2020){Dhawan}, {Bulla}, {Goobar}, {Sagu{\'e}s
  Carracedo}, \& {Setzer}}]{2020ApJ...888...67D}
{Dhawan}, S., {Bulla}, M., {Goobar}, A., {Sagu{\'e}s Carracedo}, A., \&
  {Setzer}, C.~N. 2020, \apj, 888, 67, \dodoi{10.3847/1538-4357/ab5799}

\bibitem[{{Duncan} \& {Thompson}(1992)}]{1992ApJ...392L...9D}
{Duncan}, R.~C., \& {Thompson}, C. 1992, \apjl, 392, L9, \dodoi{10.1086/186413}

\bibitem[{{Fan} \& {Piran}(2006)}]{2006MNRAS.369..197F}
{Fan}, Y., \& {Piran}, T. 2006, \mnras, 369, 197,
  \dodoi{10.1111/j.1365-2966.2006.10280.x}

\bibitem[{{Fern{\'a}ndez} {et~al.}(2015){Fern{\'a}ndez}, {Kasen}, {Metzger}, \&
  {Quataert}}]{2015MNRAS.446..750F}
{Fern{\'a}ndez}, R., {Kasen}, D., {Metzger}, B.~D., \& {Quataert}, E. 2015,
  \mnras, 446, 750, \dodoi{10.1093/mnras/stu2112}

\bibitem[{{Fong} {et~al.}(2019){Fong}, {Blanchard}, {Alexander}, {Strader},
  {Margutti}, {Hajela}, {Villar}, {Wu}, {Ye}, {Berger}, {Chornock},
  {Coppejans}, {Cowperthwaite}, {Eftekhari}, {Giannios}, {Guidorzi},
  {Kathirgamaraju}, {Laskar}, {Macfadyen}, {Metzger}, {Nicholl}, {Paterson},
  {Terreran}, {Sand}, {Sironi}, {Williams}, {Xie}, \&
  {Zrake}}]{2019ApJ...883L...1F}
{Fong}, W., {Blanchard}, P.~K., {Alexander}, K.~D., {et~al.} 2019, \apjl, 883,
  L1, \dodoi{10.3847/2041-8213/ab3d9e}

\bibitem[{{Fox} {et~al.}(2005){Fox}, {Frail}, {Price}, {Kulkarni}, {Berger},
  {Piran}, {Soderberg}, {Cenko}, {Cameron}, {Gal-Yam}, {Kasliwal}, {Moon},
  {Harrison}, {Nakar}, {Schmidt}, {Penprase}, {Chevalier}, {Kumar}, {Roth},
  {Watson}, {Lee}, {Shectman}, {Phillips}, {Roth}, {McCarthy}, {Rauch},
  {Cowie}, {Peterson}, {Rich}, {Kawai}, {Aoki}, {Kosugi}, {Totani}, {Park},
  {MacFadyen}, \& {Hurley}}]{2005Natur.437..845F}
{Fox}, D.~B., {Frail}, D.~A., {Price}, P.~A., {et~al.} 2005, \nat, 437, 845,
  \dodoi{10.1038/nature04189}

\bibitem[{{Fraija}(2014)}]{2014MNRAS.437.2187F}
{Fraija}, N. 2014, \mnras, 437, 2187, \dodoi{10.1093/mnras/stt2036}

\bibitem[{{Fraija} {et~al.}(2019{\natexlab{a}}){Fraija}, {De Colle}, {Veres},
  {Dichiara}, {Barniol Duran}, {Galvan-Gamez}, \&
  {Pedreira}}]{2019ApJ...871..123F}
{Fraija}, N., {De Colle}, F., {Veres}, P., {et~al.} 2019{\natexlab{a}}, \apj,
  871, 123, \dodoi{10.3847/1538-4357/aaf564}

\bibitem[{{Fraija} {et~al.}(2021{\natexlab{a}}){Fraija}, {Kamenetskaia},
  {Dainotti}, {Duran}, {G{\'a}lvan G{\'a}mez}, {Dichiara}, \& {Caligula do
  E.~S. Pedreira}}]{2021ApJ...907...78F}
{Fraija}, N., {Kamenetskaia}, B.~B., {Dainotti}, M.~G., {et~al.}
  2021{\natexlab{a}}, \apj, 907, 78, \dodoi{10.3847/1538-4357/abcaf6}

\bibitem[{{Fraija} {et~al.}(2020){Fraija}, {Laskar}, {Dichiara}, {Beniamini},
  {Duran}, {Dainotti}, \& {Becerra}}]{2020ApJ...905..112F}
{Fraija}, N., {Laskar}, T., {Dichiara}, S., {et~al.} 2020, \apj, 905, 112,
  \dodoi{10.3847/1538-4357/abc41a}

\bibitem[{{Fraija} {et~al.}(2019{\natexlab{b}}){Fraija}, {Lopez-Camara},
  {Pedreira}, {Betancourt Kamenetskaia}, {Veres}, \&
  {Dichiara}}]{2019ApJ...884...71F}
{Fraija}, N., {Lopez-Camara}, D., {Pedreira}, A.~C. C. d. E.~S., {et~al.}
  2019{\natexlab{b}}, \apj, 884, 71, \dodoi{10.3847/1538-4357/ab40a9}

\bibitem[{{Fraija} {et~al.}(2019{\natexlab{c}}){Fraija}, {Pedreira}, \&
  {Veres}}]{2019ApJ...871..200F}
{Fraija}, N., {Pedreira}, A.~C.~C.~d.~E.~S., \& {Veres}, P. 2019{\natexlab{c}},
  \apj, 871, 200, \dodoi{10.3847/1538-4357/aaf80e}

\bibitem[{{Fraija} {et~al.}(2021{\natexlab{b}}){Fraija}, {Veres}, {Beniamini},
  {Galvan-Gamez}, {Metzger}, {Barniol Duran}, \&
  {Becerra}}]{2021ApJ...918...12F}
{Fraija}, N., {Veres}, P., {Beniamini}, P., {et~al.} 2021{\natexlab{b}}, \apj,
  918, 12, \dodoi{10.3847/1538-4357/ac0aed}

\bibitem[{Gal-Yam(2017)}]{2017hsn..book..195G}
Gal-Yam, A. 2017, Observational and Physical Classification of Supernovae
  (Cham: Springer International Publishing), 1--43,
  \dodoi{10.1007/978-3-319-20794-0_35-1}

\bibitem[{{Galama} {et~al.}(1998){Galama}, {Vreeswijk}, {van Paradijs},
  {Kouveliotou}, {Augusteijn}, {B{\"o}hnhardt}, \& {et
  al.}}]{1998Natur.395..670G}
{Galama}, T.~J., {Vreeswijk}, P.~M., {van Paradijs}, J., {et~al.} 1998, \nat,
  395, 670, \dodoi{10.1038/27150}

\bibitem[{{Gehrels} {et~al.}(2006){Gehrels}, {Norris}, {Barthelmy}, {Granot},
  {Kaneko}, {Kouveliotou}, {Markwardt}, {M{\'e}sz{\'a}ros}, {Nakar}, {Nousek},
  {O'Brien}, {Page}, {Palmer}, {Parsons}, {Roming}, {Sakamoto}, {Sarazin},
  {Schady}, {Stamatikos}, \& {Woosley}}]{2006Natur.444.1044G}
{Gehrels}, N., {Norris}, J.~P., {Barthelmy}, S.~D., {et~al.} 2006, \nat, 444,
  1044, \dodoi{10.1038/nature05376}

\bibitem[{{Goldstein} {et~al.}(2017){Goldstein}, {Veres}, {Burns}, {Briggs},
  {Hamburg}, {Kocevski}, \& {et al.}}]{2017ApJ...848L..14G}
{Goldstein}, A., {Veres}, P., {Burns}, E., {et~al.} 2017, \apjl, 848, L14,
  \dodoi{10.3847/2041-8213/aa8f41}

\bibitem[{{Goriely} {et~al.}(2011){Goriely}, {Bauswein}, \&
  {Janka}}]{2011ApJ...738L..32G}
{Goriely}, S., {Bauswein}, A., \& {Janka}, H.-T. 2011, \apjl, 738, L32,
  \dodoi{10.1088/2041-8205/738/2/L32}

\bibitem[{{Gottlieb} {et~al.}(2018{\natexlab{a}}){Gottlieb}, {Nakar}, \&
  {Piran}}]{2018MNRAS.473..576G}
{Gottlieb}, O., {Nakar}, E., \& {Piran}, T. 2018{\natexlab{a}}, \mnras, 473,
  576, \dodoi{10.1093/mnras/stx2357}

\bibitem[{{Gottlieb} {et~al.}(2018{\natexlab{b}}){Gottlieb}, {Nakar}, {Piran},
  \& {Hotokezaka}}]{2018MNRAS.479..588G}
{Gottlieb}, O., {Nakar}, E., {Piran}, T., \& {Hotokezaka}, K.
  2018{\natexlab{b}}, \mnras, 479, 588, \dodoi{10.1093/mnras/sty1462}

\bibitem[{{Grossman} {et~al.}(2014){Grossman}, {Korobkin}, {Rosswog}, \&
  {Piran}}]{2014MNRAS.439..757G}
{Grossman}, D., {Korobkin}, O., {Rosswog}, S., \& {Piran}, T. 2014, \mnras,
  439, 757, \dodoi{10.1093/mnras/stt2503}

\bibitem[{{Hajela} {et~al.}(2019){Hajela}, {Margutti}, {Alexander},
  {Kathirgamaraju}, {Baldeschi}, {Guidorzi}, {Giannios}, {Fong}, {Wu},
  {MacFadyen}, {Paggi}, {Berger}, {Blanchard}, {Chornock}, {Coppejans},
  {Cowperthwaite}, {Eftekhari}, {Gomez}, {Hosseinzadeh}, {Laskar}, {Metzger},
  {Nicholl}, {Paterson}, {Radice}, {Sironi}, {Terreran}, {Villar}, {Williams},
  {Xie}, \& {Zrake}}]{2019ApJ...886L..17H}
{Hajela}, A., {Margutti}, R., {Alexander}, K.~D., {et~al.} 2019, \apjl, 886,
  L17, \dodoi{10.3847/2041-8213/ab5226}

\bibitem[{{Hajela} {et~al.}(2021){Hajela}, {Margutti}, {Bright}, {Alexander},
  {Metzger}, {Nedora}, {Kathirgamaraju}, {Margalit}, {Radice}, {Berger},
  {MacFadyen}, {Giannios}, {Chornock}, {Heywood}, {Sironi}, {Gottlieb},
  {Coppejans}, {Laskar}, {Cendes}, {Barniol Duran}, {Eftekhari}, {Fong},
  {McDowell}, {Nicholl}, {Xie}, {Zrake}, {Bernuzzi}, {Broekgaarden},
  {Kilpatrick}, {Terreran}, {Villar}, {Blanchard}, {Gomez}, {Hosseinzadeh},
  {Matthews}, \& {Rastinejad}}]{2021arXiv210402070H}
{Hajela}, A., {Margutti}, R., {Bright}, J.~S., {et~al.} 2021, arXiv e-prints,
  arXiv:2104.02070.
\newblock \doarXiv{2104.02070}

\bibitem[{{Hasco{\"e}t} {et~al.}(2017){Hasco{\"e}t}, {Beloborodov}, {Daigne},
  \& {Mochkovitch}}]{2017MNRAS.472L..94H}
{Hasco{\"e}t}, R., {Beloborodov}, A.~M., {Daigne}, F., \& {Mochkovitch}, R.
  2017, \mnras, 472, L94, \dodoi{10.1093/mnrasl/slx143}

\bibitem[{{Hjorth} {et~al.}(2005){Hjorth}, {Watson}, {Fynbo}, {Price},
  {Jensen}, {J{\o}rgensen}, {Kubas}, {Gorosabel}, {Jakobsson}, {Sollerman},
  {Pedersen}, \& {Kouveliotou}}]{2005Natur.437..859H}
{Hjorth}, J., {Watson}, D., {Fynbo}, J. P.~U., {et~al.} 2005, \nat, 437, 859,
  \dodoi{10.1038/nature04174}

\bibitem[{{Hotokezaka} {et~al.}(2018){Hotokezaka}, {Kiuchi}, {Shibata},
  {Nakar}, \& {Piran}}]{2018ApJ...867...95H}
{Hotokezaka}, K., {Kiuchi}, K., {Shibata}, M., {Nakar}, E., \& {Piran}, T.
  2018, \apj, 867, 95, \dodoi{10.3847/1538-4357/aadf92}

\bibitem[{{Hotokezaka} {et~al.}(2013){Hotokezaka}, {Kyutoku}, {Tanaka},
  {Kiuchi}, {Sekiguchi}, {Shibata}, \& {Wanajo}}]{2013ApJ...778L..16H}
{Hotokezaka}, K., {Kyutoku}, K., {Tanaka}, M., {et~al.} 2013, \apjl, 778, L16,
  \dodoi{10.1088/2041-8205/778/1/L16}

\bibitem[{{Hotokezaka} \& {Piran}(2015)}]{2015MNRAS.450.1430H}
{Hotokezaka}, K., \& {Piran}, T. 2015, \mnras, 450, 1430,
  \dodoi{10.1093/mnras/stv620}

\bibitem[{{Huang} \& {Cheng}(2003)}]{2003MNRAS.341..263H}
{Huang}, Y.~F., \& {Cheng}, K.~S. 2003, \mnras, 341, 263,
  \dodoi{10.1046/j.1365-8711.2003.06430.x}

\bibitem[{{Huang} {et~al.}(1998){Huang}, {Dai}, \& {Lu}}]{1998A&A...336L..69H}
{Huang}, Y.~F., {Dai}, Z.~G., \& {Lu}, T. 1998, \aap, 336, L69.
\newblock \doarXiv{astro-ph/9807061}

\bibitem[{{Huang} {et~al.}(1999){Huang}, {Dai}, \& {Lu}}]{1999MNRAS.309..513H}
---. 1999, \mnras, 309, 513, \dodoi{10.1046/j.1365-8711.1999.02887.x}

\bibitem[{{Ioka} {et~al.}(2006){Ioka}, {Toma}, {Yamazaki}, \&
  {Nakamura}}]{2006A&A...458....7I}
{Ioka}, K., {Toma}, K., {Yamazaki}, R., \& {Nakamura}, T. 2006, \aap, 458, 7,
  \dodoi{10.1051/0004-6361:20064939}

\bibitem[{{Izzo} {et~al.}(2020){Izzo}, {Auchettl}, {Hjorth}, {De Colle},
  {Gall}, {Angus}, {Raimundo}, \& {Ramirez-Ruiz}}]{2020arXiv200405941I}
{Izzo}, L., {Auchettl}, K., {Hjorth}, J., {et~al.} 2020, arXiv e-prints,
  arXiv:2004.05941.
\newblock \doarXiv{2004.05941}

\bibitem[{{Izzo} {et~al.}(2019){Izzo}, {de Ugarte Postigo}, {Maeda},
  {Th{\"o}ne}, {Kann}, {Della Valle}, {Sagues Carracedo}, {Micha{\l}owski},
  {Schady}, {Schmidl}, {Selsing}, {Starling}, {Suzuki}, {Bensch}, {Bolmer},
  {Campana}, {Cano}, {Covino}, {Fynbo}, {Hartmann}, {Heintz}, {Hjorth},
  {Japelj}, {Kami{\'n}ski}, {Kaper}, {Kouveliotou}, {Kru{\.Z}y{\'n}ski},
  {Kwiatkowski}, {Leloudas}, {Levan}, {Malesani}, {Micha{\l}owski},
  {Piranomonte}, {Pugliese}, {Rossi}, {S{\'a}nchez-Ram{\'\i}rez}, {Schulze},
  {Steeghs}, {Tanvir}, {Ulaczyk}, {Vergani}, \&
  {Wiersema}}]{2019Natur.565..324I}
{Izzo}, L., {de Ugarte Postigo}, A., {Maeda}, K., {et~al.} 2019, \nat, 565,
  324, \dodoi{10.1038/s41586-018-0826-3}

\bibitem[{{Janiuk} {et~al.}(2004){Janiuk}, {Perna}, {Di Matteo}, \&
  {Czerny}}]{2004MNRAS.355..950J}
{Janiuk}, A., {Perna}, R., {Di Matteo}, T., \& {Czerny}, B. 2004, \mnras, 355,
  950, \dodoi{10.1111/j.1365-2966.2004.08377.x}

\bibitem[{{Jin} {et~al.}(2007){Jin}, {Yan}, {Fan}, \&
  {Wei}}]{2007ApJ...656L..57J}
{Jin}, Z.~P., {Yan}, T., {Fan}, Y.~Z., \& {Wei}, D.~M. 2007, \apjl, 656, L57,
  \dodoi{10.1086/512971}

\bibitem[{{Jin} {et~al.}(2016){Jin}, {Hotokezaka}, {Li}, {Tanaka}, {D'Avanzo},
  {Fan}, {Covino}, {Wei}, \& {Piran}}]{2016NatCo...712898J}
{Jin}, Z.-P., {Hotokezaka}, K., {Li}, X., {et~al.} 2016, Nature Communications,
  7, 12898, \dodoi{10.1038/ncomms12898}

\bibitem[{{Kasen} {et~al.}(2013){Kasen}, {Badnell}, \&
  {Barnes}}]{2013ApJ...774...25K}
{Kasen}, D., {Badnell}, N.~R., \& {Barnes}, J. 2013, \apj, 774, 25,
  \dodoi{10.1088/0004-637X/774/1/25}

\bibitem[{{Kasen} {et~al.}(2015){Kasen}, {Fern{\'a}ndez}, \&
  {Metzger}}]{2015MNRAS.450.1777K}
{Kasen}, D., {Fern{\'a}ndez}, R., \& {Metzger}, B.~D. 2015, \mnras, 450, 1777,
  \dodoi{10.1093/mnras/stv721}

\bibitem[{{Kasliwal} {et~al.}(2017{\natexlab{a}}){Kasliwal}, {Korobkin}, {Lau},
  {Wollaeger}, \& {Fryer}}]{2017ApJ...843L..34K}
{Kasliwal}, M.~M., {Korobkin}, O., {Lau}, R.~M., {Wollaeger}, R., \& {Fryer},
  C.~L. 2017{\natexlab{a}}, \apjl, 843, L34, \dodoi{10.3847/2041-8213/aa799d}

\bibitem[{{Kasliwal} {et~al.}(2017{\natexlab{b}}){Kasliwal}, {Nakar}, {Singer},
  {Kaplan}, {Cook}, {Van Sistine}, \& {et al.}}]{2017Sci...358.1559K}
{Kasliwal}, M.~M., {Nakar}, E., {Singer}, L.~P., {et~al.} 2017{\natexlab{b}},
  Science, 358, 1559, \dodoi{10.1126/science.aap9455}

\bibitem[{{Kathirgamaraju} {et~al.}(2016){Kathirgamaraju}, {Barniol Duran}, \&
  {Giannios}}]{2016MNRAS.461.1568K}
{Kathirgamaraju}, A., {Barniol Duran}, R., \& {Giannios}, D. 2016, \mnras, 461,
  1568, \dodoi{10.1093/mnras/stw1441}

\bibitem[{{Kathirgamaraju} {et~al.}(2019){Kathirgamaraju}, {Giannios}, \&
  {Beniamini}}]{2019MNRAS.487.3914K}
{Kathirgamaraju}, A., {Giannios}, D., \& {Beniamini}, P. 2019, \mnras, 487,
  3914, \dodoi{10.1093/mnras/stz1564}

\bibitem[{{King} {et~al.}(2005){King}, {O'Brien}, {Goad}, {Osborne}, {Olsson},
  \& {Page}}]{2005ApJ...630L.113K}
{King}, A., {O'Brien}, P.~T., {Goad}, M.~R., {et~al.} 2005, \apjl, 630, L113,
  \dodoi{10.1086/496881}

\bibitem[{{Kisaka} {et~al.}(2017){Kisaka}, {Ioka}, {Kashiyama}, \&
  {Nakamura}}]{2017arXiv171100243K}
{Kisaka}, S., {Ioka}, K., {Kashiyama}, K., \& {Nakamura}, T. 2017, ArXiv
  e-prints.
\newblock \doarXiv{1711.00243}

\bibitem[{{Kiuchi} {et~al.}(2014){Kiuchi}, {Kyutoku}, {Sekiguchi}, {Shibata},
  \& {Wada}}]{2014PhRvD..90d1502K}
{Kiuchi}, K., {Kyutoku}, K., {Sekiguchi}, Y., {Shibata}, M., \& {Wada}, T.
  2014, \prd, 90, 041502, \dodoi{10.1103/PhysRevD.90.041502}

\bibitem[{{Kouveliotou} {et~al.}(1993){Kouveliotou}, {Meegan}, {Fishman},
  {Bhat}, {Briggs}, {Koshut}, {Paciesas}, \& {Pendleton}}]{1993ApJ...413L.101K}
{Kouveliotou}, C., {Meegan}, C.~A., {Fishman}, G.~J., {et~al.} 1993, \apjl,
  413, L101, \dodoi{10.1086/186969}

\bibitem[{{Kulkarni} {et~al.}(1998){Kulkarni}, {Frail}, {Wieringa}, {Ekers},
  {Sadler}, {Wark}, {Higdon}, {Phinney}, \& {Bloom}}]{1998Natur.395..663K}
{Kulkarni}, S.~R., {Frail}, D.~A., {Wieringa}, M.~H., {et~al.} 1998, \nat, 395,
  663, \dodoi{10.1038/27139}

\bibitem[{{Kumar} {et~al.}(2008{\natexlab{a}}){Kumar}, {Narayan}, \&
  {Johnson}}]{2008Sci...321..376K}
{Kumar}, P., {Narayan}, R., \& {Johnson}, J.~L. 2008{\natexlab{a}}, Science,
  321, 376, \dodoi{10.1126/science.1159003}

\bibitem[{{Kumar} {et~al.}(2008{\natexlab{b}}){Kumar}, {Narayan}, \&
  {Johnson}}]{2008MNRAS.388.1729K}
---. 2008{\natexlab{b}}, \mnras, 388, 1729,
  \dodoi{10.1111/j.1365-2966.2008.13493.x}

\bibitem[{{Kumar} \& {Piran}(2000)}]{2000ApJ...532..286K}
{Kumar}, P., \& {Piran}, T. 2000, \apj, 532, 286, \dodoi{10.1086/308537}

\bibitem[{{Kyutoku} {et~al.}(2014){Kyutoku}, {Ioka}, \&
  {Shibata}}]{2014MNRAS.437L...6K}
{Kyutoku}, K., {Ioka}, K., \& {Shibata}, M. 2014, \mnras, 437, L6,
  \dodoi{10.1093/mnrasl/slt128}

\bibitem[{{Lamb} \& {Kobayashi}(2017)}]{2017MNRAS.472.4953L}
{Lamb}, G.~P., \& {Kobayashi}, S. 2017, \mnras, 472, 4953,
  \dodoi{10.1093/mnras/stx2345}

\bibitem[{{Lamb} \& {Kobayashi}(2018)}]{2018MNRAS.478..733L}
---. 2018, \mnras, 478, 733, \dodoi{10.1093/mnras/sty1108}

\bibitem[{{Lamb} {et~al.}(2019){Lamb}, {Tanvir}, {Levan}, {de Ugarte Postigo},
  {Kawaguchi}, {Corsi}, {Evans}, {Gompertz}, {Malesani}, {Page}, {Wiersema},
  {Rosswog}, {Shibata}, {Tanaka}, {van der Horst}, {Cano}, {Fynbo}, {Fruchter},
  {Greiner}, {Heintz}, {Higgins}, {Hjorth}, {Izzo}, {Jakobsson}, {Kann},
  {O'Brien}, {Perley}, {Pian}, {Pugliese}, {Starling}, {Th{\"o}ne}, {Watson},
  {Wijers}, \& {Xu}}]{2019ApJ...883...48L}
{Lamb}, G.~P., {Tanvir}, N.~R., {Levan}, A.~J., {et~al.} 2019, \apj, 883, 48,
  \dodoi{10.3847/1538-4357/ab38bb}

\bibitem[{{Laskar} {et~al.}(2015){Laskar}, {Berger}, {Margutti}, {Perley},
  {Zauderer}, {Sari}, \& {Fong}}]{2015ApJ...814....1L}
{Laskar}, T., {Berger}, E., {Margutti}, R., {et~al.} 2015, \apj, 814, 1,
  \dodoi{10.1088/0004-637X/814/1/1}

\bibitem[{{Laskar} {et~al.}(2018){Laskar}, {Alexander}, {Berger}, {Guidorzi},
  {Margutti}, {Fong}, {Kilpatrick}, {Milne}, {Drout}, {Mundell}, {Kobayashi},
  {Lunnan}, {Barniol Duran}, {Menten}, {Ioka}, \&
  {Williams}}]{2018ApJ...862...94L}
{Laskar}, T., {Alexander}, K.~D., {Berger}, E., {et~al.} 2018, \apj, 862, 94,
  \dodoi{10.3847/1538-4357/aacbcc}

\bibitem[{{Lattimer} \& {Schutz}(2005)}]{2005ApJ...629..979L}
{Lattimer}, J.~M., \& {Schutz}, B.~F. 2005, \apj, 629, 979,
  \dodoi{10.1086/431543}

\bibitem[{{Lazzati} {et~al.}(2017){Lazzati}, {L{\'o}pez-C{\'a}mara},
  {Cantiello}, {Morsony}, {Perna}, \& {Workman}}]{2017ApJ...848L...6L}
{Lazzati}, D., {L{\'o}pez-C{\'a}mara}, D., {Cantiello}, M., {et~al.} 2017,
  \apjl, 848, L6, \dodoi{10.3847/2041-8213/aa8f3d}

\bibitem[{{Lazzati} {et~al.}(2018){Lazzati}, {Perna}, {Morsony},
  {Lopez-Camara}, {Cantiello}, {Ciolfi}, {Giacomazzo}, \&
  {Workman}}]{2018PhRvL.120x1103L}
{Lazzati}, D., {Perna}, R., {Morsony}, B.~J., {et~al.} 2018, \prl, 120, 241103,
  \dodoi{10.1103/PhysRevLett.120.241103}

\bibitem[{{Lee} {et~al.}(2000){Lee}, {Wijers}, \&
  {Brown}}]{2000PhR...325...83L}
{Lee}, H.~K., {Wijers}, R.~A.~M.~J., \& {Brown}, G.~E. 2000, \physrep, 325, 83,
  \dodoi{10.1016/S0370-1573(99)00084-8}

\bibitem[{{Lei} {et~al.}(2013){Lei}, {Zhang}, \& {Liang}}]{2013ApJ...765..125L}
{Lei}, W.-H., {Zhang}, B., \& {Liang}, E.-W. 2013, \apj, 765, 125,
  \dodoi{10.1088/0004-637X/765/2/125}

\bibitem[{{Li} \& {Paczy{\'n}ski}(1998)}]{1998ApJ...507L..59L}
{Li}, L.-X., \& {Paczy{\'n}ski}, B. 1998, \apjl, 507, L59,
  \dodoi{10.1086/311680}

\bibitem[{{Liang} {et~al.}(2013){Liang}, {Li}, {Gao}, {Zhang}, {Liang}, {Wu},
  {Yi}, {Dai}, {Tang}, {Chen}, {L{\"u}}, {Zhang}, {Lu}, {L{\"u}}, \&
  {Wei}}]{2013ApJ...774...13L}
{Liang}, E.-W., {Li}, L., {Gao}, H., {et~al.} 2013, \apj, 774, 13,
  \dodoi{10.1088/0004-637X/774/1/13}

\bibitem[{{Livio} \& {Waxman}(2000)}]{2000ApJ...538..187L}
{Livio}, M., \& {Waxman}, E. 2000, \apj, 538, 187, \dodoi{10.1086/309120}

\bibitem[{{Lyman} {et~al.}(2018){Lyman}, {Lamb}, {Levan}, {Mandel}, {Tanvir},
  {Kobayashi}, {Gompertz}, {Hjorth}, {Fruchter}, {Kangas}, {Steeghs}, {Steele},
  {Cano}, {Copperwheat}, {Evans}, {Fynbo}, {Gall}, {Im}, {Izzo}, {Jakobsson},
  {Milvang-Jensen}, {O'Brien}, {Osborne}, {Palazzi}, {Perley}, {Pian},
  {Rosswog}, {Rowlinson}, {Schulze}, {Stanway}, {Sutton}, {Th{\"o}ne}, {de
  Ugarte Postigo}, {Watson}, {Wiersema}, \& {Wijers}}]{2018NatAs...2..751L}
{Lyman}, J.~D., {Lamb}, G.~P., {Levan}, A.~J., {et~al.} 2018, Nature Astronomy,
  2, 751, \dodoi{10.1038/s41550-018-0511-3}

\bibitem[{{MacFadyen} \& {Woosley}(1999)}]{1999ApJ...524..262M}
{MacFadyen}, A.~I., \& {Woosley}, S.~E. 1999, \apj, 524, 262,
  \dodoi{10.1086/307790}

\bibitem[{{MacFadyen} {et~al.}(2001){MacFadyen}, {Woosley}, \&
  {Heger}}]{2001ApJ...550..410M}
{MacFadyen}, A.~I., {Woosley}, S.~E., \& {Heger}, A. 2001, \apj, 550, 410,
  \dodoi{10.1086/319698}

\bibitem[{{Margalit} \& {Piran}(2020)}]{2020arXiv200413028M}
{Margalit}, B., \& {Piran}, T. 2020, arXiv e-prints, arXiv:2004.13028.
\newblock \doarXiv{2004.13028}

\bibitem[{{Margutti} {et~al.}(2014){Margutti}, {Milisavljevic}, {Soderberg},
  {Guidorzi}, {Morsony}, {Sanders}, {Chakraborti}, {Ray}, {Kamble}, {Drout},
  {Parrent}, {Zauderer}, \& {Chomiuk}}]{2014ApJ...797..107M}
{Margutti}, R., {Milisavljevic}, D., {Soderberg}, A.~M., {et~al.} 2014, \apj,
  797, 107, \dodoi{10.1088/0004-637X/797/2/107}

\bibitem[{{Margutti} {et~al.}(2017{\natexlab{a}}){Margutti}, {Berger}, {Fong},
  {Guidorzi}, {Alexander}, {Metzger}, {Blanchard}, {Cowperthwaite}, {Chornock},
  {Eftekhari}, {Nicholl}, {Villar}, {Williams}, {Annis}, {Brown}, {Chen},
  {Doctor}, {Frieman}, {Holz}, {Sako}, \&
  {Soares-Santos}}]{2017ApJ...848L..20M}
{Margutti}, R., {Berger}, E., {Fong}, W., {et~al.} 2017{\natexlab{a}}, \apjl,
  848, L20, \dodoi{10.3847/2041-8213/aa9057}

\bibitem[{{Margutti} {et~al.}(2017{\natexlab{b}}){Margutti}, {Kamble},
  {Milisavljevic}, {Zapartas}, {de Mink}, {Drout}, {Chornock}, {Risaliti},
  {Zauderer}, {Bietenholz}, {Cantiello}, {Chakraborti}, {Chomiuk}, {Fong},
  {Grefenstette}, {Guidorzi}, {Kirshner}, {Parrent}, {Patnaude}, {Soderberg},
  {Gehrels}, \& {Harrison}}]{2017ApJ...835..140M}
{Margutti}, R., {Kamble}, A., {Milisavljevic}, D., {et~al.} 2017{\natexlab{b}},
  \apj, 835, 140, \dodoi{10.3847/1538-4357/835/2/140}

\bibitem[{{Margutti} {et~al.}(2018){Margutti}, {Alexander}, {Xie}, {Sironi},
  {Metzger}, {Kathirgamaraju}, {Fong}, {Blanchard}, {Berger}, {MacFadyen},
  {Giannios}, {Guidorzi}, {Hajela}, {Chornock}, {Cowperthwaite}, {Eftekhari},
  {Nicholl}, {Villar}, {Williams}, \& {Zrake}}]{2018ApJ...856L..18M}
{Margutti}, R., {Alexander}, K.~D., {Xie}, X., {et~al.} 2018, \apjl, 856, L18,
  \dodoi{10.3847/2041-8213/aab2ad}

\bibitem[{{M{\'e}sz{\'a}ros} \& {Waxman}(2001)}]{2001PhRvL..87q1102M}
{M{\'e}sz{\'a}ros}, P., \& {Waxman}, E. 2001, \prl, 87, 171102,
  \dodoi{10.1103/PhysRevLett.87.171102}

\bibitem[{{Metzger}(2017)}]{2017LRR....20....3M}
{Metzger}, B.~D. 2017, Living Reviews in Relativity, 20, 3,
  \dodoi{10.1007/s41114-017-0006-z}

\bibitem[{{Metzger}(2019)}]{2019LRR....23....1M}
---. 2019, Living Reviews in Relativity, 23, 1,
  \dodoi{10.1007/s41114-019-0024-0}

\bibitem[{{Metzger} {et~al.}(2015){Metzger}, {Bauswein}, {Goriely}, \&
  {Kasen}}]{2015MNRAS.446.1115M}
{Metzger}, B.~D., {Bauswein}, A., {Goriely}, S., \& {Kasen}, D. 2015, \mnras,
  446, 1115, \dodoi{10.1093/mnras/stu2225}

\bibitem[{{Metzger} {et~al.}(2018){Metzger}, {Beniamini}, \&
  {Giannios}}]{2018ApJ...857...95M}
{Metzger}, B.~D., {Beniamini}, P., \& {Giannios}, D. 2018, \apj, 857, 95,
  \dodoi{10.3847/1538-4357/aab70c}

\bibitem[{{Metzger} \& {Berger}(2012)}]{2012ApJ...746...48M}
{Metzger}, B.~D., \& {Berger}, E. 2012, \apj, 746, 48,
  \dodoi{10.1088/0004-637X/746/1/48}

\bibitem[{{Metzger} \& {Fern{\'a}ndez}(2014)}]{2014MNRAS.441.3444M}
{Metzger}, B.~D., \& {Fern{\'a}ndez}, R. 2014, \mnras, 441, 3444,
  \dodoi{10.1093/mnras/stu802}

\bibitem[{{Metzger} {et~al.}(2011){Metzger}, {Giannios}, {Thompson},
  {Bucciantini}, \& {Quataert}}]{2011MNRAS.413.2031M}
{Metzger}, B.~D., {Giannios}, D., {Thompson}, T.~A., {Bucciantini}, N., \&
  {Quataert}, E. 2011, \mnras, 413, 2031, \dodoi{10.1111/j.
  1365-2966.2011.18280.x}

\bibitem[{{Metzger} {et~al.}(2010){Metzger}, {Mart{\'{\i}}nez-Pinedo},
  {Darbha}, {Quataert}, {Arcones}, {Kasen}, {Thomas}, {Nugent}, {Panov}, \&
  {Zinner}}]{2010MNRAS.406.2650M}
{Metzger}, B.~D., {Mart{\'{\i}}nez-Pinedo}, G., {Darbha}, S., {et~al.} 2010,
  \mnras, 406, 2650, \dodoi{10.1111/j.1365-2966.2010.16864.x}

\bibitem[{{Miller} {et~al.}(2019){Miller}, {Ryan}, {Dolence}, {Burrows},
  {Fontes}, {Fryer}, {Korobkin}, {Lippuner}, {Mumpower}, \&
  {Wollaeger}}]{2019PhRvD.100b3008M}
{Miller}, J.~M., {Ryan}, B.~R., {Dolence}, J.~C., {et~al.} 2019, \prd, 100,
  023008, \dodoi{10.1103/PhysRevD.100.023008}

\bibitem[{{Modjaz} {et~al.}(2020){Modjaz}, {Bianco}, {Siwek}, {Huang},
  {Perley}, {Fierroz}, {Liu}, {Arcavi}, {Gal-Yam}, {Filippenko},
  {Blagorodnova}, {Cenko}, {Kasliwal}, {Kulkarni}, {Schulze}, {Taggart}, \&
  {Zheng}}]{2020ApJ...892..153M}
{Modjaz}, M., {Bianco}, F.~B., {Siwek}, M., {et~al.} 2020, \apj, 892, 153,
  \dodoi{10.3847/1538-4357/ab4185}

\bibitem[{{Mooley} {et~al.}(2018{\natexlab{a}}){Mooley}, {Nakar}, {Hotokezaka},
  {Hallinan}, {Corsi}, {Frail}, \& {et al.}}]{2018Natur.554..207M}
{Mooley}, K.~P., {Nakar}, E., {Hotokezaka}, K., {et~al.} 2018{\natexlab{a}},
  \nat, 554, 207, \dodoi{10.1038/nature25452}

\bibitem[{{Mooley} {et~al.}(2018{\natexlab{b}}){Mooley}, {Deller}, {Gottlieb},
  {Nakar}, {Hallinan}, {Bourke}, {Frail}, {Horesh}, {Corsi}, \&
  {Hotokezaka}}]{2018Natur.561..355M}
{Mooley}, K.~P., {Deller}, A.~T., {Gottlieb}, O., {et~al.} 2018{\natexlab{b}},
  \nat, 561, 355, \dodoi{10.1038/s41586-018-0486-3}

\bibitem[{{Murguia-Berthier} {et~al.}(2014){Murguia-Berthier}, {Montes},
  {Ramirez-Ruiz}, {De Colle}, \& {Lee}}]{2014ApJ...788L...8M}
{Murguia-Berthier}, A., {Montes}, G., {Ramirez-Ruiz}, E., {De Colle}, F., \&
  {Lee}, W.~H. 2014, \apjl, 788, L8, \dodoi{10.1088/2041-8205/788/1/L8}

\bibitem[{{Nagakura} {et~al.}(2014){Nagakura}, {Hotokezaka}, {Sekiguchi},
  {Shibata}, \& {Ioka}}]{2014ApJ...784L..28N}
{Nagakura}, H., {Hotokezaka}, K., {Sekiguchi}, Y., {Shibata}, M., \& {Ioka}, K.
  2014, \apjl, 784, L28, \dodoi{10.1088/2041-8205/784/2/L28}

\bibitem[{{Nakar} \& {Piran}(2017)}]{2017ApJ...834...28N}
{Nakar}, E., \& {Piran}, T. 2017, \apj, 834, 28,
  \dodoi{10.3847/1538-4357/834/1/28}

\bibitem[{{Nicholl} {et~al.}(2017){Nicholl}, {Berger}, {Kasen}, {Metzger},
  {Elias}, {Brice{\~n}o}, \& {et al.}}]{2017ApJ...848L..18N}
{Nicholl}, M., {Berger}, E., {Kasen}, D., {et~al.} 2017, \apjl, 848, L18,
  \dodoi{10.3847/2041-8213/aa9029}

\bibitem[{{Nicholl} {et~al.}(2020){Nicholl}, {Blanchard}, {Berger}, {Chornock},
  {Margutti}, {Gomez}, \& {et al.}}]{2020NatAs.tmp...78N}
{Nicholl}, M., {Blanchard}, P.~K., {Berger}, E., {et~al.} 2020, Nature
  Astronomy, \dodoi{10.1038/s41550-020-1066-7}

\bibitem[{{Nousek} {et~al.}(2006){Nousek}, {Kouveliotou}, {Grupe}, {Page},
  {Granot}, {Ramirez-Ruiz}, {Patel}, {Burrows}, {Mangano}, {Barthelmy},
  {Beardmore}, {Campana}, {Capalbi}, {Chincarini}, {Cusumano}, {Falcone},
  {Gehrels}, {Giommi}, {Goad}, {Godet}, {Hurkett}, {Kennea}, {Moretti},
  {O'Brien}, {Osborne}, {Romano}, {Tagliaferri}, \&
  {Wells}}]{2006ApJ...642..389N}
{Nousek}, J.~A., {Kouveliotou}, C., {Grupe}, D., {et~al.} 2006, \apj, 642, 389,
  \dodoi{10.1086/500724}

\bibitem[{{Paczy{\'n}ski}(1998)}]{1998ApJ...494L..45P}
{Paczy{\'n}ski}, B. 1998, \apjl, 494, L45, \dodoi{10.1086/311148}

\bibitem[{{Perego} {et~al.}(2014){Perego}, {Rosswog}, {Cabez{\'o}n},
  {Korobkin}, {K{\"a}ppeli}, {Arcones}, \&
  {Liebend{\"o}rfer}}]{2014MNRAS.443.3134P}
{Perego}, A., {Rosswog}, S., {Cabez{\'o}n}, R.~M., {et~al.} 2014, \mnras, 443,
  3134, \dodoi{10.1093/mnras/stu1352}

\bibitem[{{Pereyra} {et~al.}(2022){Pereyra}, {Fraija}, {Watson}, {Becerra},
  {Butler}, {De Colle}, {Troja}, {Dichiara}, {Fraire-Bonilla}, {Lee},
  {Ramirez-Ruiz}, {Bloom}, {Prochaska}, {Kutyrev}, {Gonz{\'a}lez}, \&
  {Richer}}]{2022MNRAS.511.6205P}
{Pereyra}, M., {Fraija}, N., {Watson}, A.~M., {et~al.} 2022, \mnras, 511, 6205,
  \dodoi{10.1093/mnras/stac389}

\bibitem[{{Perna} {et~al.}(2006){Perna}, {Armitage}, \&
  {Zhang}}]{2006ApJ...636L..29P}
{Perna}, R., {Armitage}, P.~J., \& {Zhang}, B. 2006, \apjl, 636, L29,
  \dodoi{10.1086/499775}

\bibitem[{{Piran} {et~al.}(2013){Piran}, {Nakar}, \&
  {Rosswog}}]{2013MNRAS.430.2121P}
{Piran}, T., {Nakar}, E., \& {Rosswog}, S. 2013, \mnras, 430, 2121,
  \dodoi{10.1093/mnras/stt037}

\bibitem[{{Piro} \& {Ott}(2011)}]{2011ApJ...736..108P}
{Piro}, A.~L., \& {Ott}, C.~D. 2011, \apj, 736, 108,
  \dodoi{10.1088/0004-637X/736/2/108}

\bibitem[{{Planck Collaboration} {et~al.}(2016){Planck Collaboration}, {Ade},
  {Aghanim}, {Arnaud}, {Ashdown}, {Aumont}, \& {et al.}}]{2016A&A...594A..13P}
{Planck Collaboration}, {Ade}, P.~A.~R., {Aghanim}, N., {et~al.} 2016, \aap,
  594, A13, \dodoi{10.1051/0004-6361/201525830}

\bibitem[{{Popham} {et~al.}(1999){Popham}, {Woosley}, \&
  {Fryer}}]{1999ApJ...518..356P}
{Popham}, R., {Woosley}, S.~E., \& {Fryer}, C. 1999, \apj, 518, 356,
  \dodoi{10.1086/307259}

\bibitem[{{Price} \& {Rosswog}(2006)}]{2006Sci...312..719P}
{Price}, D.~J., \& {Rosswog}, S. 2006, Science, 312, 719,
  \dodoi{10.1126/science.1125201}

\bibitem[{{Proga} \& {Zhang}(2006)}]{2006MNRAS.370L..61P}
{Proga}, D., \& {Zhang}, B. 2006, \mnras, 370, L61,
  \dodoi{10.1111/j.1745-3933.2006.00189.x}

\bibitem[{{Quataert} \& {Kasen}(2012)}]{2012MNRAS.419L...1Q}
{Quataert}, E., \& {Kasen}, D. 2012, \mnras, 419, L1,
  \dodoi{10.1111/j.1745-3933.2011.01151.x}

\bibitem[{{Rees} \& {M{\'e}sz{\'a}ros}(1998)}]{1998ApJ...496L...1R}
{Rees}, M.~J., \& {M{\'e}sz{\'a}ros}, P. 1998, \apjl, 496, L1,
  \dodoi{10.1086/311244}

\bibitem[{{Resmi} {et~al.}(2018){Resmi}, {Schulze}, {Ishwara-Chandra}, {Misra},
  {Buchner}, {De Pasquale}, {S{\'a}nchez-Ram{\'\i}rez}, {Klose}, {Kim},
  {Tanvir}, \& {O'Brien}}]{2018ApJ...867...57R}
{Resmi}, L., {Schulze}, S., {Ishwara-Chandra}, C.~H., {et~al.} 2018, \apj, 867,
  57, \dodoi{10.3847/1538-4357/aae1a6}

\bibitem[{{Rosswog}(2005)}]{2005ApJ...634.1202R}
{Rosswog}, S. 2005, \apj, 634, 1202, \dodoi{10.1086/497062}

\bibitem[{{Rosswog}(2007)}]{2007MNRAS.376L..48R}
---. 2007, \mnras, 376, L48, \dodoi{10.1111/j.1745-3933.2007.00284.x}

\bibitem[{{Rosswog} {et~al.}(1999){Rosswog}, {Liebend{\"o}rfer}, {Thielemann},
  {Davies}, {Benz}, \& {Piran}}]{1999A&A...341..499R}
{Rosswog}, S., {Liebend{\"o}rfer}, M., {Thielemann}, F.~K., {et~al.} 1999,
  \aap, 341, 499.
\newblock \doarXiv{astro-ph/9811367}

\bibitem[{{Ruffert} {et~al.}(1997){Ruffert}, {Janka}, {Takahashi}, \&
  {Schaefer}}]{1997A&A...319..122R}
{Ruffert}, M., {Janka}, H.~T., {Takahashi}, K., \& {Schaefer}, G. 1997, \aap,
  319, 122.
\newblock \doarXiv{astro-ph/9606181}

\bibitem[{{Sari} \& {M{\'e}sz{\'a}ros}(2000)}]{2000ApJ...535L..33S}
{Sari}, R., \& {M{\'e}sz{\'a}ros}, P. 2000, \apjl, 535, L33,
  \dodoi{10.1086/312689}

\bibitem[{{Sari} {et~al.}(1998){Sari}, {Piran}, \&
  {Narayan}}]{1998ApJ...497L..17S}
{Sari}, R., {Piran}, T., \& {Narayan}, R. 1998, \apjl, 497, L17,
  \dodoi{10.1086/311269}

\bibitem[{{Savchenko} {et~al.}(2017){Savchenko}, {Ferrigno}, {Kuulkers},
  {Bazzano}, {Bozzo}, {Brandt}, \& {et al.}}]{2017ApJ...848L..15S}
{Savchenko}, V., {Ferrigno}, C., {Kuulkers}, E., {et~al.} 2017, \apjl, 848,
  L15, \dodoi{10.3847/2041-8213/aa8f94}

\bibitem[{{Shibata} \& {Hotokezaka}(2019)}]{2019ARNPS..69...41S}
{Shibata}, M., \& {Hotokezaka}, K. 2019, Annual Review of Nuclear and Particle
  Science, 69, 41, \dodoi{10.1146/annurev-nucl-101918-023625}

\bibitem[{{Shibata} \& {Taniguchi}(2006)}]{2006PhRvD..73f4027S}
{Shibata}, M., \& {Taniguchi}, K. 2006, \prd, 73, 064027,
  \dodoi{10.1103/PhysRevD.73.064027}

\bibitem[{{Siegel} {et~al.}(2013){Siegel}, {Ciolfi}, {Harte}, \&
  {Rezzolla}}]{2013PhRvD..87l1302S}
{Siegel}, D.~M., {Ciolfi}, R., {Harte}, A.~I., \& {Rezzolla}, L. 2013, \prd,
  87, 121302, \dodoi{10.1103/PhysRevD.87.121302}

\bibitem[{{Siegel} \& {Metzger}(2017)}]{2017PhRvL.119w1102S}
{Siegel}, D.~M., \& {Metzger}, B.~D. 2017, \prl, 119, 231102,
  \dodoi{10.1103/PhysRevLett.119.231102}

\bibitem[{{Sironi} \& {Giannios}(2013)}]{2013ApJ...778..107S}
{Sironi}, L., \& {Giannios}, D. 2013, \apj, 778, 107,
  \dodoi{10.1088/0004-637X/778/2/107}

\bibitem[{{Smartt} {et~al.}(2017){Smartt}, {Chen}, {Jerkstrand}, {Coughlin},
  {Kankare}, {Sim}, \& {et al.}}]{2017Natur.551...75S}
{Smartt}, S.~J., {Chen}, T.~W., {Jerkstrand}, A., {et~al.} 2017, \nat, 551, 75,
  \dodoi{10.1038/nature24303}

\bibitem[{{Soares-Santos} {et~al.}(2017){Soares-Santos}, {Holz}, {Annis},
  {Chornock}, {Herner}, {Berger}, \& {Dark Energy Camera GW-EM
  Collaboration}}]{2017ApJ...848L..16S}
{Soares-Santos}, M., {Holz}, D.~E., {Annis}, J., {et~al.} 2017, \apjl, 848,
  L16, \dodoi{10.3847/2041-8213/aa9059}

\bibitem[{{Sobacchi} {et~al.}(2017){Sobacchi}, {Granot}, {Bromberg}, \&
  {Sormani}}]{2017MNRAS.472..616S}
{Sobacchi}, E., {Granot}, J., {Bromberg}, O., \& {Sormani}, M.~C. 2017, \mnras,
  472, 616, \dodoi{10.1093/mnras/stx2083}

\bibitem[{{Tan} {et~al.}(2001){Tan}, {Matzner}, \&
  {McKee}}]{2001ApJ...551..946T}
{Tan}, J.~C., {Matzner}, C.~D., \& {McKee}, C.~F. 2001, \apj, 551, 946,
  \dodoi{10.1086/320245}

\bibitem[{{Tanaka} \& {Hotokezaka}(2013)}]{2013ApJ...775..113T}
{Tanaka}, M., \& {Hotokezaka}, K. 2013, \apj, 775, 113,
  \dodoi{10.1088/0004-637X/775/2/113}

\bibitem[{{Tanvir} {et~al.}(2013){Tanvir}, {Levan}, {Fruchter}, {Hjorth},
  {Hounsell}, {Wiersema}, \& {Tunnicliffe}}]{2013Natur.500..547T}
{Tanvir}, N.~R., {Levan}, A.~J., {Fruchter}, A.~S., {et~al.} 2013, \nat, 500,
  547, \dodoi{10.1038/nature12505}

\bibitem[{{Tanvir} {et~al.}(2017){Tanvir}, {Levan},
  {Gonz{\'a}lez-Fern{\'a}ndez}, {Korobkin}, {Mandel}, {Rosswog}, \& {et
  al.}}]{2017ApJ...848L..27T}
{Tanvir}, N.~R., {Levan}, A.~J., {Gonz{\'a}lez-Fern{\'a}ndez}, C., {et~al.}
  2017, \apjl, 848, L27, \dodoi{10.3847/2041-8213/aa90b6}

\bibitem[{{Tchekhovskoy} {et~al.}(2008){Tchekhovskoy}, {McKinney}, \&
  {Narayan}}]{2008MNRAS.388..551T}
{Tchekhovskoy}, A., {McKinney}, J.~C., \& {Narayan}, R. 2008, \mnras, 388, 551,
  \dodoi{10.1111/j.1365-2966.2008.13425.x}

\bibitem[{{Thompson}(1994)}]{1994MNRAS.270..480T}
{Thompson}, C. 1994, \mnras, 270, 480, \dodoi{10.1093/mnras/270.3.480}

\bibitem[{{Thompson} {et~al.}(2004){Thompson}, {Chang}, \&
  {Quataert}}]{2004ApJ...611..380T}
{Thompson}, T.~A., {Chang}, P., \& {Quataert}, E. 2004, \apj, 611, 380,
  \dodoi{10.1086/421969}

\bibitem[{{Toma} {et~al.}(2006){Toma}, {Ioka}, {Yamazaki}, \&
  {Nakamura}}]{2006ApJ...640L.139T}
{Toma}, K., {Ioka}, K., {Yamazaki}, R., \& {Nakamura}, T. 2006, \apjl, 640,
  L139, \dodoi{10.1086/503384}

\bibitem[{{Totani}(2003)}]{2003ApJ...598.1151T}
{Totani}, T. 2003, \apj, 598, 1151, \dodoi{10.1086/378936}

\bibitem[{{Troja} {et~al.}(2019){Troja}, {Castro-Tirado}, {Becerra
  Gonz{\'a}lez}, {Hu}, {Ryan}, {Cenko}, \& {et al.}}]{2019MNRAS.489.2104T}
{Troja}, E., {Castro-Tirado}, A.~J., {Becerra Gonz{\'a}lez}, J., {et~al.} 2019,
  \mnras, 489, 2104, \dodoi{10.1093/mnras/stz2255}

\bibitem[{{Troja} {et~al.}(2017){Troja}, {Lipunov}, {Mundell}, \&
  et~al.}]{2017Natur.547..425T}
{Troja}, E., {Lipunov}, V.~M., {Mundell}, C.~G., \& et~al. 2017, \nat, 547, 425

\bibitem[{Troja {et~al.}(2017)Troja, Piro, van Eerten, \& et~al.}]{troja2017a}
Troja, E., Piro, L., van Eerten, H., \& et~al. 2017, Nature, 000, 1,
  \dodoi{10.1038/nature24290}

\bibitem[{{Troja} {et~al.}(2017){Troja}, {Piro}, {van Eerten}, {Wollaeger},
  {Im}, {Fox}, {Butler}, {Cenko}, {Sakamoto}, {Fryer}, {Ricci}, {Lien}, {Ryan},
  {Korobkin}, {Lee}, {Burgess}, {Lee}, {Watson}, {Choi}, {Covino}, {D'Avanzo},
  {Fontes}, {Gonz{\'a}lez}, {Khandrika}, {Kim}, {Kim}, {Lee}, {Lee}, {Kutyrev},
  {Lim}, {S{\'a}nchez-Ram{\'\i}rez}, {Veilleux}, {Wieringa}, \&
  {Yoon}}]{2017Natur.551...71T}
{Troja}, E., {Piro}, L., {van Eerten}, H., {et~al.} 2017, \nat, 551, 71,
  \dodoi{10.1038/nature24290}

\bibitem[{{Troja} {et~al.}(2020){Troja}, {van Eerten}, {Zhang}, {Ryan}, {Piro},
  {Ricci}, {O'Connor}, {Wieringa}, {Cenko}, \&
  {Sakamoto}}]{2020arXiv200601150T}
{Troja}, E., {van Eerten}, H., {Zhang}, B., {et~al.} 2020, arXiv e-prints,
  arXiv:2006.01150.
\newblock \doarXiv{2006.01150}

\bibitem[{{Urrutia} {et~al.}(2021){Urrutia}, {De Colle}, {Murguia-Berthier}, \&
  {Ramirez-Ruiz}}]{2021MNRAS.503.4363U}
{Urrutia}, G., {De Colle}, F., {Murguia-Berthier}, A., \& {Ramirez-Ruiz}, E.
  2021, \mnras, 503, 4363, \dodoi{10.1093/mnras/stab723}

\bibitem[{{Usov}(1992)}]{1992Natur.357..472U}
{Usov}, V.~V. 1992, \nat, 357, 472, \dodoi{10.1038/357472a0}

\bibitem[{{Valenti} {et~al.}(2008){Valenti}, {Benetti}, {Cappellaro}, {Patat},
  {Mazzali}, {Turatto}, {Hurley}, {Maeda}, {Gal-Yam}, {Foley}, {Filippenko},
  {Pastorello}, {Challis}, {Frontera}, {Harutyunyan}, {Iye}, {Kawabata},
  {Kirshner}, {Li}, {Lipkin}, {Matheson}, {Nomoto}, {Ofek}, {Ohyama}, {Pian},
  {Poznanski}, {Salvo}, {Sauer}, {Schmidt}, {Soderberg}, \&
  {Zampieri}}]{2008MNRAS.383.1485V}
{Valenti}, S., {Benetti}, S., {Cappellaro}, E., {et~al.} 2008, \mnras, 383,
  1485, \dodoi{10.1111/j.1365-2966.2007.12647.x}

\bibitem[{{van der Horst} {et~al.}(2007){van der Horst}, {Kamble}, {Wijers},
  {Resmi}, {Bhattacharya}, {Rol}, {Strom}, {Kouveliotou}, {Oosterloo}, \&
  {Ishwara-Chandra}}]{2007RSPTA.365.1241V}
{van der Horst}, A.~J., {Kamble}, A., {Wijers}, R.~A.~M.~J., {et~al.} 2007,
  Philosophical Transactions of the Royal Society of London Series A, 365,
  1241, \dodoi{10.1098/rsta.2006.1993}

\bibitem[{{van der Horst} {et~al.}(2014){van der Horst}, {Paragi}, {de Bruyn},
  {Granot}, {Kouveliotou}, {Wiersema}, {Starling}, {Curran}, {Wijers},
  {Rowlinson}, {Anderson}, {Fender}, {Yang}, \& {Strom}}]{2014MNRAS.444.3151V}
{van der Horst}, A.~J., {Paragi}, Z., {de Bruyn}, A.~G., {et~al.} 2014, \mnras,
  444, 3151, \dodoi{10.1093/mnras/stu1664}

\bibitem[{{Vaughan} {et~al.}(2006){Vaughan}, {Goad}, {Beardmore}, {O'Brien},
  {Osborne}, {Page}, {Barthelmy}, {Burrows}, {Campana}, {Cannizzo}, {Capalbi},
  {Chincarini}, {Cummings}, {Cusumano}, {Giommi}, {Godet}, {Hill}, {Kobayashi},
  {Kumar}, {La Parola}, {Levan}, {Mangano}, {M{\'e}sz{\'a}ros}, {Moretti},
  {Morris}, {Nousek}, {Pagani}, {Palmer}, {Racusin}, {Romano}, {Tagliaferri},
  {Zhang}, \& {Gehrels}}]{2006ApJ...638..920V}
{Vaughan}, S., {Goad}, M.~R., {Beardmore}, A.~P., {et~al.} 2006, \apj, 638,
  920, \dodoi{10.1086/499069}

\bibitem[{{Villasenor} {et~al.}(2005){Villasenor}, {Lamb}, {Ricker}, {Atteia},
  {Kawai}, {Butler}, {Nakagawa}, {Jernigan}, {Boer}, {Crew}, {Donaghy}, {Doty},
  {Fenimore}, {Galassi}, {Graziani}, {Hurley}, {Levine}, {Martel}, {Matsuoka},
  {Olive}, {Prigozhin}, {Sakamoto}, {Shirasaki}, {Suzuki}, {Tamagawa},
  {Vanderspek}, {Woosley}, {Yoshida}, {Braga}, {Manchanda}, {Pizzichini},
  {Takagishi}, \& {Yamauchi}}]{2005Natur.437..855V}
{Villasenor}, J.~S., {Lamb}, D.~Q., {Ricker}, G.~R., {et~al.} 2005, \nat, 437,
  855, \dodoi{10.1038/nature04213}

\bibitem[{{Wanajo} {et~al.}(2014){Wanajo}, {Sekiguchi}, {Nishimura}, {Kiuchi},
  {Kyutoku}, \& {Shibata}}]{2014ApJ...789L..39W}
{Wanajo}, S., {Sekiguchi}, Y., {Nishimura}, N., {et~al.} 2014, \apjl, 789, L39,
  \dodoi{10.1088/2041-8205/789/2/L39}

\bibitem[{{Wheeler} {et~al.}(2000){Wheeler}, {Yi}, {H{\"o}flich}, \&
  {Wang}}]{2000ApJ...537..810W}
{Wheeler}, J.~C., {Yi}, I., {H{\"o}flich}, P., \& {Wang}, L. 2000, \apj, 537,
  810, \dodoi{10.1086/309055}

\bibitem[{{Wijers} {et~al.}(1997){Wijers}, {Rees}, \&
  {Meszaros}}]{1997MNRAS.288L..51W}
{Wijers}, R.~A.~M.~J., {Rees}, M.~J., \& {Meszaros}, P. 1997, \mnras, 288, L51,
  \dodoi{10.1093/mnras/288.4.L51}

\bibitem[{{Woosley}(1993)}]{1993ApJ...405..273W}
{Woosley}, S.~E. 1993, \apj, 405, 273, \dodoi{10.1086/172359}

\bibitem[{{Woosley} \& {Bloom}(2006{\natexlab{a}})}]{Woosley2006ARA&A}
{Woosley}, S.~E., \& {Bloom}, J.~S. 2006{\natexlab{a}}, \araa, 44, 507,
  \dodoi{10.1146/annurev.astro.43.072103.150558}

\bibitem[{{Woosley} \& {Bloom}(2006{\natexlab{b}})}]{2006ARA&A..44..507W}
---. 2006{\natexlab{b}}, \araa, 44, 507,
  \dodoi{10.1146/annurev.astro.43.072103.150558}

\bibitem[{{Woosley} \& {Heger}(2012)}]{2012ApJ...752...32W}
{Woosley}, S.~E., \& {Heger}, A. 2012, \apj, 752, 32,
  \dodoi{10.1088/0004-637X/752/1/32}

\bibitem[{{Wu} {et~al.}(2013){Wu}, {Hou}, \& {Lei}}]{2013ApJ...767L..36W}
{Wu}, X.-F., {Hou}, S.-J., \& {Lei}, W.-H. 2013, \apjl, 767, L36,
  \dodoi{10.1088/2041-8205/767/2/L36}

\bibitem[{{Xiao} \& {Dai}(2019)}]{2019ApJ...878...62X}
{Xiao}, D., \& {Dai}, Z.-G. 2019, \apj, 878, 62,
  \dodoi{10.3847/1538-4357/ab12da}

\bibitem[{{Yang} {et~al.}(2015){Yang}, {Jin}, {Li}, {Covino}, {Zheng},
  {Hotokezaka}, \& {et al.}}]{2015NatCo...6.7323Y}
{Yang}, B., {Jin}, Z.-P., {Li}, X., {et~al.} 2015, Nature Communications, 6,
  7323, \dodoi{10.1038/ncomms8323}

\bibitem[{{Yi} {et~al.}(2013){Yi}, {Wu}, \& {Dai}}]{2013ApJ...776..120Y}
{Yi}, S.-X., {Wu}, X.-F., \& {Dai}, Z.-G. 2013, \apj, 776, 120,
  \dodoi{10.1088/0004-637X/776/2/120}

\bibitem[{{Zhang} {et~al.}(2006){Zhang}, {Fan}, {Dyks}, {Kobayashi},
  {M{\'e}sz{\'a}ros}, {Burrows}, {Nousek}, \& {Gehrels}}]{2006ApJ...642..354Z}
{Zhang}, B., {Fan}, Y.~Z., {Dyks}, J., {et~al.} 2006, \apj, 642, 354,
  \dodoi{10.1086/500723}

\bibitem[{{Zhang} \& {M{\'e}sz{\'a}ros}(2001)}]{2001ApJ...552L..35Z}
{Zhang}, B., \& {M{\'e}sz{\'a}ros}, P. 2001, \apjl, 552, L35,
  \dodoi{10.1086/320255}

\bibitem[{{Zhang} \& {M{\'e}sz{\'a}ros}(2002)}]{2002ApJ...566..712Z}
---. 2002, \apj, 566, 712, \dodoi{10.1086/338247}

\bibitem[{{Zhang} {et~al.}(2008){Zhang}, {Woosley}, \&
  {Heger}}]{2008ApJ...679..639Z}
{Zhang}, W., {Woosley}, S.~E., \& {Heger}, A. 2008, \apj, 679, 639,
  \dodoi{10.1086/526404}

\bibitem[{{Zhao} {et~al.}(2021){Zhao}, {Gao}, {Lei}, {Lan}, \&
  {Liu}}]{2021ApJ...906...60Z}
{Zhao}, L., {Gao}, H., {Lei}, W., {Lan}, L., \& {Liu}, L. 2021, \apj, 906, 60,
  \dodoi{10.3847/1538-4357/abc8ec}

\end{thebibliography}
\addcontentsline{toc}{chapter}{Bibliography}


\clearpage
\appendix
\section{Deceleration timescales and Sychrotron light curves}

\subsection{Different deceleration timescales}\label{DecelerationTimescales}

Due to the fact that materials ejected from the merger of two NSs and the gravitational collapse have an ample range of velocities, we consider in addition for this subsection the trans-relativistic (TR) and the Deep Newtonian (DN) regimes.

\paragraph{Trans-relativistic regime}
In this regime,   the kinetic energy of the shock in a stratified medium is given by $E=\frac{4\pi}{3}\sigma m_p \beta^2\Gamma^2\,A_{\rm k}\, r^{3-{\rm k}} $ \citep{1976PhFl...19.1130B} where the parameter $\sigma$ is a function of velocity  \citep[$\sigma=0.73-0.38\beta$;][]{1998A&A...336L..69H}.  The deceleration time becomes 
\bary\label{t_dec_trans}
t_{\rm dec, TR}&\simeq& 23.1\,{ \rm days}\,\left(\frac{1+z}{1.022}\right)^{\frac{3-k}{2+q-k}}\,A^{-\frac{1}{2+q-k}}_{\rm 0}\,E^{\frac{1}{2+q-k}}_{49}\,\beta^{-\frac{\alpha+5-k}{2+q-k}}_{-0.1}\Gamma^{-\frac{\alpha+2}{2+q-k}}.
\eary
When the bulk Lorentz factor $\Gamma \to 1$, the TR timescale approaches the Newtonian timescale.

\paragraph{Sub-relativistic (Newtonian) regime}

In this regime, the kinetic energy of the shock follows the sub-relativistic spherically-symmetric Sedov–Taylor solution. The deceleration time (from Eq. \ref{beta_dec}) becomes

\bary\label{t_dec_SubRel}
t_{\rm dec}&\simeq& 1.3\times10^3\,{\rm days} \left(\frac{1+z}{1.022}\right)^{\frac{k-3}{k-(q+2)}}\,\beta_{-0.3}^{\frac{\alpha+5-k}{k-(q+2)}}\, A_{0}^{\frac{1}{k-(q+2)}}E_{51}^{-\frac{1}{k-(q+2)}}\,.
\eary

\paragraph{Deep Newtonian regime}

In this regime, the Lorentz factor of the lowest-energy electrons is $\gamma_{\rm m}\simeq 2$ \citep{2013ApJ...778..107S, 2016MNRAS.461.1568K, 2020arXiv200413028M}.   Using eq. (\ref{gamma_dec}), the deceleration time in this regime becomes 
\bary\label{t_dec_DN}
t_{\rm dec, DN} &\simeq& 1.1\times 10^5\,{\rm days}\,  \left(\frac{1+z}{1.022}\right)^{\frac{3-k}{2+q-k}}\,\epsilon^{\frac{\alpha+5-k}{2(q+2-k)}}_{\rm e, -1}\, A^{-\frac{1}{q+2-k}}_{\rm 0}\, E_{51}^{\frac{1}{q+2-k}}\,.
\eary
In this case, the value of velocity is $\beta\approx 0.05$. The deceleration timescales in Eqs. \ref{t_dec_trans},  \ref{t_dec_SubRel} and \ref{t_dec_DN} were estimated for $q=0$ and ${\rm k=0}$ with $A_{\rm 0}=1\,{\rm cm^{-3}}$.

\subsection{Synchrotron emission}

During the deceleration phase,  the post-shock magnetic field evolves as, $B'\propto \,t^{-\frac{2(2+q)+k(1-q+\alpha)}{2(\alpha+5-k)}}$.   The Lorentz factors of the lowest-energy electrons and of the higher energy electrons, which are efficiently cooled by synchrotron emission are
{\small
\bary\label{gamma_dec}
\gamma_{\rm m}&=&\gamma^0_{\rm m}\,\left(\frac{1+z}{1.022}\right)^{\frac{2(3-k)}{\alpha+5-k}}\,g(p)\, \epsilon_{\rm e,-1}\, A^{-\frac{2}{\alpha+5-k}}_{\rm k}\,\tilde{E}_{51}^{\frac{2}{\alpha+5-k}}\, t_7^{\frac{2(k-(q+2))}{\alpha+5-k}}\cr
\gamma_{\rm c}&=&\gamma^0_{\rm c}\left(\frac{1+z}{1.022}\right)^{-\frac{k+1+\alpha(k-1)}{\alpha+5-k}}\, (1+Y)^{-1} \epsilon^{-1}_{\rm B,-2}\, A^{-\frac{\alpha+3}{\alpha+5-k}}_{\rm k}\, \tilde{E}_{51}^{\frac{k-2}{\alpha+5-k}}\, t_7^{\frac{-1+2q-k(-2+q-\alpha)-\alpha}{\alpha+5-k}}\,.
\eary
}

The corresponding synchrotron break frequencies are given by
{\small
\bary\label{nu_syn_de}
\nu_{\rm m}&=&\nu^{\rm 0}_{\rm m}\,\left(\frac{1+z}{1.022}\right)^{\frac{20+k(\alpha-6)-2\alpha }{2(\alpha+5-k)}}\,g(p)^{2} \epsilon^2_{\rm e,-1}\,\epsilon^\frac12_{\rm B,-2}\,  A^{\frac{\alpha-5}{2(\alpha+5-k)}}_{\rm k}\, \tilde{E}_{51}^{\frac{10-k}{2(\alpha+5-k)}}\,t_7^{-\frac{10(2+q)+k(-7-q+\alpha)}{2(\alpha+5-k)}}\cr
\nu_{\rm c}&=&\nu^{\rm 0}_{\rm c}\,\left(\frac{1+z}{1.022}\right)^{-\frac{8-2\alpha + k(3\alpha+2)}{2(\alpha +5-k)}}\, \epsilon^{-\frac32}_{\rm B,-2}\, (1+Y)^{-2} \, A^{-\frac{3(\alpha+3)}{2(\alpha+5-k)}}_{\rm k}\tilde{E}_{51}^{\frac{3(k-2)}{2(\alpha+5-k)}}\,t_7^{-\frac{8-6q+k(-7+3q-3\alpha)+4\alpha}{2(\alpha+5-k)}}\,.
\eary
}

In the self-absorption regime, the synchrotron break frequencies are 
{\small
\bary\label{nua_syn_de}
\nu_{\rm a,1}&=& \nu^{\rm 0}_{\rm a,1} \left(\frac{1+z}{1.022}\right)^{-\frac{55+8\alpha-k(21+4\alpha)}{5(\alpha+5-k)}} g(p)^{-1} \epsilon_{\rm e,-1}^{-1} \epsilon_{\rm B,-2}^{\frac15}\, A^{\frac{25+4\alpha}{5(\alpha+5-k)}}_{\rm k}\tilde{E}_{51}^{-\frac{4k+5}{5(\alpha+5-k)}}  t_6^{\frac{25+5q+4k(-5+q-\alpha)+3\alpha}{5(\alpha+5-k)}}\,\cr
\nu_{\rm a,2}&=& \nu^{\rm 0}_{\rm a,2} \left(\frac{1+z}{1.022}\right)^{\frac{-2p(-10+\alpha)+kp(-6+\alpha)+6k(4+\alpha)-12(5+\alpha)}{2(p+4)(\alpha+5-k)}}g(p)^{\frac{2(p-1)}{p+4}} \epsilon_{\rm e,-1}^{\frac{2(p-1)}{p+4}}\, A^{\frac{p(\alpha-5)+6(\alpha+5)}{2(p+4)(\alpha+5-k)}}_{\rm k} \epsilon_{\rm B,-2}^{\frac{p+2}{2(p+4)}}\,\tilde{E}_{51}^{\frac{10p-k(p+6)}{2(p+4)(\alpha+5-k)}}\,\cr
&&\hspace{9cm}\times t_6^{-\frac{10p(2+q)-4(5+\alpha)+k(22-6q+6\alpha+p(-7-q+\alpha))}{2(p+4)(\alpha+5-k)}},\cr
\nu_{\rm a, 3}&=& \nu^{\rm 0}_{\rm a,3} \left(\frac{1+z}{1.022}\right)^{-\frac{20+13\alpha-k(16+9\alpha)}{5(\alpha+5-k)}}\,(1+Y)\, \epsilon_{\rm B,-2}^{\frac65}\, A^{\frac{3(3\alpha+10)}{ 5(\alpha+5-k)}}_{\rm k} \tilde{E}_{51}^{\frac{15-9k}{5(\alpha+5-k)}} t_6^{\frac{10-15q+k(-20+9q-9\alpha)+8\alpha}{5(\alpha+5-k)}}.
\eary
}

The spectral peak flux density becomes
{\small
\bary\label{f_syn_de}
F^{\rm syn}_{\rm \nu,max}&=&F^{\rm syn,0}_{\rm \nu,max}\,\left(\frac{1+z}{1.022}\right)^{\frac{4(1-\alpha)+k(2+3\alpha)}{2(\alpha+5-k)}}\, \epsilon^{\frac12}_{\rm B,-2}\, d_{\rm z,26.5}^{-2}\, A^{\frac{3\alpha+7}{2(\alpha+5-k)}}_{\rm k}\,  \tilde{E}_{51}^{\frac{8-3k}{2(\alpha+5-k)}}\,t_7^{\frac{14-8q+k(-7+3q-3\alpha)+6\alpha}{2(\alpha+5-k)}}.\,\,\,\,\,
\eary
}

The quantities $\gamma^0_{\rm m}$, $\gamma^0_{\rm c}$, $\nu^{\rm 0}_{\rm m}$, $\nu^{\rm 0}_{\rm c}$ and $F^{\rm 0}_{\rm \nu,max}$ given in eqs. \ref{gamma_dec}, \ref{nu_syn_de}, \ref{nua_syn_de} and \ref{f_syn_de} are reported in Table \ref{table1} for ${\rm k}$=0, 1, 1.5, 2 and 2.5.\\

Using the synchrotron break frequencies (eq.~\ref{nu_syn_de}) and the spectral peak flux density (eq.~\ref{f_syn_de}),  the synchrotron light curve for $\nu_{\rm a,3}\leq \nu_{\rm c}\leq \nu_{\rm m}$ is
{\small
\begin{eqnarray}
\label{fc_dec}
F^{\rm syn}_{\rm \nu}\propto \cases{ 
t^{\frac{5+\alpha+k(2-q+\alpha)}{\alpha+5-k}}\, \nu^{2},\hspace{6.9cm} \nu<\nu_{\rm a,3}, \cr
t^{\frac{25-15q+2k(-7+3q-3\alpha)+11 \alpha}{3(\alpha+5-k)}}\, \nu^{\frac13},\hspace{5.3cm}  \nu_{\rm a,3}  < \nu<\nu_{\rm c}, \cr
t^{\frac{20-10q+k(-7+3q-3\alpha)+8\alpha}{4(\alpha+5-k)}}\, \nu^{-\frac{1}{2}},\hspace{5.3cm} \nu_{\rm c}<\nu<\nu_{\rm m},\,\,\,\,\, \cr
t^{-\frac{10p(2+q)-8(5+\alpha)+kp(-7-q+\alpha)+2k(7-q+\alpha)}{4(\alpha+5-k)}}\,\nu^{-\frac{p}{2}},\,\,\hspace{3.cm}   \nu_{\rm m}<\nu\,, \cr
}
\end{eqnarray}
}
for $\nu_{\rm a,1}\leq \nu_{\rm m}\leq \nu_{\rm c}$ is

{\small
\begin{eqnarray}
\label{sc_dec1}
F^{\rm syn}_{\rm \nu}\propto \cases{ 
t^{\frac{2(1+k-2q+\alpha)}{\alpha+5-k}}\, \nu^{2},\hspace{6.8cm}  \nu<\nu_{\rm a,1}, \cr
t^{\frac{31-7q+2k(-7+2q-2\alpha)+9\alpha}{3(\alpha+5-k)}}\, \nu^{\frac13},\hspace{5.2cm}  \nu_{\rm a,1}< \nu<\nu_{\rm m}, \cr
t^{-\frac{10p(2+q)+6(-8+q-2\alpha)+k(21-5q+5\alpha+p(-7-q+\alpha))}{4(\alpha+5-k)}}\, \nu^{-\frac{p-1}{2}},\hspace{1.7cm} \nu<\nu<\nu_{\rm c},\,\,\,\,\, \cr
t^{-\frac{10p(2+q)-8(5+\alpha)+kp(-7-q+\alpha)+
2k(7-q+\alpha)}{4(\alpha+5-k)}}\,\nu^{-\frac{p}{2}},\,\,\,\,\hspace{2.5cm}   \nu_{\rm c}<\nu\,, \cr
}
\end{eqnarray}
}
and for $\nu_{\rm m}\leq \nu_{\rm a, 2}\leq \nu_{\rm c}$ is
{\small
\begin{eqnarray}
\label{sc_dec2}
F^{\rm syn}_{\rm \nu}\propto \cases{
t^{\frac{2(1+k-2q+\alpha)}{\alpha+5-k}}\, \nu^{2},\hspace{5.6cm} \nu<\nu_{\rm m}, \cr
t^{\frac{28+k-6q-kq+8\alpha+k\alpha}{4(\alpha+5-k)}}\, \nu^{\frac52},\hspace{4.6cm}  \nu_{\rm m} <  \nu<\nu_{\rm a,2}, \cr
t^{-\frac{10p(2+q)+6(-8+q-2\alpha)+k(21-5q+5\alpha+p(-7-q+\alpha))}{4(\alpha+5-k)}}\, \nu^{-\frac{p-1}{2}},\hspace{0.5cm} \nu_{\rm a,2}<\nu<\nu_{\rm c},\,\,\,\,\, \cr
t^{-\frac{10p(2+q)-8(5+\alpha)+kp(-7-q+\alpha)+2k(7-q+\alpha)}{4(\alpha+5-k)}}\,\nu^{-\frac{p}{2}},\,\,\,\,\hspace{1.3cm}   \nu_{\rm c}<\nu\,, \cr
}
\end{eqnarray}
}
respectively.

\clearpage
\begin{table}
\centering \renewcommand{\arraystretch}{1.85}\addtolength{\tabcolsep}{1pt}
\caption{Quantities associated with synchrotron afterglow model with energy injection.}
\label{table1}
\begin{tabular}{l   c  c  c c c}
 \hline \hline
\scriptsize{} &  \scriptsize{${\bf k=0}$  }  &\hspace{0.5cm}   \scriptsize{${\bf k=1.0}$ } &\hspace{0.5cm}   \scriptsize{${\bf k=1.5}$ } &\hspace{0.5cm}   \scriptsize{${\bf k=2.0}$ }  &\hspace{0.5cm}   \scriptsize{${\bf k=2.5}$ }  \\ 
\hline
$A_{\rm k}$ & $1\,{\rm cm^{-3}}$ & $1.5\times 10^{19}\,{\rm cm^{-2}}$ & $2.7\times 10^{28}\,{\rm cm^{-\frac32}}$ & $3\times 10^{36}\,{\rm cm^{-1}}$  & $1.3\times 10^{45}\,{\rm cm^{-\frac12}}$ \\
\hline\hline
$\gamma^0_{\rm m}$& $1.9\times10^3$ & $1.2\times10^3$ & $1.0\times10^3$ & $2.1\times10^3$ & $2.5\times10^3$\\
$\gamma^0_{\rm c}\,(\times 10^3)$& $7.2$ & $1.3$ & $0.8$ & $6.4$& $1.3\times10$\\
$\nu^{0}_{\rm a,1}\,(\rm Hz)$& $1.5\times10^8$ & $1.2\times10^{9}$ & $2.2\times10^{9}$ & $2.0\times10^{8}$ & $1.2\times10^{8}$\\
$\nu^{0}_{\rm a,2}\,(\rm Hz)$& $6.6\times10^{9}$ & $1.8\times10^{10}$ & $2.5\times10^{10}$ & $8.8\times10^{9}$ & $6.9\times10^{9}$\\
$\nu^{0}_{\rm a,3}\,(\rm Hz)$& $8.2\times10^{7}$ & $1.1\times10^9$ & $2.5\times10^{9}$ & $1.2\times10^{8}$ & $6.0\times10^{7}$\\
$\nu^{0}_{\rm m}\,(\rm Hz)$& $3.2\times10^{11}$ & $2.9\times10^{11}$ & $2.9\times10^{11}$ & $4.2\times10^{11}$ & $4.3\times10^{11}$\\
$\nu^{0}_{\rm c}\,(\rm Hz)$&$9.8\times10^{12}$ & $3.9\times10^{11}$ & $1.5\times10^{11}$ & $7.3\times10^{12}$    & $2.8\times10^{13}$\\
$F^{0}_{\rm \nu,max}\,({\rm mJy})    $ & $8.3\times10^{5}$  &  $4.2\times10^{6}$ & $5.5\times10^{6}$ & $2.3\times10^5$       &  $5.0\times10^4$   \\


\hline \hline

\end{tabular}
\end{table}

Quantities: $\tilde{E}=10^{49}\ \mathrm{erg}$, $d_{z}\approx 100\ \mathrm{Mpc}$ ($z=0.022$), $p=2.6$, $\epsilon_{B}=10^{-3}$, $\epsilon_{e}=10^{-1}$, $\alpha=3.0$, $q=0.7$, $t=10^7\ \mathrm{sec}$



\begin{table}[h!]
\caption{PL indexes of the density parameter in each cooling condition of  the synchrotron afterglow model.}
\centering \renewcommand{\arraystretch}{1.5}\addtolength{\tabcolsep}{1.6pt}
\label{table:dens}
\begin{tabular}{cccccc}
\hline \hline
 & $k=0$ & $k=1$ & $k=1.5$ & $k=2$ & $k=2.5$  \\
\hline \hline

{\large $\nu_{\rm a,3}\leq \nu_{\rm c}\leq \nu_{\rm m}$} \\ \hline
$\nu < \nu_{\rm a,3}$ &  $-1$                 & $-\frac{\alpha+5}{\alpha+4}$                 & $-\frac{2(\alpha+5)}{2\alpha+7}$                  & $-\frac{\alpha+5}{\alpha+3}$ & $-\frac{2(\alpha+5)}{2\alpha+5}$ \\
$\nu_{\rm a,3} < \nu < \nu_{\rm c}$ &  $\frac{2\alpha+5}{\alpha+5}$                 & $\frac{2\alpha+5}{\alpha+4}$                 & $\frac{2(2\alpha+5)}{2\alpha+7}$                  & $\frac{2\alpha+5}{\alpha+3}$ & $2$ \\
$\nu_{\rm c} < \nu < \nu_{\rm m}$ &  $\frac{3\alpha+5}{4(\alpha+5)}$                 & $\frac{3\alpha+5}{4(\alpha+4)}$                 & $\frac{3\alpha+5}{2(2\alpha+7)}$                  & $\frac{3\alpha+5}{4(\alpha+3)}$ & $\frac{3\alpha+5}{2(2\alpha+5)}$ \\
$ \nu_{\rm m} < \nu $ &  $\frac{\alpha(p+2)-5(p-2)}{4(\alpha+5)}$                 & $\frac{p(\alpha-5)+2(\alpha+5)}{4(\alpha+4)}$                 & $\frac{p(\alpha-5)+2(\alpha+5)}{2(2\alpha+7)}$                  & $\frac{p(\alpha-5)+2(\alpha+5)}{4(\alpha+3)}$ & $\frac{p(\alpha-5)+2(\alpha+5)}{2(2\alpha+5)}$ \\
\hline \hline
{\large $\nu_{\rm a,1}\leq \nu_{\rm m}\leq \nu_{\rm c}$} \\ \hline
$\nu < \nu_{\rm a,1}$ &  $-\frac{4}{\alpha+5}$                 & $-\frac{4}{\alpha+4}$                 & $-\frac{8}{2\alpha+7}$                  & $-\frac{4}{\alpha+3}$ & $-\frac{8}{2\alpha+5}$ \\
$\nu_{\rm a,1} < \nu < \nu_{\rm m}$ &  $\frac{4\alpha+13}{3(\alpha+5)}$                 & $\frac{4\alpha+13}{3(\alpha+4)}$                 & $\frac{2(4\alpha+13)}{3(2\alpha+7)}$                  & $\frac{4\alpha+13}{3(\alpha+3)}$ & $\frac{2(4\alpha+13)}{3(2\alpha+5)}$ \\
$\nu_{\rm m} < \nu < \nu_{\rm c}$ &  $\frac{5\alpha+19+p(\alpha-5)}{4(\alpha+5)}$                 & $\frac{5\alpha+19+p(\alpha-5)}{4(\alpha+4)}$                 & $\frac{5\alpha+19+p(\alpha-5)}{2(2\alpha+7)}$                  & $\frac{5\alpha+19+p(\alpha-5)}{4(\alpha+3)}$ & $\frac{5\alpha+19+p(\alpha-5)}{2(2\alpha+5)}$ \\
$ \nu_{\rm c} < \nu $ &  $\frac{\alpha(p+2)-5(p-2)}{4(\alpha+5)}$                 & $\frac{p(\alpha-5)+2(\alpha+5)}{4(\alpha+4)}$                 & $\frac{5(2-p)+\alpha(p+2)}{2(2\alpha+7)}$                  & $\frac{p(\alpha-5)+2(\alpha+5)}{4(\alpha+3)}$ & $\frac{p(\alpha-5)+2(\alpha+5)}{2(2\alpha+5)}$ \\
\hline \hline
{\large $\nu_{\rm m}\leq \nu_{\rm a,2}\leq \nu_{\rm c}$} \\ \hline
$\nu < \nu_{\rm m}$ &  $-\frac{4}{\alpha+5}$                 & $-\frac{4}{\alpha+4}$                 & $-\frac{8}{2\alpha+7}$                  & $-\frac{4}{\alpha+3}$ & $-\frac{8}{2\alpha+5}$ \\
$\nu_{\rm m} < \nu < \nu_{\rm a,2}$ &  $-\frac{\alpha+11}{4(\alpha+5)}$                 & $-\frac{\alpha+11}{4(\alpha+4)}$                 & $-\frac{\alpha+11}{2(2\alpha+7)}$                  & $-\frac{\alpha+11}{4(\alpha+3)}$ & $-\frac{\alpha+11}{2(2\alpha+5)}$ \\
$\nu_{\rm a,2} < \nu < \nu_{\rm c}$ &  $\frac{19+p(\alpha-5)+5\alpha}{4(\alpha+5)}$                 & $\frac{19+p(\alpha-5)+5\alpha}{4(\alpha+4)}$                 & $\frac{19+p(\alpha-5)+5\alpha}{2(2\alpha+7)}$                  & $\frac{19+p(\alpha-5)+5\alpha}{4(\alpha+3)}$ & $\frac{19+p(\alpha-5)+5\alpha}{2(2\alpha+5)}$ \\
$ \nu_{\rm c} < \nu $ &  $\frac{\alpha(p+2)-5(p-2)}{4(\alpha+5)}$                 & $\frac{p(\alpha-5)+2(\alpha+5)}{4(\alpha+4)}$                 & $\frac{p(\alpha-5)+2(\alpha+5)}{2(2\alpha+7)}$                  & $\frac{p(\alpha-5)+2(\alpha+5)}{4(\alpha+3)}$ & $\frac{p(\alpha-5)+2(\alpha+5)}{2(2\alpha+5)}$ \\
\hline
\end{tabular}
\end{table}

\end{document}